%% file: main.tex
\documentclass[lettersize,journal,preprint]{IEEEtran}

\usepackage[table]{xcolor} % Loads xcolor once with the 'table' option
\usepackage{graphicx}
\usepackage{colortbl}
\usepackage{pgfplots}
\usepackage{pgf-pie}
\usepackage{soul}

\usepackage{xcolor}
\usepackage{etoolbox}  % for toggles (optional, for clean version later)

\usepackage{verbatim}
\usepackage{booktabs}
\usepackage{soul}
\usepackage{subcaption}
\usepackage{tabularx}
\usepackage{amsmath}
\usepackage{array}
\usepackage{mdframed}
\usepackage{dblfloatfix}
\usepackage{ragged2e}
\usepackage{adjustbox}
\usepackage{multirow}
\usepackage{subfiles}
\usepackage{xr}
\usepackage{fontawesome5}
\usepackage{algorithmic}
\usepackage{listings}

\usepackage[numbers,sort&compress]{natbib}

\usepackage[colorlinks=true,linkcolor=blue,citecolor=blue,urlcolor=blue]{hyperref}

\usepackage[breakable, skins]{tcolorbox}

\newcolumntype{M}[1]{>{\centering\arraybackslash}m{#1}}

\definecolor{naturalColor}{RGB}{8,230,0}
\definecolor{LatentColor}{RGB}{69,150,66}
\definecolor{externalColor}{RGB}{164,224,34}
\definecolor{featureColor}{RGB}{71,179,255}
\definecolor{inferenceColor}{RGB}{56,112,232}
\definecolor{applicationColor}{RGB}{107,47,247}
\definecolor{mixedColor}{RGB}{66, 227, 192}

\newlength{\taxonomyIndent}
\newcommand{\LevelOne}[1]{#1}
\newcommand{\LevelTwo}[1]{\hspace*{\taxonomyIndent}\hangindent=\taxonomyIndent #1} % Indented and allows hanging indent

\pgfplotsset{compat=1.18}
\def\BibTeX{{\rm B\kern-.05em{\sc i\kern-.025em b}\kern-.08em
    T\kern-.1667em\lower.7ex\hbox{E}\kern-.125emX}}

\usepackage{tikz}
\usetikzlibrary{trees,shadows,positioning,shapes.geometric,decorations.pathreplacing,calc,fit,arrows}

\usetikzlibrary{backgrounds}

\tcbuselibrary{skins}

\newtcolorbox{walkthroughbox}[2]{
  title={\faSearch\ #1},                    % This explicitly enables the title argument
  colback=gray!10,
  colframe=gray!75,
  fonttitle=\itshape\normalsize,
  fontupper=\small,
  breakable,
  arc=4pt,
  boxrule=1pt,
  coltitle=black,
  colbacktitle=#2
}

\begin{document}

\title{From Tea Leaves to System Maps: A Survey and Framework on Context-aware Machine Learning Monitoring}
% \title{From Tea Leaves to System Maps: Context-awareness in  Monitoring Operational Machine Learning Models}

\author{
\IEEEauthorblockN{Joran Leest\IEEEauthorrefmark{1}}
\and
\IEEEauthorblockN{Claudia Raibulet\IEEEauthorrefmark{1}\IEEEauthorrefmark{2}}
\and
\IEEEauthorblockN{Patricia Lago\IEEEauthorrefmark{1}}
\and
\IEEEauthorblockN{Ilias Gerostathopoulos\IEEEauthorrefmark{1}}
\vspace{0.3cm}\\
\IEEEauthorblockA{\IEEEauthorrefmark{1}Vrije Universiteit Amsterdam, The Netherlands\\
Email: \{j.g.leest, c.raibulet, p.lago, i.g.gerostathopoulos\}@vu.nl}\\
\IEEEauthorblockA{\IEEEauthorrefmark{2}Universita' degli Studi di Milano-Bicocca, Italy\\
Email: claudia.raibulet@unimib.it}
}

\maketitle

\begingroup
  \renewcommand\thefootnote{}
  \footnote{This work has been submitted to the IEEE for possible publication.
  Copyright may be transferred without notice, after which this version may
  no longer be accessible.}
  \addtocounter{footnote}{-1}
\endgroup

\begin{abstract}
Machine learning (ML) models in production fail when their broader systems -- from data pipelines to deployment environments -- deviate from training assumptions, not merely due to statistical anomalies in input data. Despite extensive work on data drift, data validation, and out-of-distribution detection, ML monitoring research remains largely model-centric while neglecting \textit{contextual information}: auxiliary signals about the system around the model (external factors, data pipelines, downstream applications). Incorporating this context turns statistical anomalies into actionable alerts and structured root-cause analysis. Drawing on a systematic review of 94 primary studies, we identify three dimensions of contextual information for ML monitoring: the \textit{system} element concerned (natural environment or technical infrastructure); the \textit{aspect} of that element (runtime states, structural relationships, prescriptive properties); and the \textit{representation} used (formal constructs or informal formats). This forms the \textit{Contextual System-Aspect-Representation} (C-SAR) framework, a descriptive model synthesizing our findings. We identify 20 recurring triplets across these dimensions and map them to the monitoring activities they support. This study provides a holistic perspective on ML monitoring: from interpreting ``tea leaves'' (i.e., isolated data and performance statistics) to constructing and managing ``system maps'' (i.e., end-to-end views that connect data, models, and operating context).

\end{abstract}

% \begin{IEEEkeywords}
% monitoring, software engineering, machine learning, artificial intelligence
% \end{IEEEkeywords}

\subfile{sections/Introduction.tex}

\subfile{sections/background.tex}

\subfile{sections/related_work.tex}

\subfile{sections/methodology.tex}

\subfile{sections/results.tex}

\subfile{sections/discussion.tex}

\subfile{sections/conclusion.tex}

\input{main.bbl}

% \bibliographystyle{IEEEtran}
% \bibliography{primary_studies, bibliography} 

\end{document}

%% file: sections/introduction.tex
\section{Introduction} \label{Introduction}
A Machine Learning (ML) system relies on the stability of the past -- where data is assumed to behave as during training, pipelines pass unit tests, and training metrics suggest the model performs well. 
However, it operates in a future where distributional assumptions no longer hold \cite{vela2022temporal}, high-velocity MLOps workflows drive frequent changes \cite{shankars2024we}, feedback loops amplify bias \cite{chaney2018algorithmic}, and failures often emerge beyond what evaluation once suggested was safe \cite{ebreck2017the}. Monitoring is critical to managing these risks in ML deployments, enabling detection of issues as models are exposed to real-world data. Multiple research communities have sought to automate ML monitoring -- understood here as continuous verification and validation under real operating conditions -- through techniques such as drift detection \cite{gama2014survey, lu2018, hu2020no}, performance estimation without labels \cite{garg2022leveraging, guillory2021predicting, baek2022agreement}, and data validation \cite{breck2019data, schelter2018automating, shankar2023automatic, redyuk2021automating}. These efforts identify changes in feature distributions or input characteristics to raise alerts that a model might be diverging from its expected behavior. However, most approaches remain model-centric -- monitoring only inputs, predictions, and (when available) ground-truth labels -- while largely ignoring the systems around the model.

The problem is that ML models often fail in subtle ways that cannot be properly monitored without considering the broader system context. Such context can indeed be decisive in spotting and diagnosing real model failures. This is confirmed by recent practitioner reports: \textbf{not all distribution shifts are harmful, and not all failures manifest as detectable shifts}; discerning harmful shifts from harmless ones requires observability of the broader system and understanding of the application domain \cite{shankars2024we, shergadwalamn2022a, sculley2015hidden,redyuk2021automating,bernardil2019150,amershi2019software,paleyes2022challenges}. Consider the following example: 

\textit{Page visits from younger users suddenly drop in the logs that feed the site’s churn-risk model, triggering a drift alert on the \emph{page\_visits} feature. Three explanations compete: a marketing campaign that targets older shoppers and temporarily skews traffic (harmless); a competitor’s flash sale that genuinely shifts younger users’ behaviour and breaks the model’s assumptions (requires retraining); or a silent schema change plus stale features that corrupt the input stream (pipeline repair). If the engineer relies on the statistical alert, they cannot tell whether the observed drift is (a) harmless and transient, (b) a genuine market shift, or (c) a self-inflicted data-quality bug. Only by correlating feature-store freshness metrics, ETL error logs, marketing-campaign events, and external market signals can the on-call ML- or data-quality engineer decide whether to retrain, roll back, or simply wait.}

Organizations widely employ contextual monitoring for their ML systems \cite{ebreck2017the, shankars2024we,shergadwalamn2022a,sculley2015hidden,lelwakatare2021on}, but implementation often remains ad hoc and unsystematic. When deploying new models, engineers typically create monitoring rules based on their individual judgment, such as ``raise an alert if the fraction of premium accounts changes by more than ten percent'' \cite{lelwakatare2021on, shankars2024we,ebreck2017the}. These monitoring decisions, often documented only in Slack conversations or personal notes, create knowledge silos that hinder the transfer of monitoring practices across models and teams \cite{ebreck2017the,xxu2022dependency,shankars2024we,shergadwalamn2022a}.

Through interviews and case studies, practitioners have identified the following three critical challenges in ML monitoring, each highlighting the importance of contextual information \cite{shankars2024we, shergadwalamn2022a, sculley2015hidden, muirurid2022practices, lelwakatare2021on, bernardil2019150,huyen2022}.

\textbf{Alert Fatigue:} Without domain knowledge, thresholds are poorly calibrated and two-sample tests over-alert \cite{huyen2022, shankars2024we, lelwakatare2021on}. Encoding what ``normal'' looks like reduces noise and highlights real failures.

\textbf{Root-Cause Analysis:} When alerts are triggered, engineers face the task of reconstructing system state to identify whether the cause stems from benign ''natural'' changes, critical data quality issues, or model misuse \cite{huyen2022, xxu2022dependency, shankars2024we, cavenesse2020tensorflow}. With a prior understanding of information such as the data lineage or causal relations in the domain, this time-consuming manual process can becomes more systematic \cite{xxu2022dependency}.

\textbf{Information Management:} Organizations operating multiple ML models struggle with scattered domain insights about feature constraints, seasonal patterns, and web of dependencies involving Extract-Transform-Load (ETL) processes and model entanglement across Slack threads, outdated documentation, and individual engineers' knowledge \cite{xxu2022dependency, ebreck2017the, shankars2024we, shergadwalamn2022a}. Systematically organizing contextual information -- including valid value ranges, expected distributions, known correlations, and interdependencies between ETL pipelines and models -- helps build a shared knowledge base that promotes reuse and prevents reinventing monitoring logic for each new model deployment.

Researchers have begun to highlight the potential of richer ``context-aware'' strategies, from the specification of nominal data properties \cite{rc2020overton, cavenesse2020tensorflow}, graphs that capture the causal structure of the data \cite{budhathokik2021why, schrouffj2022diagnosing}, to scenario-based models of expected distribution changes \cite{sobolewskip2017scr, leestj2024expert}. Through these efforts, prior work has explored the signals that can be extracted from the rich context of real-world deployment environments.
Yet such contextual elements remain scattered across research communities and lack a unifying framework and vocabulary to structure and organize them. They therefore remain buried within specialized research silos, appearing as narrow contributions to specific monitoring tasks like drift detection or performance estimation, while their broader significance lies in enabling holistic ML monitoring in production environments.

We believe that understanding contextual elements -- information about the broader system beyond the model itself -- is essential for developing robust and standardized monitoring practices in production ML systems. By systematically reviewing how prior works conceptualize, structure, and deploy contextual information, we aim to establish patterns that can guide more effective monitoring practices. Our analysis focuses on studies that explicitly incorporate contextual elements in ML monitoring workflows, where we define context as any auxiliary information beyond a model's features, predictions, and (target) labels.

This study yields three distinct, yet interconnected, contributions: (1) We provide a holistic mapping of ML monitoring terminology across multiple research communities, along with a systematic study selection approach for identifying relevant works. (2) We synthesize the results of the mapping study in a conceptual framework called \textit{Contextual System-Aspect-Representations} (C-SAR) that characterizes how contextual information is structured across different research domains. (3) We elicit recurring patterns from the ML monitoring literature by analyzing interactions across the dimensions defined in the C-SAR framework, capturing how context manifests across different monitoring approaches. Figure~\ref{fig:high_overview} provides an overview of the contributions.

\subfile{figures/roadmap.tex}

Together, these contributions build a deeper understanding of context-aware monitoring, giving researchers and practitioners a solid foundation for integrating contextual information into ML monitoring workflows. 

The rest of the paper is structured as follows: Section \ref{background} lays out the foundations of ML monitoring, followed by a discussion of related work in Section \ref{related_work}. After presenting our three contributions, we discuss the impact of our findings from both practical and academic perspectives in  Section \ref{discussion}, showing how the C-SAR framework can shape real-world monitoring practices and open new directions for research in context-aware ML monitoring, and conclude in Section~\ref{conclusion}.

%% file: sections/figures/roadmap.tex
\definecolor{featureColor}{RGB}{220,220,220}
\definecolor{inferenceColor}{RGB}{169,169,169}
\definecolor{applicationColor}{RGB}{128,128,128}
\definecolor{naturalColor}{RGB}{112,128,144}

\newcommand{\stackedboxes}[8]{%
  \newlength{\mygap}%
  \setlength{\mygap}{\columnwidth}%
  \addtolength{\mygap}{-#1}%
  \addtolength{\mygap}{-#2}%
  \addtolength{\mygap}{-#4}%
  \setlength{\mygap}{0.3\mygap}%
  
  \begin{figure}[htbp]
  \small
    \centering
    \hspace{-0.6em}
    \begin{tikzpicture}[
     xscale=1,
      main box/.style={
        draw,
        rectangle,
        rounded corners=5pt,
        text width=#2-4pt,  % Subtract some padding
        minimum height=#3,
        align=center
        },
        side box/.style={
            draw,
            rectangle,
            rounded corners=5pt,
            text width=#4-4pt,  % Subtract some padding
            minimum height=#5,
            align=center
        },
      ref circle/.style={
          draw,
          circle,
          minimum size=#1,
          align=center,
          inner sep=0
      },
      header/.style={
          font=\bfseries,
          align=center
      }
    ]
      % Row 1
      \node[ref circle, anchor=west, fill=featureColor!20] (c1) at (0,0) {\ref{methodology}};
      \node[main box, anchor=west, fill=featureColor!20] (b1) at ($(c1.east)+(\mygap,0)$) {\textbf{Systematic review}};
      \node[side box, anchor=west, fill=featureColor!20] (s1) at ($(b1.east)+(\mygap,0)$) {\textit{Holistic mapping of ML monitoring research across communities using a novel semantic filtering method to create an 8000+ study corpus}};
      
      % Row 2
      \node[ref circle, anchor=west, fill=inferenceColor!20] (c2) at (0,-#6) {\ref{sec:theory}};
      \node[main box, anchor=west, fill=inferenceColor!20] (b2) at ($(c2.east)+(\mygap,0)$) {\textbf{C-SAR Framework}};
      \node[side box, anchor=west, fill=inferenceColor!20] (s2) at ($(b2.east)+(\mygap,0)$) {\textit{Unified framework with taxonomies organizing context information in ML monitoring in three dimensions}};
      
      % Row 3
      \node[ref circle, anchor=west, fill=applicationColor!20] (c3) at (0,-2*#6) {\ref{sec:analysis}};
      \node[main box, anchor=west, fill=applicationColor!20] (b3) at ($(c3.east)+(\mygap,0)$) {\textbf{C-SAR interaction patterns}};
      \node[side box, anchor=west, fill=applicationColor!20] (s3) at ($(b3.east)+(\mygap,0)$) {\textit{A pattern catalog of common context information patterns used in ML monitoring literature}};

    % Row 4
      % \node[ref circle, anchor=west, fill=naturalColor!20] (c4) at (0,-3*#6) {\ref{context_aware_monitoring}};
      % \node[main box, anchor=west, fill=naturalColor!20] (b4) at ($(c4.east)+(\mygap,0)$) {\textbf{Practical guidance}};
      % \node[side box, anchor=west, fill=naturalColor!20] (s4) at ($(b4.east)+(\mygap,0)$) {\textit{Mapping of ML monitoring tasks enabled though context integration, providing practical guidance for context-aware ML monitoring}};
      
      \begin{scope}[on background layer]
          \foreach \i in {1,2,3} {
              \draw[#7, line width=#8] 
                  ($(c\i.west)+(0.2,0)$) -- ($(s\i.east)+(-0.2,0)$);
          }
      \end{scope}
      
    \end{tikzpicture}
    \caption{Study roadmap outlining contributions per section.}
    \label{fig:high_overview}
  \end{figure}
}

\stackedboxes{0.7cm}{2cm}{1.2cm}{5.2cm}{1.5cm}{1.8cm}{gray}{2pt}

%% file: sections/background.tex
\section{Background} \label{background}

We define \textit{model monitoring} as the interpretation of collected metrics to detect model failures, diagnose underlying causes, perform impact assessment, and prescribe corrective actions \cite{ebreck2017the, shankars2024we, huyen2022}. 
Broadly, we consider monitoring from two perspectives: \emph{verification} and \emph{validation}, concepts that are closely tied to ML evaluation and testing \cite{schroder2022monitoring}. 
Monitoring operates continuously during deployment, working with live data under real-world constraints, detecting issues in real time, and adhering to production resource and latency requirements \cite{shankars2024we, huyen2022}. This distinguishes it from evaluation and testing, which typically occur at discrete points, pre-deployment, under controlled scenarios \cite{cloudera2021, shankars2024we}. 

\subsection{Common Challenges in ML Monitoring}
\label{sec:monitoring_challenges}

Monitoring ML systems in production presents several challenges. Data drift is a leading cause of performance degradation. Drift may manifest itself as feature drift ($P(\mathbf{X}_{\text{old}}) \neq P(\mathbf{X}_{\text{new}})$), prediction drift ($P(\hat{Y}_{\text{old}}) \neq P(\hat{Y}_{\text{new}})$), or concept drift ($P(Y \mid \mathbf{X}_{\text{old}}) \neq P(Y \mid \mathbf{X}_{\text{new}})$) \cite{webb2016characterizing, huyen2022}. Natural drift arises from external influences such as user or market behavior, while unnatural drift stems from data integrity issues, such as schema mismatches or pipeline errors \cite{schelter2018automating, foidl2022data, shankars2024we}.

Another challenge is label feedback. Many real-world systems operate under conditions of delayed or partial feedback \cite{shankars2024we, guillory2021predicting}. Without immediate ground-truth labels, monitoring must rely on proxy signals or unsupervised strategies to infer the presence of a failure without directly observing model quality metrics \cite{shankars2022towards, garg2022leveraging}. Finally, monitoring granularity presents difficulties. Failures often occur in localized subsets of the input space, remaining invisible to global metrics unless monitoring incorporates subgroup or slice-based analyses \cite{shankars2024we, ghosha2022fair}.

\subsection{Monitoring Activities}
\label{sec:monitoring_activities}

Monitoring activities can be broadly categorized as \textit{continuous verification} or \textit{continuous validation} \cite{cloudera2021, huyen2022}. Verification checks whether the deployment environment continues to satisfy the same assumptions used during model development, often through distribution comparisons. Validation checks whether the ML model has the desired impact and achieves intended goals in real-world settings.

\subsubsection{Continuous Verification}

Verification is cause-focused monitoring that aims to detect meaningful changes in the model's input data attributable to specific ``events'' or ``instances'' underlying a model failure. Key activities include:

\textit{Drift detection} has traditionally been the core focus of ML monitoring, aiming to identify "natural" environmental changes that cause persistent model performance degradation. Rooted in data mining research \cite{gama2014survey,lu2018,webb2016characterizing,huyen2022,han2022survey}, these methods detect changes in external factors such as user behavior, market dynamics, or broader contextual influences, manifesting as feature drift, prediction drift, or concept drift. While feature and prediction drift might be benign, concept drift ($P(Y \mid \mathbf{X})$ changing) usually leads to performance degradation and is difficult to detect, as it involves changes in an unobserved, hidden variable \cite{cloudera2021,zheng2019labelless}.

\textbf{Data validation} targets data integrity issues, often referred to as unnatural drift \cite{shankars2022towards, shankars2022towards, schelter2018automating}. These issues arise from schema mismatches, missing values, or pipeline malfunctions rather than genuine changes in the environment \cite{schelter2018automating, ehrlingerl2019a}. By checking domain or schema rules, validation ensures that input data aligns with assumptions made during development. Violations may signal broken transformations, incorrect encodings, or even malicious data poisoning attacks \cite{lin2021adversarial}.

\textbf{Out-of-distribution (OOD) detection} flags individual samples that lie far outside the training data distribution \cite{jang2022sequential, olson2021contrastive}. This ensures that the model does not make overconfident predictions on unknown or rare inputs. OOD detection is traditionally studied in computer vision and representation learning, where the aim is to flag images containing concepts outside of the model's scope. It shares similarities with drift detection, but instead focuses on identifying problematic instances, rather than batches, and the cause of failure is often not attributed to ''events'', but rather to sampling bias.

\subsubsection{Continuous Validation}

Validation involves result-focused monitoring to track how well the model is performing and whether it delivers its intended real-world impact. It aims to ensure the ongoing effectiveness, fairness, and compliance with intended impact metrics -- such as KPIs -- of the deployed model. Key activities include:

\textbf{Performance monitoring} regards the traditional view of monitoring, focusing on tracking key metrics such as accuracy, recall, and other evaluation measures to detect persistent issues. This approach -- particularly common in software engineering and machine learning research -- relies on runtime metrics to assess model performance and identify potential failures \cite{sculley2015hidden,amershi2019software,paleyes2022challenges}. Beyond accuracy-based monitoring, fairness monitoring has gained attention as a way to track biases and disparate impacts between subgroups \cite{bacelar2021monitoring, ghosha2022fair, singh2021fairness}. This has led to the development of fairness metrics that measure real-time disparities and shifts in model behavior over time.

However, performance monitoring becomes challenging when ground truth labels are delayed or unavailable, making it difficult to directly evaluate model accuracy. In such cases, performance estimation methods approximate model performance under data drift -- when the data distribution shifts and ground truth labels are not immediately available \cite{shankars2022towards,garg2022leveraging, chenm2021mandoline}.

\textbf{Model validation} focuses on ensuring that the system continues to meet its desired goals, independent of quality metrics on which the model was trained. This area has been explored in autonomous systems literature, and from a general software engineering perspective, where models are considered components of larger systems acting as decision-making agents \cite{klsm2019uncertainty, binderf2022putting, myllyahol2022on, bernardil2019150}. It also includes the study of how model outputs shape future inputs, creating feedback loops of algorithmic influence on the domain \cite{chaney2018algorithmic, ensign2018runaway, sculley2015hidden}. Model validation is inherently contextual, as it involves evaluating aspects that extend beyond traditional ML metrics and instead looks at the real-world impact of the model once deployed.

%% file: sections/related_work.tex
\section{Related Work} \label{related_work}
This section examines two key areas of related work. First, we review existing literature surveys  focused on ML monitoring, highlighting the fragmented nature of current research in this space. Second, we explore works that, while not directly addressing ML monitoring, investigate the role of context in ML workflows -- a crucial aspect that informs our research direction. Through this analysis, we identify gaps in current literature regarding the integration of contextual information in ML monitoring systems. Lastly, we discuss related work on context modeling in related software engineering disciplines.

\subsection{Existing Reviews and Taxonomies of ML Monitoring}

Several reviews have addressed aspects of ML monitoring, though each focuses on specific concerns, contributing taxonomies, identifying challenges, and cataloging metrics and practices.

Schröder and Schulz present a foundational taxonomy distinguishing data-level monitoring (e.g., data quality, drift, validity) from model-level monitoring (e.g., performance, confidence) \cite{schroder2022monitoring}. They highlight core challenges such as high dimensionality, scalability, and interdependencies across system components.

Other works focus on metrics and performance monitoring. Bodor et al. review quality assurance metrics in MLOps, covering both distribution-based metrics and time-tracked model metrics like precision and recall \cite{bodor2023machine}. Karval et al. explore metrics at the intersection of monitoring and explainability, specifically targeting data drift, anomalies, and adversarial attacks \cite{karval2023catching}. Both studies address important technical aspects but offer limited discussion of contextual or systemic factors.

Several reviews take a broader MLOps perspective. Paleyes et al. synthesize challenges from real-world deployments, identifying issues like data drift and degenerate feedback loops, though monitoring is only one focus among many \cite{paleyes2022challenges}. Kreuzberger et al. combine interviews and literature to map monitoring within MLOps workflows, offering a high-level overview without detailing specific techniques \cite{kreuzberger2023machine}.

Reviews of ML testing also touch on monitoring. Zhang et al. classify monitoring as part of “online testing” in a broader testing taxonomy but devote little attention to runtime failures or monitoring-specific literature \cite{zhang2020machine}. Similarly, Chandrasekaran et al. include runtime evaluation and monitoring in their review of testing ML-enabled systems, with particular attention to handling data drift \cite{chandrasekaran2023test}.

In sum, existing reviews offer valuable but partial views of ML monitoring. They variously emphasize data and model metrics, systemic challenges, and workflow integration -- but no single study provides a comprehensive synthesis across these dimensions.

\subsection{Contextualization of ML Systems}
While no prior work directly addresses the use of context in ML monitoring, several studies inform how context is treated more broadly in ML systems.

Beckh et al. review how prior knowledge is used in explainable ML, identifying representations such as knowledge graphs, logical rules, algebraic structures, probabilistic relations, and human feedback \cite{beckh2023harnessing}. Though focused on design-time explainability, these categories overlap with context representations we identify in this study.

Chen et al. examine expert involvement across the ML lifecycle, particularly for runtime feedback and adaptation \cite{chen2023perspectives}. While monitoring is not their focus, the work suggests mechanisms for incorporating expert understanding at runtime.

Other studies emphasize the need for contextual awareness in evaluation. Garcia et al. argue that ML models cannot be meaningfully evaluated without context, particularly when developing evaluation metrics \cite{garciacontext}. Gerke et al., analyzing ML incidents in medical settings, show how neglecting the surrounding system context can lead to critical failures at runtime \cite{gerke2020need}.

Overall, there is a growing recognition of the importance of context in ML systems, yet its explicit application in the monitoring stage remains largely unexplored.

\subsection{Context Modeling in Software Engineering}  
Early work on context modelling in software systems focused on enabling context-aware behaviour through general-purpose abstractions. Dey et al. introduced the \emph{Context Toolkit}, defining context as any information that characterises the situation of an entity, and separating its acquisition, interpretation, and use through a widget–interpreter–aggregator pattern \cite{dey2001conceptual}.  
Baldauf et al.\ surveyed context-aware computing in pervasive systems, classifying context types and proposing a layered reference architecture that highlights the role of modelling and reasoning in enabling adaptive behaviour \cite{baldauf2007survey}.

Context modelling has since been specialized to several software-engineering sub-fields.
Elyasaf et al. embedded context constructs in model-driven‐engineering metamodels to link structural variability with behavioural adaptation \cite{elyasaf2022modeling}.  
Bedjeti et al. elevated environmental assumptions to first-class architectural artefacts through context description viewpoints \cite{bedjeti2017modeling}.  
Rodrigues et al. refined contextual variables with design-time data mining to increase dependability in self-adaptive systems \cite{rodrigues2019enhancing}.  
Cabrera et al. employed ontology-based models to enable cross-service reasoning and adaptation in service-oriented computing \cite{cabrera2017ontology}.

These works show that the notion of concept can be very diverse, and what is modelled as context depends on the goals and concerns of the system at hand. Our study complements this line of work by focusing on context and its modelling for ML systems, and examining how context can be systematically captured for the purpose of monitoring.

%% file: sections/methodology.tex
\section{Study Design} \label{methodology}
The following describes the research method used in this study. Our goal is to systematically identify and categorize the role of contextual information in ML monitoring.

\subsection{Research Questions}\label{research_questions}
Our research questions investigate how contextual information -- defined as any auxiliary information beyond observations of the model's features, ground-truth labels, and predictions -- is described and used in ML monitoring literature.

\subfile{figures/search_figure.tex}

\begin{quote}
\textbf{RQ1:} How is contextual information in ML monitoring described in scientific literature?
\end{quote}

This question is motivated by the need to consolidate definitions and conceptualizations of contextual information found across different studies. By synthesizing these descriptions, we aim to develop a unified vocabulary that clarifies how context is framed in ML monitoring

\begin{quote}
\textbf{RQ2:} How is contextual information used to support ML monitoring activities, according to the literature?
\end{quote}

This question examines the role of contextual information in ML monitoring by identifying recurring usage patterns and providing an actionable perspective on how context supports ML monitoring activities.

\subsection{Search Strategy} \label{sec:search_string_and_terms}
Our search strategy consists of two main stages: an exploratory search followed by a systematic search via the construction and use of a search string (Fig.~\ref{fig:search_string_and_terms}).

\subsubsection{Exploratory Search} \label{sec:exploratory_search}
ML monitoring spans various research communities with a diverse vocabulary in describing ML monitoring related aspects. In our study, we aim for high coverage of these aspects to mitigate the threat of limited scope (if we 
relied only on contemporary terminology like ``model monitoring''). To do so, we map key terminology of relevant literature on ML monitoring. 

We began the search with a set of influential (highly cited) works on operational ML \cite{sculley2015hidden, paleyes2022challenges, shankars2024we, amershi2019software, ebreck2017the, bernardil2019150}, spanning several communities (including machine learning, software engineering, database management systems, and human-computer interaction), and works focused on ML monitoring \cite{schroder2022monitoring, karval2023catching, bodor2023machine, bacelar2021monitoring, shergadwalamn2022a}. Building on this set, the search process consisted of the following steps:

\begin{enumerate}
    \item We analyzed the initial set of papers -- involving full text reading and selective backward snowballing -- to identify research interests related to ML monitoring activities (e.g., model monitoring, data validation, performance estimation) and ML failures (e.g., concept drift, out-of-distribution data, algorithmic influence, fairness).
    \item Using the identified terms and concepts, we conducted targeted and broad searches on Google Scholar.
    \item We iteratively refined our search terms, considering saturation reached when new searches consistently led to previously identified works and the same themes were repeatedly encountered across different sources -- attempts to broaden the search scope did not yield new insights.
\end{enumerate}

Through this process we identified relevant literature as: \textit{technical studies}, proposing or implementing specific ML monitoring techniques, and \textit{in-practice}, discussing ML monitoring in holistic, production-oriented contexts. Moreover, the exploratory process yielded: (1) a set of key terms and related keywords that appeared in the titles and abstracts of relevant ML monitoring literature (Fig. \ref{fig:key_terms}); (2) a collection of 27 (pilot) studies selected on explicit discussions of context in ML monitoring. 

The key terms follow from RQ1 (\textit{ML} and \textit{Monitoring}) and the two primary study categories as identified from the pilot studies: \textit{technical studies} (\textit{ML Failures}) and \textit{in-practice} (\textit{ML Operations}). The related keywords for the key terms were selected for high coverage of common ML monitoring vocabulary -- validated on the pilot studies' titles and abstracts. We did this in an iterative manner, balancing keyword relevance with the number of search results returned. The result was a structured taxonomy of ML monitoring vocabulary. Note that this taxonomy is general to ML monitoring -- does not include \textit{context}, the core focus of this study. We did identify context-related keyword groups from our pilot study, in particular: (1) operational aspects of deployed ML monitoring, (2) human involvement in monitoring, and (3) contextual modeling techniques. 
% To ensure validity, we deliberately excluded (2) and (3) from the taxonomy -- and our subsequent search in Sec. \ref{sec:search_string} -- opting for broad coverage of ML monitoring literature and avoiding premature categorization.

The pilot studies were further used to refine the search string (Sec. \ref{sec:search_string}), inform study selection criteria (Sec. \ref{sec:study_select}), and pilot the synthesis (Sec. \ref{sec:synth}).

\subsubsection{Search String} \label{sec:search_string}
Informed by our exploratory search, we developed a search string that captures a wide set of relevant terminology of ML monitoring literature, as three main components (Fig. \ref{fig:search_string}).

\textbf{Activity-focused Monitoring}: Focuses on studies explicitly discussing ML monitoring activities -- controlled phrases of specific research interests ($\bullet$ $\circ$ $\circ$) -- in conjunction with broad ML related terminology ($\bullet$ $\bullet$). This component aims to catch technical studies without further conditioning, allowing for high recall of studies within this space.
    
\textbf{Operations-focused Monitoring}: Captures broader terminology of monitoring ($\circ$ $\bullet$ $\circ$), conditioned on common phrases in operational ML related literature, designed for high coverage of ML monitoring related aspects and concerns to catch \textit{in-practice}. We utilized a word proximity operator (NEAR) to include common phrasing variations phrases, like ``ML-enabled systems'', `operationalizing ML'', and `ML deployment'', while avoiding irrelevant matches. 
    
\textbf{Failure-focused Monitoring}: Targets technical studies addressing monitoring with a narrow focus on specific ML failures. This component captures the full range of monitoring terminology ($\circ$ $\bullet$ $\bullet$) in proximity of specific ML failures. % mention threat here; without conditining many irrelevatn studies.
Here, we used a more relaxed proximity constraint to catch studies that mention a monitoring-related keyword within six words of mentioning an ML failure, allowing for more varied phrasing while maintaining relevance.

This structure serves multiple purposes: (1) ensure coverage across the multidisciplinary field, with each component capturing studies others might miss, (2) balance precision and recall, with the \textit{activity-focused} component providing high precision while the \textit{operations-focused} component ensures high recall, and (3) allow for the capture of in-depth discussions of specific monitoring challenges through the \textit{failure-focused} component.

To evaluate the effectiveness of our search strategy, we tested it against our set of 27 pilot studies. The results, visualized in the Venn diagram in Fig. \ref{fig:search_string_and_terms}, show both unique contributions and overlap between the components.

\subsubsection{Threats to Validity in Search Strategy} \label{sec:threats_to_validity_search}
We identified and mitigated several potential threats to the validity of the search strategy.

\textbf{Misalignment of the search process.}
A threat to study validity is whether our search process accurately captured the intended construct -- contextual information in ML monitoring. To mitigate this threat, we validated our search string and database selection against the pre-identified set of 27 pilot studies drawn from a variety of research communities. The successful retrieval of this set provided confidence that our search process was well-aligned with the construct under investigation.
 
\textbf{Bias from context-specific keywords}. As mentioned in Sec.~\ref{sec:exploratory_search}, while examining the pilot studies, we identified three context-related keyword groups: (1) operational aspects of deployed ML monitoring, (2) human involvement in monitoring, and (3) contextual modeling techniques. To ensure validity, we deliberately excluded (2) and (3) from the search query to mitigate the threat of premature categorization.

\textbf{Coverage limitations}: The \textit{activity-focused} and \textit{failure-focused} components might not exhaustively cover all relevant terms. We addressed this threat by: (1) conducting an independent mapping to ensure comprehensive coverage of monitoring practices and concerns across various communities, instead of relying on existing reviews with inconsistent coverage discussed in Sec.~\ref{related_work}; (2) rigor in the exploratory search process, involving iterative refinement, saturation-based termination, and cross-disciplinary investigation detailed in Sec.~\ref{sec:exploratory_search}; (3) including the \textit{operational monitoring} component in our search string as a holistic component.

\subsection{Study Selection} \label{sec:study_select}
We developed a set of selection criteria and a pipeline to identify relevant primary studies from our initial search results. 

\subsubsection{Selection Criteria}
Study selection follows the criteria outlined in Table~\ref{tab:select_criteria}. A primary study must meet at least one relevance criterion (\textit{RC-1} targeting relevant \textit{in-practice} and \textit{RC-2} targeting relevant \textit{technical studies}), all inclusion criteria, and none exclusion criterion. The application of these criteria is further discussed in Sec. \ref{sec:select_pipe}.

\subfile{figures/filtering_fig.tex}

\subfile{figures/select_criteria.tex}

\subsubsection{Selection Pipeline} \label{sec:select_pipe}
We identified relevant primary studies from our initial search results through a systematic selection process (Fig. \ref{fig:study_select} outlines the selection pipeline). This process followed a pilot-driven approach, validating exclusions by ensuring the inclusion of pilot studies.

\subfile{figures/study_select2.tex}

\setcounter{figure}{4}

\textbf{Database search}. We applied the search string to three databases (retrieved on 24th of July 2024): Scopus, IEEEXplore, and ACM Digital Library. These databases were selected based on: (1) coverage of pilot studies; (2) coverage of high ranking conferences in identified relevant fields of: machine learning, software engineering, database management systems, data mining, security, and human-computer interaction. We chose not to include preprint servers (e.g. arXiv), maintaining quality by only including peer-reviewed work (\textit{IC-1}). We recognize that this may limit the inclusion of recent works.

\textbf{Pre-processing}. First, duplicate studies across databases are identified and removed based on DOI uniqueness and title similarity (first step applying \textit{EC-2}). Furthermore, through manual screening and n-gram analysis we identified common phrases and textual patterns of irrelevant works, and developed exclusion rules of those irrelevant works, focusing on enforcing \textit{EC-1}, \textit{EC-3}, and \textit{EC-4}. 

\textbf{Semantic filter.} To further reduce the corpus prior to manual review, we developed a semantic filtering approach, similar to semantic search methods used within information retrieval \cite{bordawekar2017using} and retrieval-augmented generation \cite{cuconasu2024power}, to remove studies unrelated to ML monitoring (\textit{EC-1}). Fig. \ref{fig:filtering_process} provides a detailed overview.

We used the \textit{text-embedding-3-large} model from the OpenAI API \cite{openai} to embed the concatenated titles and abstracts (metadata) of all studies. Open-source alternatives were considered, but their token limit (512) proved too restrictive for our needs, and the alternative, \textit{BGE-M3} \cite{chen2024bge}, was less effective at filtering relevant papers.

\textbf{Metadata screening}. The titles of the remaining studies were screened -- most were excluded at this point -- followed by the screening of their abstracts. We applied all IC and \textit{EC-1} to \textit{EC-4} during screening and checked for fit with either \textit{RC-1} or \textit{RC-2}. 
A relaxed form of \textit{RC-1} was used, where an explicit mention of context in the metadata was not required for inclusion. The selection criteria applied at this stage were conservative to ensure manageability. We acknowledge the limitation of potentially missing relevant studies, particularly \textit{technical studies} where context is not central to the research.

\textbf{Full text read} involved two stages. First, we screened the full text of studies (those that passed the metadata screening) to verify they met at least one of the relevance criteria (\textit{RC-1} or \textit{RC-2}), and all inclusion (\textit{IC-1} to \textit{IC-3}) and exclusion (\textit{EC-1} to \textit{EC-5}) criteria. Then, the remaining studies were fully read and data were extracted (discussed in Sec. \ref{sec:synth}). 
% Ambiguous cases were discussed with all authors.

\subsubsection{Threats to Validity in Study Selection}

\textbf{Semantic filter accuracy}. While our semantic filtering approach improved efficiency, it risked excluding relevant studies. We mitigated this by: (1) tuning the filter to use a diverse set of known relevant studies, (2) manually reviewing a sample of filtered-out studies to assess false negatives, and (3) maintaining a lower (forgiving) threshold for inclusion.

\subsection{Extraction and Synthesis} \label{sec:synth}
While we aimed to identify general patterns in how contextual information is used through hierarchical categorization, we acknowledged that papers often present multiple independent context information usages that cannot be reduced to a single theme. Therefore, we extracted at the level of individual context usage traces.
% while using more general categories where possible.

For each usage of contextual information, we extracted the type of contextual information, or \textit{contextual aspect}, used, its \textit{representation} format, and the \textit{system components} it describes or connects to (RQ1). Additionally, we extract the \textit{monitoring activity} for which the information is used. Our data extraction protocol was developed iteratively through pilot studies. 
% The first author conducted the extraction, with regular discussions of extracted codes and specific instances within the research team to ensure consistent interpretation.

We applied an open coding approach to extract key elements from the primary studies -- adding new categories when results cannot be accurately coded using existing ones, while maintaining mutual exclusivity.

All data and code used for the database search, study selection, and synthesis are available in our replication package\footnote{https://edu.nl/uwua7}.

%% file: sections/figures/search_figure.tex
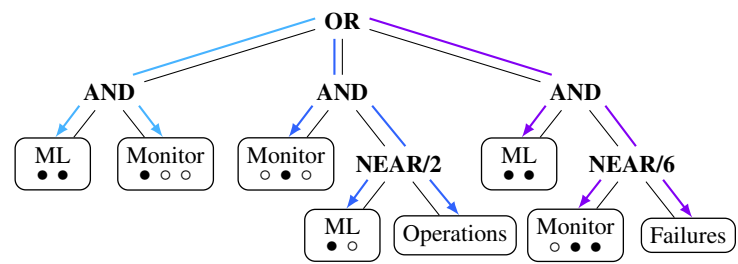
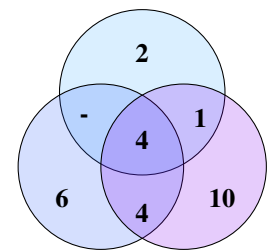
\begin{figure*}[!b]
\centering

% Define colors
\definecolor{contextcolor}{RGB}{173, 216, 230}
\definecolor{operationalcolor}{RGB}{2, 145, 217}
\definecolor{concernscolor}{RGB}{130, 0, 243}
\definecolor{excolor}{RGB}{100, 100, 170}
\definecolor{modelcolor}{RGB}{125, 135, 165}

\definecolor{path1c}{RGB}{71, 179, 255}
\definecolor{path2c}{RGB}{48, 99, 252}
\definecolor{path3c}{RGB}{130, 0, 243}

\begin{subfigure}{\textwidth}
\centering
\small

\begin{tabularx}{\linewidth}{
    >{\columncolor{white}[\tabcolsep]}p{1.78cm}
    @{}p{0.65cm}|
    >{\raggedright\arraybackslash}X|
    >{\raggedright\arraybackslash}p{3.4cm}|
    >{\raggedright\arraybackslash}p{3.6cm}
}
\toprule[1pt]
\textbf{Key term} & \multicolumn{1}{c|}{} & \textbf{Keywords} & \textbf{Description} & \textbf{Occurrence in pilot papers} \\
\midrule
\textit{ML}
 & $\bullet$ $\circ$ & machine learning, artificial intelligence, ML, AI
   & Core ML concepts & All papers except \cite{borchanih2015modeling,masegosaar2020analyzing,pwelinder2013a} \\
\cmidrule{2-5}
 & $\circ$ $\bullet$ & learning
   & General learning concept & \cite{borchanih2015modeling,masegosaar2020analyzing,pwelinder2013a} \\
\midrule[1pt]
\textit{Monitoring}
 & $\bullet$ $\circ$ $\circ$ & model monitoring, model validation, data validation, performance monitoring, performance estimat*
   & Activity-focused monitoring terms & \cite{bachingerf2024data,ebreck2017the,chenm2021mandoline,fengj2024designing,ghosha2022fair,lelwakatare2021on,shergadwalamn2022a} \\
\cmidrule{2-5}
 & \vspace{-0.35em} $\circ$ $\bullet$ $\circ$ & monitor*, validat*, verif*, root-cause, observability, quality assurance, audit*
   & Core monitoring activities and related tasks & \cite{amarco2023out,bachingerf2024data,bernardil2019150,ebreck2017the,chenm2021mandoline,cobbo2022contextaware,ehrlingerl2019a,fengj2024designing,galewis2022augur,ghosha2022fair,heynhm2023automotive,jayalathh2022enhancing,leestj2024expert,lelwakatare2021on,onnesa2023bayesian,paleyesandreiandlawrenceneilda,shankars2024we,shergadwalamn2022a,zhoux2019a} \\
\cmidrule{2-5}
 & \vspace{-0.35em} $\circ$ $\circ$ $\bullet$ & detect*, diagnos*, alert*, metric, explain*, interpret*, track*, address, mitigate, attribut*, debug*, defens*
   & Task-specific monitoring terminology & \cite{amarco2023out,bachingerf2024data,borchanih2015modeling,ebreck2017the,budhathokik2021why,chenm2021mandoline,cobbo2022contextaware,ehrlingerl2019a,fengj2024designing,galewis2022augur,ghosha2022fair,heynhm2023automotive,jayalathh2022enhancing,jbgomes2014mining,leestj2024expert,lelwakatare2021on,masegosaar2020analyzing,onnesa2023bayesian,paleyesandreiandlawrenceneilda,sackerman2021machine,schrouffj2022diagnosing,shankars2024we,shergadwalamn2022a,zhangh2023why,zhoux2019a} \\
\midrule[1pt]
%\rowcolor{operationalcolor!15}
\textit{Operations} & & operation*, deploy*, system, pipeline, project
   & Coverage of MLOps and ML systems vocabulary & \cite{bernardil2019150,ebreck2017the,chenm2021mandoline,cobbo2022contextaware,fengj2024designing,galewis2022augur,ghosha2022fair,heynhm2023automotive,jayalathh2022enhancing,jbgomes2014mining,leestj2024expert,lelwakatare2021on,onnesa2023bayesian,paleyesandreiandlawrenceneilda,sackerman2021machine,schrouffj2022diagnosing,shankars2024we,shergadwalamn2022a,zhangh2023why,zhoux2019a} \\
\midrule[1pt]
%\rowcolor{concernscolor!10}
\textit{Failures} & 
 & \small drift; data/label (quality, integrity, error);
model (failure, fault, error, health, bias, quality); performance (deterioration, degradation, decay, drop); (concept, distribution, covariate, data, dataset, model, label) shift/change;
out-of-(distribution, domain); OOD; stale*; responsible AI; fairness; feedback loop; sampling bias; performativity; algorithmic influence; spurious correlat*;
(adversarial, evasion, poisoning) attack;
   & Specific ML monitoring failures, challenges, and data-related issues & 
\cite{amarco2023out,bachingerf2024data,borchanih2015modeling,ebreck2017the,budhathokik2021why,chenm2021mandoline,cobbo2022contextaware,ehrlingerl2019a,fengj2024designing,galewis2022augur,ghosha2022fair,heynhm2023automotive,jayalathh2022enhancing,jbgomes2014mining,leestj2024expert,lelwakatare2021on,masegosaar2020analyzing,paleyesandreiandlawrenceneilda,sackerman2021machine,schrouffj2022diagnosing,shankars2024we,shergadwalamn2022a,zhangh2023why,zhoux2019a}
 \\
\bottomrule[1pt]
\end{tabularx}

\caption{Results of exploratory analysis: taxonomy of key terms and keywords in ML monitoring literature. (*) denotes wildcards.}
\label{fig:key_terms}

\vspace{0.3em}
\hrule height 0.4pt width \textwidth

\end{subfigure}

\hfill

\vspace{-0.8em}
\centering
\begin{subfigure}{0.54\textwidth}
\centering
\begin{tikzpicture}[
    path1/.style={path1c, line width=0.8pt},
    path2/.style={path2c, line width=0.8pt},
    path11/.style={path1c, ->, >=latex, line width=0.8pt},
    path22/.style={path2c, ->, >=latex, line width=0.8pt},
    path3/.style={path3c, line width=0.8pt},
    path33/.style={path3c, ->, >=latex, line width=0.8pt},
    scale=1, 
    level distance=0.95cm,
    level 1/.style={sibling distance=3.1cm},
    level 2/.style={sibling distance=1.5cm},
    level 3/.style={sibling distance=1.5cm},
    box2/.style = {draw, black, rounded corners, line width=0.5pt, minimum width=1cm, minimum height=0.3cm, align=center, font=\small},
    box1/.style = {draw, black, rounded corners, line width=0.5pt, minimum width=1cm, minimum height=0.3cm, align=center, font=\small},
    operator/.style = {font=\small\bfseries}
]
\node[operator] (or1) {OR}
    child {
        node[operator] (and1) {AND}
        child {
            node[box2] (ML2) {ML\\[-0.75ex]$\bullet$ $\bullet$}
        }
        child {
            node[box2] (monitoring1) {Monitor\\[-0.75ex]$\bullet$ $\circ$ $\circ$}
        }
    }
    child {
        node[operator] (and2) {AND}
        child {
            node[box2] (monitoring2) {Monitor\\[-0.75ex]$\circ$ $\bullet$ $\circ$}
        }
        child {
            node[operator] (near1) {NEAR/2}
            child {
                node[box2] (ML1) {ML\\[-0.75ex]$\bullet$ $\circ$}
            }
            child {
                node[box1] (system) {Operations}
            }
        }
    }
    child {
        node[operator] (and3) {AND}
        child {
            node[box2] (ML2_2) {ML\\[-0.75ex]$\bullet$ $\bullet$}
        }
        child {
            node[operator] (near2) {NEAR/6}
            child {
                node[box2] (monitoring3) {Monitor\\[-0.75ex]$\circ$ $\bullet$ $\bullet$}
            }
            child {
                node[box1] (concern) {Failures}
            }
        }
    };

\draw[path1] ([xshift=0.0em, yshift=0.1em]or1.west) -- ([xshift=0.9em, yshift=0em]and1.north);
\draw[path11] ([xshift=0.2em, yshift=-0.5em]and1.west) -- ([xshift=0.1em, yshift=0em]ML2.north);
\draw[path11] ([xshift=-0.2em, yshift=-0.5em]and1.east) -- ([xshift=-0.1em, yshift=0em]monitoring1.north);

\draw[path2] ([xshift=-0.3em, yshift=-0.0em]or1.south) -- ([xshift=-0.3em, yshift=0em]and2.north);
\draw[path22] ([xshift=0.2em, yshift=-0.5em]and2.west) -- ([xshift=0.1em, yshift=0em]monitoring2.north);
\draw[path2] ([xshift=-0.2em, yshift=-0.5em]and2.east) -- ([xshift=0.25em, yshift=0em]near1.north);
\draw[path22] ([xshift=0.8em, yshift=-0.6em]near1.west) -- ([xshift=0.15em, yshift=0em]ML1.north);
\draw[path22] ([xshift=-0.8em, yshift=-0.6em]near1.east) -- ([xshift=0.15em, yshift=0em]system.north);

\draw[path3] ([xshift=0.0em, yshift=0.1em]or1.east) -- ([xshift=-0.9em, yshift=0em]and3.north);
\draw[path33] ([xshift=0.2em, yshift=-0.5em]and3.west) -- ([xshift=0.1em, yshift=0em]ML2_2.north);
\draw[path3] ([xshift=-0.2em, yshift=-0.5em]and3.east) -- ([xshift=0.15em, yshift=0em]near2.north);
\draw[path33] ([xshift=0.8em, yshift=-0.6em]near2.west) -- ([xshift=0.1em, yshift=0em]monitoring3.north);
\draw[path33] ([xshift=-0.8em, yshift=-0.6em]near2.east) -- ([xshift=0.15em, yshift=0em]concern.north);

\end{tikzpicture}

\caption{Systematic search logic: 
\textcolor{path1c}{Activity-focused Monitoring} (high precision), 
\textcolor{path2c}{Operations-focused Monitoring} (high coverage), and 
\textcolor{path3c}{Failure-focused Monitoring} (high depth)}
\label{fig:search_string}
\vspace{0.3em}
\hrule height 0.4pt width \textwidth
\end{subfigure}
\hfill
\vrule width 0.4pt height 4.4cm
\hfill
\begin{subfigure}{0.4\textwidth}

\begin{minipage}{0.75\textwidth}
    \centering
    \begin{tabular}{lll}
        \tikz\draw[fill=path1c,fill opacity=0.2] (0,0) circle (0.3em); {\small\selectfont Activity-focused} &
        \tikz\draw[fill=path2c,fill opacity=0.2] (0,0) circle (0.3em); {\small\selectfont Operations} &
        \tikz\draw[fill=path3c,fill opacity=0.2] (0,0) circle (0.3em); {\small\selectfont Failure-focused}
    \end{tabular}
\end{minipage}

    \vspace{1em} 
    \centering
    \begin{tikzpicture}[scale=0.55, set/.style={draw, fill opacity=0.2}]
        \begin{scope}
            \draw[set, fill=path1c] (0,0.8) circle (2cm);
            \draw[set, fill=path2c] (-1,-1) circle (2cm);
            \draw[set, fill=path3c] (1,-1) circle (2cm);

            \node[font=\bfseries] at (0,5em) {2};
            \node[font=\bfseries] at (-5.5em,-5em) {6};
            \node[font=\bfseries] at (5.5em,-5em) {10};
            
            \node[font=\bfseries] at (0,-1em) {4};
            \node[font=\bfseries] at (-4em,0.5em) {-};
            \node[font=\bfseries] at (4em,0.5em) {1};
            \node[font=\bfseries] at (0,-6em) {4};
        \end{scope}
    \end{tikzpicture}
    \caption{Distribution of pilot papers across search themes}
    \vspace{0.3em}
\hrule height 0.4pt width \textwidth
\end{subfigure}

\caption{Search strategy development based on keyword analysis of ML monitoring literature.}
\label{fig:search_string_and_terms}
\end{figure*}

%% file: sections/figures/filtering_fig.tex
\setcounter{figure}{3}
\begin{figure*}[!b]
    \definecolor{darkgreen}{RGB}{0,130,0}

    \centering
    \begin{minipage}[t]{0.485\textwidth}
        \centering\textbf{(a) Analysis of semantic similarity and cluster patterns.}
        \vspace{0.5em}
        \hrule height 0.5pt
        \vspace{0.3em}
        
        \begin{minipage}[t]{\textwidth}
            \centering
            \normalsize\textbf{Target description}
            
            \normalsize{\textit{``The core focus of this paper is \{\textcolor{black}{\textbf{monitoring keywords*}}\} of issues in machine learning including \{\textcolor{black}{\textbf{ML concern keywords*}}\} in operational, deployment, or real-world scenarios''}} 

            \vspace{0.3em}
            
            \includegraphics[width=\textwidth]{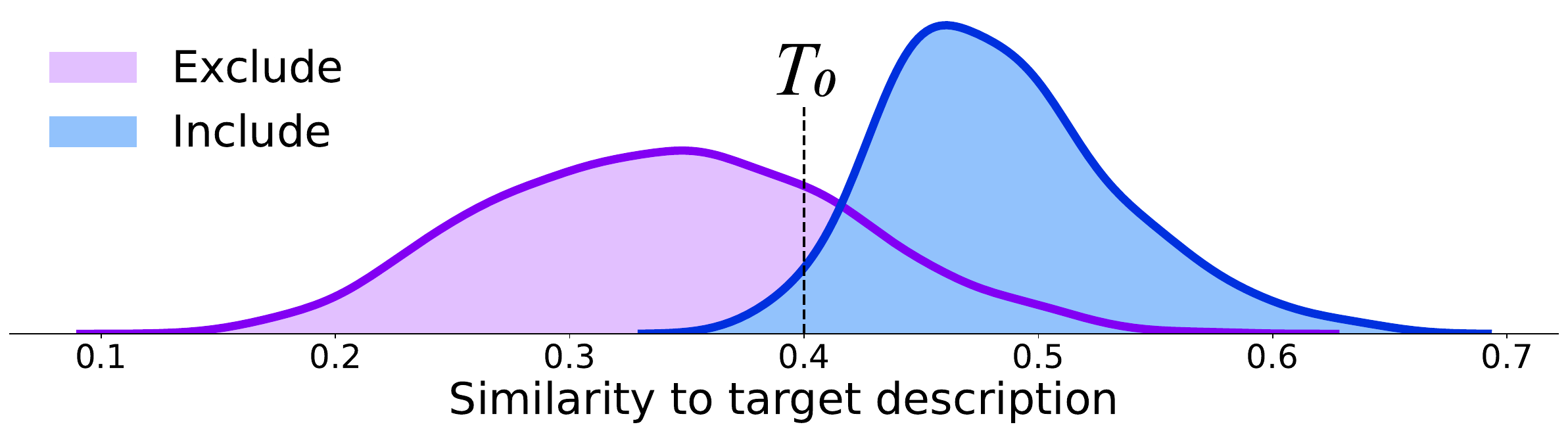}
            \raggedright\normalsize\textbf{(I)} \justifying{We derived a target description from our search string, and manually annotated 1,500 papers ($D_{test}$) on their relevance to ML monitoring (\textit{EC-1}). We observe a separation boundary between the similarity score distributions of included and excluded papers around $T_0$, with minimal loss of relevant papers.}
        \end{minipage}
        
        \vspace{0.5em}
        \hrule height 0.5pt
        \vspace{0.5em}
        
        \begin{minipage}[t]{\textwidth}
            \centering
            \includegraphics[width={1\textwidth}]{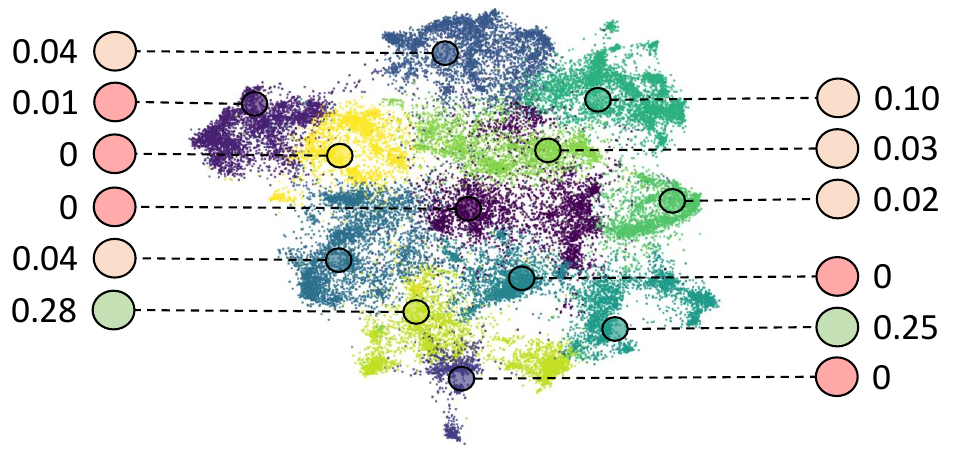}
            \raggedright\normalsize\textbf{(II)} \justifying{The clustering shows distinct groups of papers with similar semantic content. Some clusters show a high relevance on $P(i)_{test}$; manual inspection of samples drawn from the clusters yielded a similar conclusion.}
        \end{minipage}
        
        \vspace{0.5em}
        \hrule height 0.5pt
    \end{minipage}%
    \hfill
    \vrule width 0.5pt
    \hfill
    \begin{minipage}[t]{0.485\textwidth}
    \centering\textbf{(b) The study filtering workflow based on an adaptive threshold over description similarity.}
    \vspace{0.5em}
    \hrule height 0.5pt
    \vspace{0.5em}
    
        \begin{minipage}[t]{\textwidth}
            \centering
            \[
            \text{category}(c) = 
            \begin{cases} 
                \textcolor{red}{\text{low}} & \text{if } P(i)_{test} < 0.02 \\
                \textcolor{orange}{\text{mod}} & \text{if } 0.02 \le P(i)_{test} < 0.1 \\
                \textcolor{darkgreen}{\text{high}} & \text{if } P(i)_{test} \geq 0.1
            \end{cases}
            \]
            \raggedright\normalsize\textbf{(I)} \justifying{We categorize clusters as low, moderate, or high (relevance) based on their inclusion proportion $P(i)_{test}$.}
        \end{minipage}
        
        \vspace{0.5em}
    \hrule height 0.5pt
    \vspace{0.2em}
    
    \begin{minipage}[t]{\textwidth}
        \begin{center}
        \begin{tikzpicture}[scale=0.9]
            \normalsize
            \draw[gray, ->, thick] (-0.5,0.6) -- (8,0.6) node[right]{};

            \draw[black, dashed, line width=1pt] (3.75,0.6) -- (3.75,2.2);
            \node[above] at (3.75,2) {$T^{\textcolor{orange}{\textbf{mod}}}_p=T_0$};

            \draw[black, dashed, ->, line width=1pt] (3.95,1) -- (8,1);
            \node[above] at (6,1.1) {$T^{\textcolor{red}{\textbf{low}}}_p = T_0 + \theta \cdot (1-d)$};

            \draw[black, dashed, ->, line width=1pt] (3.55,1) -- (-0.5,1);
            \node[above] at (1.5,1.1) {$T^{\textcolor{darkgreen}{\textbf{high}}}_p = T_0 - \theta \cdot (1-d)$};
            
            \node[below] at (3.75,0.5) {Similarity to target description};
        \end{tikzpicture}
        \end{center}
        \normalsize
        \raggedright\normalsize\textbf{(II)} \justifying{We adjust the threshold for each paper based on cluster relevance and its position within the cluster. $T_0$ is our base threshold, $\theta$ is the maximum adjustment we apply, and $d$ is the paper's normalized distance from its cluster centroid. We define $T^c_{p}$ as the adjusted threshold for paper $p$ in cluster $c$.}
    \end{minipage}
    
    \vspace{0.5em}
    \hrule height 0.5pt
    \vspace{-0.5em}
    \hspace{2cm}
    \begin{minipage}[t]{\textwidth}
    \hspace{1cm}
    \begin{minipage}[t]{0.45\textwidth}
        \vspace{0.5em} 
        \centering     
        \[
        \hspace{2.2em}
        \text{Filter} = 
        \begin{cases} 
            \text{Include} & \text{if } s(p) \geq T_{p} \\
            \text{Exclude} & \text{otherwise}
        \end{cases}
        \]
        \vspace{0.3em}
        \newline
        \centering
        \hfill
    \end{minipage}
    \hfill
    \vspace{0.25em} 

    \raggedright\normalsize\textbf{(III)} \justifying{We set the base threshold $T_0 = .4$ and maximum adjustment $\theta = .1$. For each paper $p$, we calculate its threshold $T_p$ using the formula in (II). The decision d to include or exclude a paper is then based on comparing its similarity $s(p)$ to $T_p$. Applying the filter on $D_{test}$ resulted in a recall of 0.99.}
\end{minipage}
    
    \vspace{0.7em}
    \hrule height 0.5pt
\end{minipage}
    \vspace{-0.5em}
    
    \caption{Paper filtering and selection process: Our analysis reveals the distribution of paper similarities and their clustering patterns. Based on these insights, we develop an adaptive thresholding method.}
    \label{fig:filtering_process}
\end{figure*}

%% file: sections/figures/select_criteria.tex
\begin{table}[htb]
\small
\begin{tabularx}{\columnwidth}{lX}
\toprule
ID & Criterion \\
\midrule
\textcolor{black}{\textit{RC-1}} & The paper is discussing the role of context in ML monitoring within operational settings \\
\textcolor{black}{\textit{RC-2}} & The paper is proposing or implementing one or more ML monitoring techniques that utilize contextual information \\
\midrule
\textcolor{black}{\textit{IC-1}} & The paper is a peer-reviewed journal or conference article \\
\textcolor{black}{\textit{IC-2}} & The paper is published between 2009 and 2024 \\
\textcolor{black}{\textit{IC-3}} & The paper is written in English \\
\midrule
\textcolor{black}{\textit{EC-1}} & The paper is not focused on ML monitoring \\ 
\textcolor{black}{\textit{EC-2}} & The paper is a duplicate or extension of an included study \\
\textcolor{black}{\textit{EC-3}} & The paper is a secondary or tertiary study \\ 
\textcolor{black}{\textit{EC-4}} & The paper is a non-research publication (e.g., editorial) \\ 
\textcolor{black}{\textit{EC-5}} & The paper's full text is not available \\

\bottomrule
\end{tabularx}
\caption{\textit{Relevance} (\textcolor{black}{\textit{RC}}), \textit{inclusion} (\textcolor{black}{\textit{IC}}), and \textit{exclusion} (\textcolor{black}{\textit{EC}}) criteria in the study selection process}
\label{tab:select_criteria}
\end{table}

%% file: sections/figures/study_select2.tex
\setcounter{figure}{2}
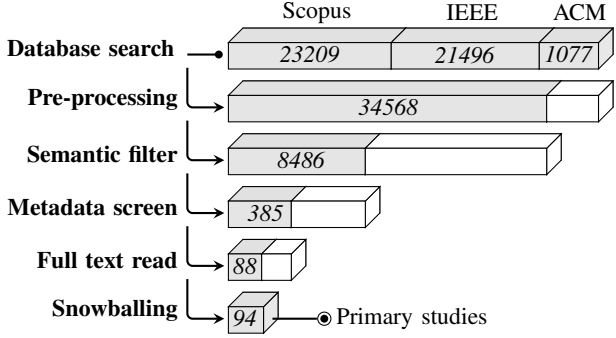
\begin{figure}
\normalsize
\setlength{\abovecaptionskip}{0.6em} 
\centering
\begin{tikzpicture}[scale=0.5, y=20, x=28]
\newcommand{\cuboid}[8][0]{
    \begin{scope}[shift={(#2,#3,#4)}]
        \draw[fill=#8] (0,0,0) -- (#7,0,0) -- (#7,1,0) -- (0,1,0) -- cycle;
        \draw[fill=#8!70] (#7,0,0) -- (#7,1,0) -- (#7,1,#5) -- (#7,0,#5) -- cycle;
        \draw[fill=#8!90] (0,1,0) -- (#7,1,0) -- (#7,1,#5) -- (0,1,#5) -- cycle;
        \draw[fill=#8!80] (0,0,#5) -- (#7,0,#5) -- (#7,1,#5) -- (0,1,#5) -- cycle;
        \ifx\relax#6\relax\else
            \node[left] at (0,0.5,{#5+#1}) {\textbf{#6}};
        \fi
    \end{scope}
}
\definecolor{keepcolor}{RGB}{222, 222, 222}
\definecolor{removecolor}{RGB}{255, 194, 194}
\definecolor{prevremovecolor}{RGB}{255, 255, 255}
\small

\node at (2.0,-1.4) {Scopus};
\node at (6.2,-1.4) {IEEE};
\node at (9.1,-1.4) {ACM};

\cuboid[0.5]{0}{-3}{0}{1}{}{4.4}{keepcolor}
\cuboid[0.5]{4.4}{-3}{0}{1}{}{4}{keepcolor}
\cuboid[0.5]{8.4}{-3}{0}{1}{}{1.6}{keepcolor}

\node at (1.8,-3) {\textit{23209}};
\node at (6.0,-3) {\textit{21496}};
\node at (8.8,-3) {\textit{1077}};

% \cuboid[0.5]{0}{-5}{0}{1}{}{9.4}{keepcolor}
% \cuboid[0.5]{9.4}{-5}{0}{1}{}{0.6}{prevremovecolor}

\cuboid[0.5]{0}{-5}{0}{1}{}{8.6}{keepcolor}
\cuboid[0.5]{8.6}{-5}{0}{1}{}{1.4}{prevremovecolor}

\cuboid[0.5]{0}{-7}{0}{1}{}{3.7}{keepcolor}
\cuboid[0.5]{3.7}{-7}{0}{1}{}{4.9}{prevremovecolor}

\cuboid[0.5]{0}{-9}{0}{1}{}{1.7}{keepcolor}
\cuboid[0.5]{1.7}{-9}{0}{1}{}{2}{prevremovecolor}

\cuboid[0.5]{0}{-11}{0}{1}{}{0.9}{keepcolor}
\cuboid[0.5]{0.9}{-11}{0}{1}{}{0.8}{prevremovecolor}

\cuboid[0.5]{0}{-13}{0}{1}{}{0.95}{keepcolor}

% \node[right] at (3.2,-5) {\textit{40631}};
\node[right] at (2.9,-5) {\textit{34568}};
\node[right] at (0.7,-7) {\textit{8486}};
\node[right] at (-0.1,-9) {\textit{385}};
\node[right] at (-0.5,-11) {\textit{88}};
\node[right] at (-0.5,-13) {\textit{94}};

\foreach \y/\text in {
    -5/Pre-processing,
    % -7/Syntactic filter,
    -7/Semantic filter,
    -9/Metadata screen,
    -11/Full text read,
    -13/Snowballing
} {
    \draw[-stealth, thick, black, rounded corners=2pt] (-1.5,\y+1.5) |- (-0.5,\y);
    \node[above, text width=2.498cm, align=right, anchor=south] at (-4.3, \y+-0.4) {\textbf{\text}};
}

\draw[thick] (-1.5,-3) -- (-0.6,-3);
\fill (-0.65,-3) circle (3pt);

\node[above, text width=2.413cm, align=right, anchor=south] at (-4.3, -3.4) {\textbf{Database search}};
\draw[thick] (0.8,-13) -- (2,-13);

\fill (2.2,-13) circle (3pt);
\draw (2.2,-13) circle (5pt);
\node[right] at (2.3,-13) {Primary studies};

\end{tikzpicture}
\caption{Study selection process, showing the number of studies remaining after each stage.}
\label{fig:study_select}
\end{figure}

%% file: sections/results.tex
\section{Study Results -- The Contextual System-Aspect-Representation Framework} \label{sec:theory}

Monitoring operational ML models inherently requires understanding more than just the inputs, outputs, and internal parameters of the models themselves. External conditions -- such as how data is generated, how models are used in downstream applications, and what constraints the domain imposes -- are all part of a broader \emph{context} that influences whether an ML model performs reliably over time. Although ``context'' can mean many things in practice, we find that contextual information in ML monitoring consistently involves identifiable system elements, their key aspects, and specific ways to represent these properties. To systematically describe how various forms of context appear and can be structured for use in ML monitoring, we propose a unifying C-SAR (Contextual System–Aspect–Representation) framework, which consolidates our understanding of contextual information across research communities. In this framework, a piece of context is formally described by three complementary dimensions:

\begin{enumerate}
    \item \textbf{System} (\(\mathbf{S}\)) -- The ``where'' of the analysis: the specific element over which some information is described. In many cases, a \emph{System} element can simply be a data variable, but it can also be an underlying natural data-generating functions or computational function within the ML pipeline. We group these elements into two high-level categories: \textbf{Natural system}, encompassing real-world data sources, external influences, and latent factors that affect features or targets; and the \textbf{Technical system}, covering all computational components from data-processing functions to downstream policies that govern how predictions are turned into outcomes.
    \item \textbf{Aspect} (\(\mathbf{A}\)) -- The ``what'' of the analysis: specifies what type of information is captured about a system element. This includes \emph{runtime states} (dynamic conditions during operation), \emph{structural} (the composition and interconnections of system elements), and \emph{prescriptive properties} (semantics and norms the system should adhere to).
    \item \textbf{Representation} (\(\mathbf{R}\)) -- The ``how'' of the analysis: describes the way we encode this information, whether as a \emph{numeric measurement}, \emph{logical rule}, \emph{probabilistic model}, \emph{textual description}, or other format.
\end{enumerate}

\subfile{figures/system_overview}

Putting these three dimensions together, we treat each (singular) usage of context in ML monitoring literature as a \((S,A,R)\) triplet. For example, if domain experts note a causal dependency between two features and a target, the \emph{System} is the underlying data-generating function between the variables, the \emph{Aspect} is the causal relation, and the \emph{Representation} can be a graphical diagram or an informal textual document. Alternatively, if we enforce ``\emph{age} must be below 100'', then the \emph{System} is the \emph{age} feature itself, the \emph{Aspect} is the nominal property of that feature, and the \emph{Representation} can be a logical predicate or probability distribution.

This section presents the taxonomies of the individual dimensions of the C-SAR framework, starting with the \textbf{system} dimension.

We illustrate the concepts introduced in the framework by using a running example based on a \emph{churn-risk prediction} case. To ensure consistency and gradually build familiarity as new concepts and properties of the framework are introduced, we selected a synthetic case that allows every element to be demonstrated within a single, cohesive scenario. While synthetic, the case reflects domain-specific details informed by academic literature \cite{panimalar2025miss, barsotti2024decade} and practitioner reports \cite{churn1,churn2,churn3,churn4}. The example centers on an e-commerce platform that predicts user churn risk based on behavioral signals, such as page visits and session activity. Model predictions trigger retention actions like discounts or targeted messaging. Failures in model behavior may stem from external events (e.g., a marketing campaign shifting user demographics), or internal issues (e.g., stale features due to pipeline bugs). This example serves as a thread throughout the paper to illustrate the different types of contextual information and how it supports monitoring decisions, with each instance marked by the \faSearch\ icon.

\subsection{Contextual Systems in ML Monitoring} \label{contextual_system}

ML models do not exist in isolation; they are embedded in and continuously interact with interdependent systems. Each unit of context is therefore tied to a specific system, which we characterize through its constituent data and process elements -- the essential elements that capture both the quantities and the data-generating processes within the system. Figure~\ref{fig:system_diagram} provides an overview of these systems -- their decomposition and inter-linkage as a structure that emerged from the primary studies -- and the distribution of how frequently they are referenced across the primary studies. Note that this overview does not necessarily represent the system as the context that is directly monitored, as one might intuitively think; rather, it represents the broader subject that can be characterized to support monitoring tasks, including (static) properties or (dynamic) observations. Specifically, this structure emerged from extracting the specific data and process elements described across the ML monitoring literature.

Two systems consistently surfaced as distinct concerns across the primary studies: the \textbf{natural system}, representing the real-world random variables and processes that generate data, and the \textbf{technical system}, representing the environment in which the ML model is deployed and executed. Each system is further decomposed into subsystems that define the scope and boundaries of their respective data elements and processes.

\subsubsection{Natural System}
The \textbf{natural system} represents the real-world environment, encapsulating the \emph{reference domain}, \emph{latent influences}, and \emph{exogenous influences} subsystems. 

The \textbf{reference domain} as the most frequently studied subsystem, encompassing the random variables modeled by ML systems -- namely, features and targets. Contextual information on this domain is commonly used in ML monitoring, often in the form of prior domain knowledge about the real-world processes. Studies frequently incorporate information about these variables themselves -- such as their distributions and semantic constraints~\cite{azonoozi2016contrack,barkerm2023feedbacklogs,cavenesse2020tensorflow,drevesm2020from,ehrlingerl2019a,fengj2024designing,ghosha2022fair,henrikssonj2023outofdistributi,henzingerthomasandkarimimahyar,lelwakatare2021on,liu2020towards,myllyahol2022on,pichlerg2024on,rc2020overton,sackerman2021machine,schelter2018automating,shankars2024we,swami2020data,zhoux2019a,henzingerta2023monitoring,fedelea2024the,mammanh2024biastrap}, as well as the relationships between them and the underlying data-generating processes they reflect~\cite{bachingerf2024data,bontempellia2022humanintheloop,budhathokik2021why,chenl2022estimating,dchen2022twostage,dreyfuspa2022databased,ehrlingerl2019a,fengj2024designing,henrikssonj2023outofdistributi,lelwakatare2021on,liu2020towards,pichlerg2024on,sasthana2021ml,schelter2018automating,schrouffj2022diagnosing,swami2020data,zhangh2023why}.

The \textbf{latent influences} subsystem represents hidden factors that affect the reference domain while remaining unmeasurable directly; it includes confounding factors that contribute to concept drift. This subsystem operates through two mechanisms: \emph{latent variables} that represent unobserved upstream factors that drive distributional changes in the \emph{reference domain}~\cite{apaul2024mlops,borchanih2015modeling,dreyfuspa2022databased,feng2022clinical,galewis2022augur,haidert2021domain,henzingerta2023monitoring,leestj2024expert,masegosaar2020analyzing,xuz2023alertiger}, and \emph{latent processes} that determine how these variables affect the \emph{reference domain}~\cite{apaul2024mlops,borchanih2015modeling,chenm2021mandoline,dreyfuspa2022databased,feng2022clinical,feng2024monitoring,galewis2022augur,godaup2023deployment,haidert2021domain,leestj2024expert,pichlerg2024on,pwelinder2013a,sibliniw2020master,sobolewskip2017scr,xuanjunyuandlujieandzhangguang}. While these influences cannot be directly observed, acknowledging and characterizing their existence helps to understand potential sources of drift in the \textbf{reference domain}.

The \textbf{exogenous influences} subsystem represents factors that influence the \emph{reference domain} in a unidirectional manner -- they affect the system but are not affected by it, following the principle of exogeneity from causal inference theory \cite{pearl2010introduction}. For example, while feature and target data in the reference domain may be affected by weather conditions or regulatory changes, these external factors evolve independently. This subsystem is composed of two elements: through \emph{external variables} that represent the data elements that has some influence on the reference domain influence~\cite{bensalems2024continuous,castelnovoa2021towards,cobbo2022contextaware,gomesjb2010calds,jbgomes2014mining,khoshravanazara2023the,kirchheimk2023towards,kkirchheim2024outofdistributio,klsm2019uncertainty,langfordma2023modalas,manerikerp2023online,schrouffj2022diagnosing,torfahh2022learning,vasudevans2020lift,yangz2021biasrv,zhoux2019a}, and through \emph{exogenous processes} that formalize both how these variables might interact and their specific mechanisms of influence on the \emph{reference domain}~\cite{fengj2024designing,kirchheimk2023towards,kkirchheim2024outofdistributio,klsm2019uncertainty,zhoux2019a}. 
% These exogenous factors are observable, originate upstream of features and targets, and lie outside the model’s representational scope.

\subsubsection{Technical system}
The \textbf{technical system} represents the system responsible for the deployment and execution of the ML model, encapsulating the \emph{processing pipeline}, \emph{inference pipeline}, and \emph{application}. In the \emph{technical system}, processes represent intentional processes implemented within the system, and data elements represent the whole lineage from data ingestion to actions.

The \textbf{Processing pipeline} traces the lineage of data from its origins to model-ready input data. Specifically, this subsystem involves three elements: (1) the \emph{source data} that represent the raw and unprocessed data often stored in relational tables~\cite{foidlh2019riskbased,muirurid2022practices,paleyesandreiandlawrenceneilda}; (2) the chain of \emph{transformation processes} that transform the raw data through specified functions ~\cite{ebreck2017the,foidlh2019riskbased,heynhm2023automotive,namakimh2020vamsa,nguyenmt2024novel,paleyesandreiandlawrenceneilda,sculley2015hidden,shankars2022towards,swami2020data,schelters2021jenga,schelters2020learning}; (3) resulting in a feature vector ready for ingestion \cite{ehrlingerl2019a, myllyahol2022on, lelwakatare2021on, cavenesse2020tensorflow,schelter2018automating, swami2020data}. These pipelines are often circular, with auxiliary ML model predictions feeding back as inputs to feature generation, creating model-to-model dependencies ~\cite{paleyesandreiandlawrenceneilda,heynhm2023automotive}.

The \textbf{inference pipeline} represents the deployment of \emph{ML models} and execution of \emph{predictions}. Studies in this area follow two main approaches: some focus on auxiliary information about the \emph{monitored ML model} and its predictions, such as metadata and non-functional qualities ~\cite{amarco2023out,csun2024ai,ebreck2017the,tzoppi2021detect,wangs2021a,xuz2023alertiger}, while others consider information of more complex inference architectures with multiple \emph{auxiliary ML models} operating independently for various applications within an organization or collaboratively for a single application \cite{baquon2022concept,bernardil2019150,binderf2022putting,ginartaa2022mldemon,guann2022fila,habdelkader2024mlonrails,heynhm2023automotive,kangdanielandguibasjohnandbail,myllyahol2022on,sculley2015hidden,vandervorstf2024claims,vishwakarmah2024taming,xuz2023alertiger,xxu2022dependency}.

The \textbf{application} translates model predictions into real-world decisions through \emph{policies} -- rules and protocols that define how predictions should be interpreted and used \cite{allenb2021evaluation,bensalems2024continuous,bernardil2019150,ebreck2017the,fengj2024designing,langfordma2023modalas,onnesa2022monitoring,onnesa2023bayesian,paleyesandreiandlawrenceneilda}. These policies formalize the decision-making logic that determines appropriate actions based on model predictions, whether through human or automated decision making. The execution of these processes produces \emph{actions} that drive the real-world impact of the ML system \cite{bensalems2024continuous,binderf2022putting,cabreraa2021discovering,csun2024ai,edsnascimento2019understanding,habdelkader2024mlonrails,hanafimf2024machineassisted,jayalathh2022enhancing,jayalathh2023continual,langfordma2023modalas,paleyesandreiandlawrenceneilda,sculley2015hidden,shergadwalamn2022a}.

\subsection{Contextual Aspects in ML Monitoring}\label{contextual_aspects}

An aspect represents a type of information we capture about a system or part thereof. From analyzing the literature, we identify three core aspects: state (runtime information)), structure (compositional relationships), and properties (prescriptive characteristics). Table~\ref{fig:context_taxonomy_updated} provides an overview of these aspects and their subcategories.

\subfile{figures/context_info_taxonomy.tex}

\subsubsection{State aspects} They are dynamic aspects that describe the runtime system conditions of a deployed model as it runs in production, and the overall system’s behavior over time.

\textbf{Conditions} are frequently considered in the literature, where environmental factors support the analysis and understanding of the condition in which the ML model operates 
\cite{bensalems2024continuous,jbgomes2014mining,kirchheimk2023towards,klsm2019uncertainty,cobbo2022contextaware,gomesjb2010calds,kkirchheim2024outofdistributio,torfahh2022learning,langfordma2023modalas,zhoux2019a}. Incorporating such knowledge allows the monitor to correlate environmental shifts with changes in the input distribution. In addition, studies discuss operational factors -- like code changes, resource constraints, or information on the context in which the ML model's predictions are used \cite{baquon2022concept,sculley2015hidden,xxu2022dependency,onnesa2023bayesian,heynhm2023automotive,xuz2023alertiger,swami2020data,muirurid2022practices,tzoppi2021detect,fengj2024designing,allenb2021evaluation,wangs2021a,nguyenmt2024novel,ebreck2017the,foidlh2019riskbased,paleyesandreiandlawrenceneilda}.

\textbf{Evaluations} refer to any runtime information that provides feedback on the system’s behavior. They are often used to monitor the actual impact of the ML deployment -- whether predictions actively contribute to strategic goals or generate tangible business value \cite{bensalems2024continuous,csun2024ai,edsnascimento2019understanding,binderf2022putting,langfordma2023modalas,bernardil2019150,hanafimf2024machineassisted,ebreck2017the,shergadwalamn2022a,paleyesandreiandlawrenceneilda}. Additionally, studies incorporate various human-in-the-loop practices, where domain experts and end-users provide feedback on the operational system -- offering validation at intermediate and final points of the ML workflow, respectively \cite{vandervorstf2024claims,jayalathh2022enhancing,ginartaa2022mldemon,binderf2022putting,jayalathh2023continual,vishwakarmah2024taming,guann2022fila,cabreraa2021discovering}.

\subsubsection{Structure aspects} They highlight how information on the underlying entities and relations in the monitored system enables targeted monitoring and a deeper understanding of where and why issues arise.

\textbf{Entities} capture the natural segmentation of the application domain into distinct units that exhibit fundamentally different characteristics and behaviors; they represent individual data-generating processes that need to be considered on a case-by-case basis. These divisions reflect how data in the domain generates differently across different types of actors and objects -- from demographic groups whose inherent differences shape how they interact with the system \cite{castelnovoa2021towards,drevesm2020from,fedelea2024the,fengj2024designing,ghosha2022fair,henzingerta2023monitoring,henzingerthomasandkarimimahyar,khoshravanazara2023the,mammanh2024biastrap,manerikerp2023online,schrouffj2022diagnosing,swami2020data,vasudevans2020lift,yangz2021biasrv}, to key actors that represent the critical segments to which some priority is assigned \cite{azonoozi2016contrack,drevesm2020from,rc2020overton,sackerman2021machine,schelter2018automating,swami2020data,zhoux2019a}.

\textbf{Relations} capture the inherent connections and dependencies between elements in the application domain. These connections form a directional flow and come in three forms: causal relations that express how elements of the domain influence each other \cite{schrouffj2022diagnosing,feng2022clinical,borchanih2015modeling,dreyfuspa2022databased,zhangh2023why,fengj2024designing,budhathokik2021why}, functional relations that describe how computational elements transform and depend on each other \cite{baquon2022concept,xxu2022dependency,shankars2022towards,heynhm2023automotive,ebreck2017the,namakimh2020vamsa,paleyesandreiandlawrenceneilda}, and semantic relations that capture the meaningful associations between elements, structuring the elements in an ontological sense \cite{sasthana2021ml,bontempellia2022humanintheloop,kkirchheim2024outofdistributio}. Together, these relations form the interconnected structure through which changes propagate in the domain.

\subsubsection{Properties} The information that describes any (behavioral) characteristic of a system. This information is typically used prescriptively to ensure that data and models align with expected behaviors, anticipated events, and established norms.

\textbf{Nominal properties} represent the inherent constraints and patterns that exist in the domain regardless of any specific implementation or goal. These natural laws define what observations are possible and typical. We see studies considering these properties as distribution properties of observed data (variables) -- like the mean, variance or maximum value \cite{cavenesse2020tensorflow,ehrlingerl2019a,lelwakatare2021on,liu2020towards,myllyahol2022on,pichlerg2024on,schelter2018automating,shankars2024we,swami2020data}, or as dependency properties that define how the observed data should respond as a function of other variables \cite{bachingerf2024data,chenl2022estimating,dchen2022twostage,ehrlingerl2019a,kirchheimk2023towards,lelwakatare2021on,liu2020towards,pichlerg2024on,schelter2018automating}.

\textbf{Event properties} focus on how the data behaves during predictable or a priori known events. The characterization of events is often considered important in the literature as they represent the underlying cause of observed changes in the application domain; they are interpretable conceptual units that drive failure modes and can be reasoned about using prior knowledge. One of these characteristics considered important in the literature is the post-event distributions, which describes how the data changes following an expected event; it can be understood as a time-dependent subspace within the nominal data space \cite{feng2024monitoring,sobolewskip2017scr,godaup2023deployment,leestj2024expert,pwelinder2013a,pichlerg2024on,chenm2021mandoline,schelters2021jenga,sibliniw2020master,xuanjunyuandlujieandzhangguang,zhoux2019a,schelters2020learning}. Some works focus on specifying information on the transition properties -- the temporal aspects that describe changes between stable data distributions, such as the slope, duration, or recurrence of an expected event \cite{xuz2023alertiger,masegosaar2020analyzing,borchanih2015modeling,dreyfuspa2022databased,henzingerta2023monitoring}. Event properties can be seen as synonymous to ``drift'' properties, as understood in data mining literature.

\subfile{figures/representation_taxonomy}

\textbf{Normative properties} define the boundaries of intended operation -- not what the system naturally does, but what we want it to do. Foremost, these properties establish the normative conditions under which the system is designed to operate -- known as the operational design domain (ODD) in autonomous system literature \cite{henrikssonj2023outofdistributi,barkerm2023feedbacklogs,klsm2019uncertainty}. It can be seen as a normative subspace that constrains the system’s behavior within acceptable and designed boundaries. Furthermore, some works define norms as the behaviors that the ML model should exhibit to comply with predefined requirements -- ensuring consistency in the intended behavior of the model \cite{myllyahol2022on,sculley2015hidden,bensalems2024continuous,onnesa2023bayesian,amarco2023out,csun2024ai,langfordma2023modalas,onnesa2022monitoring,habdelkader2024mlonrails,kangdanielandguibasjohnandbail}.

\subsection{Context Representations in ML Monitoring} \label{contextual_representations}

A representation defines the format in which context information is explicitly encoded within the ML monitoring workflow. Our analysis reveals two initial categories: formal representations that leverage mathematical and logical constructs for precise specification, and informal representations that capture contextual information in more flexible formats. Table~\ref{fig:representation_taxonomy} provides an overview of these representation types and their usage across the primary studies.

\subsubsection{Formal Representations}
Formal representations include precise formats through well-defined mathematical constructs.

\textbf{Numerical} representations encode context as quantitative measurements, appearing in two primary forms. Measures collected during runtime serves as a direct measurement of contextual information~\cite{zhoux2019a,klsm2019uncertainty,castelnovoa2021towards,binderf2022putting,torfahh2022learning,khoshravanazara2023the,kirchheimk2023towards,henzingerta2023monitoring,bensalems2024continuous,langfordma2023modalas,vandervorstf2024claims,fedelea2024the,allenb2021evaluation,tzoppi2021detect,wangs2021a,jbgomes2014mining,kkirchheim2024outofdistributio,gomesjb2010calds,vasudevans2020lift,yangz2021biasrv,xxu2022dependency,guann2022fila,paleyesandreiandlawrenceneilda,manerikerp2023online,mammanh2024biastrap,onnesa2023bayesian,ginartaa2022mldemon,cobbo2022contextaware,vishwakarmah2024taming,fengj2024designing,schrouffj2022diagnosing}. Additionally, pre-computed metrics derived from data through metric functions computed by other sources provide higher-level numerical representations~\cite{binderf2022putting,ebreck2017the,baquon2022concept,edsnascimento2019understanding,muirurid2022practices,csun2024ai,bernardil2019150,xuz2023alertiger,hanafimf2024machineassisted,nguyenmt2024novel,shergadwalamn2022a,sculley2015hidden}.

\textbf{Probabilistic} representations capture uncertainty and variability in contextual information through statistical modeling approaches. The most basic form specifies parameters of statistical distributions directly from domain knowledge~\cite{sibliniw2020master,henzingerta2023monitoring,borchanih2015modeling,sobolewskip2017scr,schelters2020learning,galewis2022augur,masegosaar2020analyzing,schelters2021jenga,pichlerg2024on}. A more fine-grained probabilistic representation is to use Bayesian priors that not only specify distribution parameters but also quantify uncertainty in these parameters through structured elicitation of domain expertise~\cite{godaup2023deployment,amarco2023out,pwelinder2013a,leestj2024expert,onnesa2023bayesian,feng2024monitoring}.

\textbf{Logical} representations use formal logic to define rules and constraints within the monitoring process, often expressed through first-order predicates that specify formal conditions~\cite{zhoux2019a,klsm2019uncertainty,ehrlingerl2019a,henrikssonj2023outofdistributi,kirchheimk2023towards,bensalems2024continuous,langfordma2023modalas,myllyahol2022on,sackerman2021machine,azonoozi2016contrack,lelwakatare2021on,csun2024ai,cavenesse2020tensorflow,drevesm2020from,xuanjunyuandlujieandzhangguang,ghosha2022fair,xuz2023alertiger,henzingerthomasandkarimimahyar,kangdanielandguibasjohnandbail,shankars2024we,habdelkader2024mlonrails,chenm2021mandoline,rc2020overton,fengj2024designing,swami2020data,schelter2018automating,liu2020towards,sculley2015hidden,heynhm2023automotive,shankars2022towards,namakimh2020vamsa} and boolean propositions that capture logical assignments or fixed conditions~\cite{dchen2022twostage,bachingerf2024data,onnesa2022monitoring,chenl2022estimating,fedelea2024the, henzingerta2023monitoring,mammanh2024biastrap}.

\subsubsection{Informal Representations}
Informal representations emphasize human interpretability and flexibility in capturing contextual information.

\textbf{Graphical} representations leverage graph structures -- commonly implemented as directed acyclic graphs (DAGs) -- to visualize relationships and dependencies in the monitored system~\cite{borchanih2015modeling,dreyfuspa2022databased,paleyesandreiandlawrenceneilda,fengj2024designing,zhangh2023why,schrouffj2022diagnosing,budhathokik2021why,feng2022clinical,bontempellia2022humanintheloop,sasthana2021ml,kkirchheim2024outofdistributio}.

\textbf{Semi-structured} representations encode context in partially structured forms that lack a fixed schema, allowing for flexibility in how information is organized and expressed. While free-form text captures contextual knowledge from domain experts and users~\cite{dreyfuspa2022databased,ebreck2017the,heynhm2023automotive,apaul2024mlops,cabreraa2021discovering,jayalathh2022enhancing,xuz2023alertiger,barkerm2023feedbacklogs,leestj2024expert,jayalathh2023continual,haidert2021domain,liu2020towards,feng2022clinical}, metadata -- such as pipeline configurations, logs, or versioning data -- embed context in a contextual, source-dependent structure ~\cite{foidlh2019riskbased,swami2020data,baquon2022concept,xxu2022dependency}. Unlike formal representations, which are standardized and machine-interpretable across settings, semi-structured representations often require human interpretation to extract their contextual relevance.

\subfile{figures/3d_matrix}

\section{Study Results -- C-SAR Interaction Patterns} \label{sec:analysis}

With the C-SAR framework established in the preceding section (see Section \ref{sec:theory}), we now examine how the \emph{system} (see Section \ref{contextual_system}), \emph{aspect} (see Section \ref{contextual_aspects}), and \emph{representation} (see Section \ref{contextual_representations}) dimensions interact across the ML monitoring literature. Our objective is to elicit and organize patterns of common C-SAR triplets, which represent the atomic units of contextual information in ML monitoring -- describing how aspects are represented and associated with different parts of monitored systems. In section \ref{csar_patterns}, we present the resulting catalog of patterns, and in Section \ref{csar_activities} we analyze at their mappings to monitoring activities.

\subsection{A Catalog of C-SAR Triplet Patterns} \label{csar_patterns}

We begin by analyzing the relations between the \emph{system}, \emph{aspect}, and \emph{representation} taxonomies based on our extraction from primary studies. Figure \ref{fig:system-properties-dist} visualizes these relations, showing co-occurrences across the three dimensions.

\subfile{figures/3d_table}

\textbf{General observations from this mapping:} \emph{System} and \emph{aspect} categories exhibit many-to-many relationships, with most \emph{system} categories spanning multiple \emph{aspects}. However, \emph{state aspects} (e.g., \emph{condition, evaluation}) are mainly linked to the \emph{technical system}, while \emph{properties aspects} (e.g., \emph{nominal, event, normative}) are concentrated in the \emph{natural system}.

Patterns also emerge between \emph{aspect} and \emph{representation}. \emph{State aspects} are predominantly \emph{numerical}, indicating their reliance on quantitative data. In contrast, \emph{properties aspects} vary: \emph{nominal} and \emph{normative} aspects are largely \emph{logical}, while \emph{event aspects} tend to be \emph{probabilistic}. \emph{Structural aspects} span \emph{graphical}, \emph{logical}, and \emph{semi-structured} representations.  

Representation preferences also differ across \emph{system} categories. \emph{State aspects} in the \emph{natural system} are exclusively \emph{numerical}, capturing environmental factors through measurements. In contrast, in the \emph{technical system}, they are often \emph{semi-structured}. Similarly, \emph{relations aspects} are mainly \emph{graphical} in the \emph{natural system} (e.g., causal graphs) but \emph{logical} or \emph{semi-structured} in the \emph{technical system}.

These initial observations motivate a more fine-grained identification of patterns that recur across the literature. \textbf{We identify patterns across the \emph{system}, \emph{aspect}, and \emph{representation} dimensions.} Pattern identification begins at the second-level categories defined in the taxonomies of the three dimensions (see Section \ref{sec:theory}), requiring a minimum support of three primary studies to ensure empirical grounding. Each pattern is defined by a unique combination of a \textbf{system} category (e.g., \textbf{reference domain}), an \textbf{aspect} category (e.g., \textbf{entities}), and a \textbf{representation} category (e.g., \textbf{logical}). These second-level patterns correspond to the cells in Figure~\ref{fig:system-properties-dist}, where each observed combination is derived from contextual information in primary studies. For example, the pattern \emph{reference domain} – \emph{entities} – \emph{logical} represents how entities in the natural environment are structured using logical slicing predicates.  

To organize these patterns, we group second-level patterns into broader categories based on the taxonomy hierarchy. When multiple second-level categories share the same first-level parent, they are generalized into a broader pattern (e.g., \emph{reference domain} and \emph{latent influences} under \emph{natural system}). If a second-level category has no siblings, it remains distinct to preserve precision.

\textbf{The resulting patterns are organized into two overarching groups based on the \emph{system} taxonomy:} the \emph{natural system map}, covering contextual information related to the natural environment, and the \emph{technical system map}, covering information related to data processing, inference, and application of the monitored ML model. Table~\ref{tabpatterns_reorg} presents the final catalog of C-SAR patterns, which we break down in this section.

\subsubsection{Natural System Map} \label{domain_map} Captures contextual information related to the environment in which an ML model is deployed. It involves \emph{natural subgroups}, \emph{natural diagrams}, \emph{natural specifications}, and \emph{exogenous diagnostics}.

\paragraph{\textbf{Natural Subgroups}} \label{domain_subgroups} They define entities, or subpopulations, in the natural system through formal representations. We observe two patterns in how these subgroups are defined.

Through \textbf{reference slices}, entities are identified as part of the reference domain and specified by defining logical or predicate-based membership \cite{azonoozi2016contrack,drevesm2020from,fengj2024designing,ghosha2022fair,henzingerthomasandkarimimahyar,rc2020overton,sackerman2021machine,schelter2018automating,swami2020data,zhoux2019a,fedelea2024the,henzingerta2023monitoring,mammanh2024biastrap}. These slices define semantic groupings in the data, identifying specific subsets that require focused analysis. In many studies, slices are constructed directly over individual features, grouping data points that share common, semantically meaningful properties. Some studies define slices through combinations of multiple features -- over the underlying function between features -- to specify entities as intersectional groups (e.g., individuals within a particular age range and from a specific country of origin) \cite{swami2020data, schelter2018automating, ghosha2022fair}. These slices help localize where issues like drift or unfairness disproportionately appear, through a deeper understanding of performance disparities across subpopulations. 

Another common pattern to natural subgroups is through \textbf{exogenous identifiers} -- integrating external variables that provide key-value mappings linking data points to their corresponding entity or group membership \cite{castelnovoa2021towards,khoshravanazara2023the,manerikerp2023online,schrouffj2022diagnosing,vasudevans2020lift,yangz2021biasrv}. A prominent example comes from fairness research: sensitive attributes (e.g., race, gender) are intentionally excluded from model inputs to prevent direct use in prediction, yet they are leveraged in monitoring to assess disparate impacts across groups. In this way, the model is ``unaware'' of protected entities, but the monitoring framework remains able to identify performance discrepancies or biases. Because these identifiers are kept outside the model’s immediate feature space, practitioners can respect legal guidelines while still preserving the capacity to audit the model for fairness issues.

\paragraph{\textbf{Natural Diagrams}} \label{domain_diagrams} They capture relational information about the natural environment using graph-based representations over three subsystems in the natural system.

Through \textbf{reference diagrams}, relationships between features and target variables in the reference domain are captured, encoding structural understanding of the process the monitored ML model aims to capture \cite{bontempellia2022humanintheloop,budhathokik2021why,dreyfuspa2022databased,fengj2024designing,sasthana2021ml,schrouffj2022diagnosing,zhangh2023why}.
Most studies use DAGs to specify the causal structure of the reference domain, representing how feature and target variables interact in the natural environment \cite{dreyfuspa2022databased,fengj2024designing,zhangh2023why,schrouffj2022diagnosing,budhathokik2021why}. Embedding domain knowledge into such diagrams aligns monitoring with domain knowledge of the data-generating processes -- something that models trained purely on observational data may lack. Some studies use graphs to specify ontological structures that capture semantic relationships between feature and target variables \cite{sasthana2021ml, bontempellia2022humanintheloop}. Notably, these semantic relations often do not require ad-hoc domain expert input; they can be extracted from shared knowledge bases \cite{sasthana2021ml}.

Beyond relationships among observed variables, \textbf{latent diagrams} represent hidden factors in the natural system, such as events or underlying processes that cannot be directly measured \cite{borchanih2015modeling,dreyfuspa2022databased,feng2022clinical}. By modeling these unobserved phenomena as latent variables, their directional, causal effects on both feature and target variables are often captured -- visualizing the ``under-the-surface'' mechanisms driving observed data shifts.

\paragraph{\textbf{Natural Specifications}} \label{domain_specifications} They encode the properties of the natural system through logical or probabilistic models.

Logical rules known as \textbf{Reference Assertions} encode properties of feature and target variables within the reference domain, including their nominal constraints and inter-relationships \cite{bachingerf2024data,cavenesse2020tensorflow,chenl2022estimating,dchen2022twostage,ehrlingerl2019a,lelwakatare2021on,liu2020towards,myllyahol2022on,schelter2018automating,shankars2024we,swami2020data}. For example, an assertion might specify that a given feature must remain within certain value ranges (e.g., age must be non-negative and below 100), or that specific feature and target variables exhibit a monotonic relationship \cite{bachingerf2024data, chenl2022estimating, dchen2022twostage}. Studies commonly employ declarative rule frameworks that enable the specification of various constraints based on domain knowledge, verifying data quality and triggering monitoring alerts to safeguard the ML system from ingesting spurious inputs \cite{ehrlingerl2019a,schelter2018automating,shankars2024we}.

When reasoning about the effects of unobserved events on features and target variables, \textbf{latent projections} describe these influences through probabilistic models \cite{borchanih2015modeling,feng2024monitoring,galewis2022augur,godaup2023deployment,henzingerta2023monitoring,leestj2024expert,masegosaar2020analyzing,pichlerg2024on,pwelinder2013a,sibliniw2020master,sobolewskip2017scr}. These models can represent the ``presence'' of an event as a binary or continuous latent variable \cite{borchanih2015modeling,masegosaar2020analyzing}. Such models are used either to represent the temporal properties of how an event unfolds over time (i.e., state transition properties) \cite{henzingerta2023monitoring,borchanih2015modeling,galewis2022augur,masegosaar2020analyzing} or to model how the event affects features and target variables as a function of its presence \cite{sibliniw2020master,godaup2023deployment,pwelinder2013a,sobolewskip2017scr,leestj2024expert,galewis2022augur,pichlerg2024on,feng2024monitoring}. The effect of an event can be formalized as the joint probability distribution over affected features and targets, conditioned on the latent event variable -- \( P(X, Y \mid L) \). In some cases, Bayesian formulations are used to define priors over uncertain events; they extend probabilistic models of conditional influence by introducing a distribution over the model’s parameters, capturing the uncertainty in how latent variables affect features and targets \cite{feng2024monitoring, godaup2023deployment, leestj2024expert}.

To describe the scope and conditions under which unobserved events affect features and target variables, \textbf{latent expectations} specify logical rules that define event boundaries \cite{chenm2021mandoline,xuanjunyuandlujieandzhangguang,xuz2023alertiger}. Studies commonly use these constraints to specify the subspace or timeframe in which a domain shift is hypothesized. Whereas \textbf{latent projections} provide a probabilistic model of event occurrence, \textbf{latent expectations} logically define the region or periods where such a shift is expected, using declarative rules. This pattern shares similarities with \textbf{reference slices} but instead captures distributional properties without necessarily assuming semantic meaning for the bounded regions. For example, based on prior knowledge of a pricing change, we might define the subset of transactions most affected, such as those involving discounted products.

\paragraph{\textbf{Latent Scenarios}} \label{latent_scenarios} They capture qualitative or semi-structured descriptions of potential shifts in unobserved influences and failure modes \cite{apaul2024mlops,dreyfuspa2022databased,feng2022clinical,haidert2021domain,leestj2024expert,xuz2023alertiger}. These scenarios are commonly elicited from domain experts, who list possible events (e.g., logistical issues, machine failures) along with approximate or textual outlines of how these events affect the observed data \cite{apaul2024mlops,leestj2024expert,haidert2021domain,feng2022clinical}. This scenario catalog may be stored in a table describing each event’s triggers, affected features, and potential consequences, providing an interpretable representation for non-technical stakeholders involved in the monitoring process, even though it is not as formally specified as \textbf{latent projections} or \textbf{latent expectations}.

\paragraph{\textbf{Exogenous Diagnostics}} \label{exogenous_diagnostics} These represent external observations beyond the model’s immediate feature space, often used to capture environmental or contextual factors that drive changes or drift in the domain \cite{bensalems2024continuous,cobbo2022contextaware,gomesjb2010calds,jbgomes2014mining,kirchheimk2023towards,klsm2019uncertainty,langfordma2023modalas,torfahh2022learning,zhoux2019a,kkirchheim2024outofdistributio}. These measurements typically reflect real-world conditions -- such as weather data, temporal indicators, or other domain-specific signals -- that can shift independently of what the model’s original features capture. While model inputs aim to encode all relevant domain factors, \emph{exogenous diagnostics} often prove critical to detect or anticipate domain shifts that do not register within the model’s internal feature space \cite{klsm2019uncertainty, kirchheimk2023towards}. Variables may also be left out of the feature set intentionally in the feature selection process, where upstream and collinear features might be removed to avoid redundancy and noise. However, the excluded variable may still represent the point where a change originates -- it is exogenous -- and propagates into the model inputs, retaining value for monitoring \cite{jbgomes2014mining, cobbo2022contextaware, gomesjb2010calds}. In computer vision deployments, external variables such as road conditions or time-of-day can help diagnose when the operational environment (e.g., roads in autonomous driving) deviates from training conditions \cite{bensalems2024continuous, langfordma2023modalas}. External variables can also be extracted from images (where the image is the input vector for the monitored model); these variables can be seen as exogenous to the image, explaining the circumstances under which the image was generated. \cite{kirchheimk2023towards,kkirchheim2024outofdistributio}. Although additional contextual variables could theoretically enhance model performance, they often remain external due to technical limitations or because the model aims to generalize within a specific modality, such as images, and therefore does not incorporate structured data.

\begin{walkthroughbox}{Constructing a natural system map}{naturalColor!15}

A team responsible for a churn-risk model begins by mapping the natural system context. They ask a series of questions to identify patterns that could inform monitoring:

\begin{enumerate}
    \item \emph{Which segments stand out by behavior, risk, or value?}

    \textbf{Reference slices}: The team defines a segment for young adults (ages 18–25) over the age feature \textbf{($x_1$)}; a business critical segmenet that shows highly variable churn behavior, making it a priority for monitoring.

    \item \emph{What are the relationships underlying churn behavior?}

    \textbf{Reference diagrams}: They sketch a causal graph hypothesizing that a user’s age \textbf{($x_1$)} influences both activity \textbf{($x_2$)} and churn-risk \textbf{($y$)}. They also note that a higher the churn risk causes the user activity pattern.

    \item \emph{What should valid data look like?}

    \textbf{Reference assertions}: They formalize constraints grounded in natural properties. This includes a bound on the average weekly user activity \textbf{($x_2$)} at 168 hours -- the total number of hours in a week.

    \item \emph{Are there any external factors we should track?}

    \textbf{Exogenous diagnostics}: They register the holiday calendar, denoted as \textbf{$u$}, to distinguish meaningful change from expected variation. This helps ensure shifts in user activity during public holidays are recognized as external influences, not mistaken for model issues.

    \item \emph{What scenarios could challenge model assumptions?}

    \textbf{Latent scenarios}: In discussion with a business analyst, the team documents a potential risk: competitor campaigns, denoted as \textbf{$z$}, are expected to increase marketing efforts in the upcoming quarter. While not explicitly tracked, it may influence churn risk and manifest as reduced user activity.
\end{enumerate}

% ------------------------------------------------------------
%  leave some white-space before the picture
\vspace{0.1\baselineskip}

%  centre the whole graph
\begin{center}
% ------------------------------------------------------------
%              reusable “circle-with-label” macro
% ------------------------------------------------------------
% \nodevar{<name>}{<x>}{<y>}{<colour>}{<centre symbol>}{<label>}
\newcommand{\nodevar}[6]{%
  \node[draw, circle, minimum size=15pt, inner sep=0pt,
        fill=#4!25,
        label=center:{\strut \small #5},                             % center symbol
        label={[font=\small, label distance=-2pt]-90:#6}]            % label below node
        (#1) at (#2,#3) {};%
}

% ------------------------------------------------------------
% distance “knobs” (keep if you like the x-/y-scaling trick)
\newcommand{\HorizontalDist}{2cm}
\newcommand{\VerticalDist}{0.5cm}
% ------------------------------------------------------------
\begin{tikzpicture}[>=stealth,
                    x=\HorizontalDist, y=-\VerticalDist]  % row-0 on top
\usetikzlibrary{positioning}

% -----------------  nodes  ----------------------------------
\nodevar{y}{2}{0}{naturalColor}   {$y$}{}
\nodevar{z}{3}{0}{LatentColor}  {$z$}{\textbf{5}}

\nodevar{x1}{1}{0}{naturalColor}{$x_1$}{\textbf{1}}

\nodevar{x2}{2}{2}{naturalColor} {$x_2$}{\textbf{3}}
\nodevar{u}{3}{2}{externalColor}{$u$}{\textbf{4}}

% -------------- floating number (centre of triangle) --------

% -----------------  arrows  ---------------------------------
\draw[->] (x1) -- (y);
\draw[->] (x1) -- (x2);
\draw[->] (z)  -- (y);
\draw[->] (u)  -- (x2);
\draw[->] (y) -- (x2);

\node[font=\bfseries, fill=gray!10, text=black, inner sep=1pt, rounded corners=2pt] at (2,0.9) {2};
\node[font=\bfseries, fill=gray!10, text=black, inner sep=1pt, rounded corners=2pt] at (1.5,0) {2};
\node[font=\bfseries, fill=gray!10, text=black, inner sep=1pt, rounded corners=2pt] at (1.5,1) {2};
\end{tikzpicture}

\end{center}

\begin{center}
\footnotesize\makebox[0.95\linewidth][c]{\textit{Simplified diagram showing step numbers for involved system elements.}}
\end{center}

\end{walkthroughbox}

\subsubsection{Technical System Map} \label{pipeline_map} It captures contextual information related to how data is processed, how inferences are generated, and how the broader application uses these predictions. It is organized into \emph{technical observations}, \emph{technical signals}, \emph{technical descriptions}, and \emph{technical specifications}.

\paragraph{\textbf{Technical Observations}} \label{pipeline_measurements} They involve numerical observations that quantify the state of the technical system. This pattern is composed of five lower-level patterns -- \emph{processing diagnostics}, \emph{inference diagnostics}, \emph{inference assessments}, \emph{application diagnostics}, and \emph{application assessments}.

Before data enters the model for inference, \textbf{processing diagnostics} support assessment of the runtime state of upstream processing pipelines using numerical indicators of processing integrity \cite{ebreck2017the,muirurid2022practices,nguyenmt2024novel,paleyesandreiandlawrenceneilda,sculley2015hidden}. Studies commonly track such indicators throughout the data lineage. Examples include staleness, which reflects the recency of the last pipeline run or transformation, as well as metadata such as model version, feature set version, or data snapshot timestamp.

As ML models operate in production, \textbf{inference diagnostics} capture quantitative measurements of their operational condition \cite{baquon2022concept, ebreck2017the, paleyesandreiandlawrenceneilda, sculley2015hidden, tzoppi2021detect, wangs2021a, xuz2023alertiger, xxu2022dependency}. These include technical metrics, such as inference latency or model staleness, which can act as proxies for functional failures \cite{tzoppi2021detect, wangs2021a, xuz2023alertiger, ebreck2017the}. They may also include observations from auxiliary models within the pipeline -- such as raw observations of their predictions or functional metrics \cite{xxu2022dependency, paleyesandreiandlawrenceneilda, baquon2022concept}. These auxiliary models can function as upstream dependencies or operate as independent components. While their outputs do not directly assess the model under observation, they can indicate the health of multi-model applications that share feature subsets, serving as proxy signals for related models \cite{baquon2022concept}.

When naturally occurring ground-truth labels are delayed or only arrive periodically in batches, \textbf{inference assessments} offer an alternative means to evaluate an ML model’s functional quality at runtime by leveraging additional information sources \cite{binderf2022putting, ginartaa2022mldemon, guann2022fila, vandervorstf2024claims, vishwakarmah2024taming}. These often rely on human-in-the-loop mechanisms, where a domain expert -- or an oracle -- is assumed to be available to provide feedback on demand, selectively validating uncertain or high-impact predictions. Unlike naturally occurring ground-truth labels, which may only arrive in batches and with set delays, queried labels serve as an auxiliary ground-truth information source.

Once predictions are consumed to produce downstream outcomes, \textbf{application diagnostics} capture the condition under which these predictions are used \cite{allenb2021evaluation, fengj2024designing, onnesa2023bayesian}. This includes tracking when and how predictions influence decisions -- such as whether they are followed, overridden, or selectively applied -- as well as protocols and usage scenarios that govern prediction-to-decision processes.

Finally, \textbf{application assessments} capture the broader impact of the monitored ML model by measuring downstream outcomes and key performance indicators (KPIs) in the real application, indirectly evaluating the model’s performance through its influence on business or operational success \cite{bensalems2024continuous, bernardil2019150, binderf2022putting, csun2024ai, ebreck2017the, edsnascimento2019understanding, hanafimf2024machineassisted, langfordma2023modalas, paleyesandreiandlawrenceneilda, shergadwalamn2022a}. Metrics like revenue, customer retention, or operational efficiency are often viewed as the ultimate signals of success, but they may be influenced by numerous external factors beyond the ML model itself \cite{bernardil2019150, ebreck2017the}. Changes in these metrics can indicate potential issues in model performance or application logic, yet attributing them directly to the model often requires additional analysis of factors influencing the metrics. In addition to high-level performance metrics, some studies also monitor raw observations of downstream actions, providing more immediate signals of shifts that could reflect changes in model behavior or the application policy \cite{bensalems2024continuous, fengj2024designing, langfordma2023modalas, paleyesandreiandlawrenceneilda}.

\paragraph{\textbf{Technical Signals}} \label{pipeline_signals} They are characterizations of the technical system state that often extend beyond structured measurements, relying instead on semi-structured indicators of \emph{processing notifications} and \emph{application comments}.

As models rely on upstream dependencies managed by different teams that continuously evolve, \textbf{processing notifications} involve semi-structured representations that capture these changes in processing conditions through textual descriptions. Studies highlight the cross-functional nature of ML deployments, emphasizing the need to monitor communication channels (e.g., Slack) for updates about code changes, data source deprecations, or modifications to transformations \cite{ebreck2017the, swami2020data}. Beyond human communication, the state of upstream processing is often encoded in the code artifacts themselves, which can be analyzed statically to reveal information about dependency quality and upstream processing health \cite{heynhm2023automotive, foidlh2019riskbased}.

Further downstream, \textbf{application comments} integrate end-users into the monitoring process by capturing their sem-structured (textual) feedback on application outcomes \cite{jayalathh2022enhancing, cabreraa2021discovering, jayalathh2023continual}. Failure mode descriptions can be embedded in these comments, offering diagnostic insight and serving as proxy signals for the model’s functional quality.

\paragraph{\textbf{Processing Traces}} \label{pipeline_architecture} 
These map the flow of data and the dependencies between transformation steps in a data pipeline \cite{heynhm2023automotive, namakimh2020vamsa, shankars2022towards}. This trace is often defined by declarative logic, such as SQL queries, embedded within scripts that map source data to derived data (e.g., features). This logical definition allows any derived or aggregated data point to be traced back to its origins, including specific source tables, columns, or even individual cells as referenced in the query. While structured diagrams of this flow may not always exist, the logical definitions can be analyzed and rerun to reconstruct the dependency graph of the data lineage.

\paragraph{\textbf{Technical Specifications}} \label{pipeline_specifications} They define formal models and rules at each stage of the data processing, inference, and application pipeline. They are composed of \emph{processing assertions}, \emph{application guardrails}, and \emph{inference guardrails}.

\textbf{Processing assertions} specify technical constraints over processed data to verify that nominal pipeline execution has occurred \cite{ehrlingerl2019a, myllyahol2022on, lelwakatare2021on, cavenesse2020tensorflow,schelter2018automating, swami2020data}. These rules are typically applied after transformations and are used to check for structural or formatting issues rather than semantic validity. Examples include checking data types (e.g., integers or strings), excessive NULLs, or string formats via regex -- ensuring the pipeline outputs technically usable, consistent data.

At model deployment, \textbf{inference guardrails} enforce constraints on predictions or inputs, ensuring outputs adhere to normative properties that define acceptable and safe behavior \cite{csun2024ai,habdelkader2024mlonrails,kangdanielandguibasjohnandbail,myllyahol2022on}.
These constraints specify what the predictions should satisfy given the system’s operational requirements -- e.g., limiting regression outputs to values that align with downstream decision thresholds or blocking classification outputs when confidence falls below a required level. These guardrails function as a final check to ensure predictions respect system-defined norms.

If not enforced at the model level, \textbf{application guardrails} specify normative properties of a system through logical rules or constraints on downstream application outcomes \cite{bensalems2024continuous,habdelkader2024mlonrails,langfordma2023modalas,onnesa2022monitoring,sculley2015hidden}. These properties define boundaries that the system must not cross, ensuring safety and compliance even when upstream models degrade or drift. For instance, studies in autonomous driving define conditions under which an ML model's control decisions must be overridden to ensure safety \cite{bensalems2024continuous,langfordma2023modalas}. Similarly, in business applications, guardrails are often imposed to prevent decisions from violating regulatory or ethical standards \cite{sculley2015hidden,onnesa2022monitoring}. Some studies define guardrails as a function of external variables. For example, in autonomous driving, normative properties may relate driving speed (an outcome) to rainfall (an external variable), ensuring safe operation when a vision-based model encounters unseen conditions that might lead the application to cross known safe conditions \cite{bensalems2024continuous,langfordma2023modalas,onnesa2022monitoring}.

\subfile{figures/activity_mappings}

\begin{walkthroughbox}{Constructing a technical system map}{featureColor!15}

The team responsible for a churn-risk model now maps their technical system context. Similarly, they ask a series of questions corresponding to understanding this context:

\begin{enumerate}
    \item \emph{How to stay informed of upstream changes?}
    
    \textbf{Processing notifications}: The team configures a Slack bot to alert on announcements about changes to upstream sessions ($s_1$) and users ($s_2$) tables, which contain the raw data used to compute the user activity features used by the churn model.
    
    \item \emph{How can we map the lineage in case a feature breaks?}

    \textbf{Processing traces}: The team instruments the SQL logic that computes feature vectors to log the lineage information for traceability. This includes joins between the sessions ($s_1$) and users ($s_2$) tables to compute features like weekly session frequency using rolling averages ($\hat{x}$).
    
    \item \emph{How to track the health of the model artefact?}
    
    \textbf{Inference diagnostics}: The team expects the churn model to retrain weekly, but label delays and failed retraining jobs can leave the churn-risk scores ($\hat{y}$) stale unnoticed. They log the training timestamp and display model age in a dashboard to spot outdated deployments during review.
    
    \item \emph{What are the operational limits on predictions?}
    
    \textbf{Application guardrails}: Retention offers, denoted as $a$, are allocated based on churn-risks, but the team enforces limits on the number of offers sent to prevent exhausting the promotion budget or leaving campaign capacity unused. These bounds ensure that offer allocation remains aligned with strategic constraints.

\end{enumerate}

    % ------------------------------------------------------------
%  leave some white-space before the picture
\vspace{0.1\baselineskip}

%  centre the whole graph
\begin{center}
% ------------------------------------------------------------
%              reusable “circle-with-label” macro
% ------------------------------------------------------------
% \nodevar{<name>}{<x>}{<y>}{<colour>}{<centre symbol>}{<label>}
\newcommand{\nodevar}[6]{%
  \node[draw, circle, minimum size=15pt, inner sep=0pt,
        fill=#4!25,
        label=center:{\strut \small #5},                             % center symbol
        label={[font=\small, label distance=-2pt]-90:#6}]            % label below node
        (#1) at (#2,#3) {};%
}

% ------------------------------------------------------------
% distance “knobs” (keep if you like the x-/y-scaling trick)
\newcommand{\HorizontalDist}{2cm}
\newcommand{\VerticalDist}{0.5cm}
% ------------------------------------------------------------
\begin{tikzpicture}[>=stealth,
                    x=\HorizontalDist, y=-\VerticalDist]  % row-0 on top
\usetikzlibrary{positioning}

% -----------------  nodes  ----------------------------------
\nodevar{x1}{0}{0}{featureColor}{$s_1$}{\textbf{1}}

\nodevar{y}{0}{2}{featureColor}   {$s_2$}{\textbf{1}}
\nodevar{z}{3}{2}{applicationColor}  {$a$}{\textbf{4}}

\nodevar{x2}{1}{2}{featureColor} {$\tilde{x}$}{\textbf{}}
\nodevar{u}{2}{2}{inferenceColor}{$\hat{y}$}{\textbf{3}}

% -----------------  arrows  ---------------------------------
\draw[->] (x1) -- (x2);
\draw[->] (y) -- (x2);
\draw[->] (x2) -- (u);
\draw[->] (u) -- (z);

\node[font=\bfseries, fill=gray!10, text=black, inner sep=1pt, rounded corners=2pt] at (0.5,1) {2};
\node[font=\bfseries, fill=gray!10, text=black, inner sep=1pt, rounded corners=2pt] at (0.5,2) {2};

\end{tikzpicture}
\end{center}

\begin{center}
\footnotesize\makebox[0.95\linewidth][c]{\textit{Simplified diagram showing step numbers for involved system elements.}}
\end{center}

\end{walkthroughbox}

\subsection{Mapping C-SAR patterns to Monitoring Activities} \label{csar_activities}

Having established the \textbf{C-SAR triplet patterns} (Section \ref{csar_patterns}) and \textbf{monitoring activities} (Section \ref{sec:monitoring_activities}), this section analyzes how these patterns are leveraged across drift detection, data validation, performance monitoring, model validation, and out-of-distribution (OOD) detection. Figure \ref{fig:bar-chart} illustrates their distribution, showing how frequently each pattern appears in different monitoring activities and highlighting recurring themes in context usage, such as reliance on data partitioning.

Each \emph{monitoring activity} has its own focus and scope: \emph{drift detection}, \emph{data validation} and \emph{OOD detection} focus on the identification of the causes of failures while \emph{performance monitoring} and \emph{model validation} focus on the effects of failures. Yet, \emph{C-SAR triplet patterns} are used across multiple activities, and each activity may leverage multiple patterns.

The widespread use of \emph{natural subgroups} across various activities is evident. In particular, \emph{reference slices} are commonly employed to isolate specific partitions of raw data observations corresponding to entities of interest. These are used for purposes such as \emph{drift detection} \cite{azonoozi2016contrack,sackerman2021machine,zhoux2019a}, \emph{data validation} \cite{schelter2018automating,swami2020data,zhoux2019a}, and \emph{performance monitoring} \cite{drevesm2020from,fedelea2024the,fengj2024designing,ghosha2022fair,henzingerta2023monitoring,henzingerthomasandkarimimahyar,mammanh2024biastrap,rc2020overton}.

Similarly, \emph{natural specifications} are widely applied across activities, supporting tasks in the monitoring workflow -- from computing metrics over selected data partitions to configuring alerts and conducting root-cause analysis. This is seen in \emph{drift detection} \cite{borchanih2015modeling,chenl2022estimating,feng2024monitoring,galewis2022augur,leestj2024expert,masegosaar2020analyzing,pichlerg2024on,sobolewskip2017scr,vandervorstf2024claims,xuanjunyuandlujieandzhangguang,xuz2023alertiger}, \emph{data validation} \cite{bachingerf2024data,cavenesse2020tensorflow,ehrlingerl2019a,lelwakatare2021on,myllyahol2022on,schelter2018automating,shankars2024we,swami2020data}, and \emph{performance monitoring} \cite{chenm2021mandoline,godaup2023deployment,henzingerta2023monitoring,pwelinder2013a,sibliniw2020master}. 

Among \emph{C-SAR triplet patterns}, \emph{technical observations} are the most pervasive in their usage across activities, appearing in \emph{drift detection} \cite{allenb2021evaluation,baquon2022concept,ginartaa2022mldemon,paleyesandreiandlawrenceneilda,tzoppi2021detect,vandervorstf2024claims,wangs2021a,xuz2023alertiger}, \emph{data validation} \cite{ebreck2017the,muirurid2022practices,nguyenmt2024novel,sculley2015hidden,xxu2022dependency}, \emph{OOD detection} \cite{bensalems2024continuous,langfordma2023modalas,vishwakarmah2024taming}, \emph{performance monitoring} \cite{binderf2022putting,fengj2024designing,guann2022fila}, and \emph{model validation} \cite{bernardil2019150,binderf2022putting,csun2024ai,ebreck2017the,edsnascimento2019understanding,hanafimf2024machineassisted,onnesa2023bayesian,shergadwalamn2022a}. They primarily integrate monitored observations and metrics to detect failures and assess model health. In some studies, they are not directly monitored but are instead incorporated as contextual information to aid in interpreting and computing metrics over monitored observations \cite{binderf2022putting,guann2022fila,ginartaa2022mldemon,vandervorstf2024claims,vishwakarmah2024taming}. Additionally, they play a role in root-cause analysis, where they support identifying the source of failures \cite{allenb2021evaluation,paleyesandreiandlawrenceneilda}.

Below, we detail how these patterns are applied within each activity, detailing common usages and supported by examples from the literature.

\subsubsection{Drift Detection}

While \emph{natural subgroups} are used similarly to data validation and performance monitoring, the use of \emph{natural specifications} in \textbf{drift detection} most often involves \emph{latent projections}. These projections are primarily employed in three ways: (1) directly configuring alerts by computing metrics that capture whether a predefined drift event is occurring \cite{feng2024monitoring,leestj2024expert,pichlerg2024on}, (2) supporting root-cause analysis after an alert is triggered by attributing the change to a specific drift event \cite{borchanih2015modeling,masegosaar2020analyzing,sobolewskip2017scr}, and (3) simulating drift events to refine alert configuration, including metric selection and threshold specification \cite{galewis2022augur,sobolewskip2017scr}.

\emph{Drift detection} also frequently relies on \emph{natural diagrams} \cite{bontempellia2022humanintheloop,borchanih2015modeling,budhathokik2021why,dreyfuspa2022databased,feng2022clinical,sasthana2021ml,schrouffj2022diagnosing,zhangh2023why}. These diagrams are most commonly used as \emph{reference diagrams} that capture the (causal) relationships between data elements directly involved in the (ML) modeling process \cite{bontempellia2022humanintheloop,budhathokik2021why,schrouffj2022diagnosing,zhangh2023why,dreyfuspa2022databased}. They also include \emph{latent diagrams} capturing relationships in an abstract feature space \cite{dreyfuspa2022databased,feng2022clinical}. Diagrams that capture the causal structure of the natural system enable root-cause analysis by attributing distributional changes in downstream variables to a change in an upstream parent.

\emph{Latent scenarios} serve a different function for \emph{drift detection} \cite{apaul2024mlops,dreyfuspa2022databased,feng2022clinical,haidert2021domain,leestj2024expert,xuz2023alertiger}. They are used exclusively for documentation and inter-stakeholder communication. Since changes in natural systems often correlate with specific drift events, domain experts (e.g., business analysts) have a nuanced understanding of these events -- an understanding that engineers responsible for ML system maintenance might lack. Eliciting potential events in the natural system, even if they are not directly monitored, supports the creation of a ``playbook'' for diagnoses of failures as they emerge at runtime.

Finally, \emph{exogenous diagnostics} \cite{cobbo2022contextaware,gomesjb2010calds,jbgomes2014mining,zhoux2019a} are employed not for direct monitoring but to aid root-cause analysis once drift is detected in monitored data elements, such as features or targets. By tracing drift back to changes in external (exogenous) variables, they help characterize new system states and diagnose downstream changes. This method is often used to identify recurring drifts, enabling either corrective actions for known drifts \cite{gomesjb2010calds,jbgomes2014mining} or the suppression of alerts for (known) benign drifts \cite{cobbo2022contextaware}.

\subsubsection{Data Validation}

When validating data quality, \emph{natural specifications} focus on defining ``what is'' as opposed to ``what can'', as seen in \emph{drift detection}. Specifically, \emph{reference assertions} are used for \textbf{data validation} to define expectations of semantic properties \cite{bachingerf2024data,cavenesse2020tensorflow,ehrlingerl2019a,lelwakatare2021on,myllyahol2022on,schelter2018automating,shankars2024we,swami2020data}, while \emph{processing assertions} are used to define technical constraints \cite{ehrlingerl2019a, myllyahol2022on, lelwakatare2021on, cavenesse2020tensorflow,schelter2018automating, swami2020data}. These assertions are commonly used to configure alerts or compute data quality metrics, allowing controlled sensitivity -- such as triggering only when more than 5\% of records violate a constraint.

When an alert indicates a data quality issue, \emph{processing traces} are assessed for root-cause analysis  \cite{heynhm2023automotive, namakimh2020vamsa, shankars2022towards}, allowing engineers to trace issues back to the upstream process that failed.

\emph{Technical signals} are used for data validation, specifically as \emph{processing notifications} that support inferring the quality of the input data based on information about the functions that process it \cite{ebreck2017the,foidlh2019riskbased,heynhm2023automotive,swami2020data}. These notifications help preemptively manage issues arising from modifications in feature logic, normalization, or scaling that could impact data integrity. Automations can be set up to monitor these message boards or communication channels and notify the team responsible for the monitored ML model of potentially problematic upstream changes \cite{ebreck2017the,heynhm2023automotive}. 

\subsubsection{OOD Detection}

Similar to \emph{data validation}, \emph{natural specifications} are applied in \textbf{OOD detection} by defining entities through \emph{reference assertions}. However, rather than flagging data quality issues, they aim to detect incorrect extrapolative predictions of the ML model by defining assertions on nominal properties, such as monotonicity in feature-target relationships \cite{liu2020towards,dchen2022twostage}. 

To detect when model inputs originate from an unseen distribution, \emph{exogenous diagnostics} \cite{bensalems2024continuous,kirchheimk2023towards,kkirchheim2024outofdistributio,klsm2019uncertainty,langfordma2023modalas,torfahh2022learning} and \emph{technical observations}, particularly of outcomes \cite{bensalems2024continuous,langfordma2023modalas,vishwakarmah2024taming}, are commonly used. These methods frequently integrate simulations to model system responses to exogenous conditions \cite{bensalems2024continuous,langfordma2023modalas,torfahh2022learning,kirchheimk2023towards,kkirchheim2024outofdistributio}. For instance, in autonomous driving, simulations help identify cases where exogenous conditions generate OOD images that lead to misclassifications. Simulation results can then be used to discover monitors that preemptively trigger alerts based on exogenous conditions, acting as proxies for failures when labeled data is unavailable. This approach ensures the computer vision model operates within its defined operational design domain (ODD).

\subsubsection{Performance Monitoring}

\emph{Natural subgroups}, \emph{natural specifications}, and \emph{technical observations} are used in \textbf{performance monitoring} similar to other monitoring activities. However, some key specializations exist.

A special case of \emph{natural subgroups} in performance monitoring is the use of \emph{exogenous identifiers} to compute fairness metrics such as demographic parity, ensuring that model performance is systematically evaluated across different demographic groups and highlighting potential biases \cite{castelnovoa2021towards,khoshravanazara2023the,manerikerp2023online,vasudevans2020lift}. Additionally, \emph{natural subgroups} are used in performance monitoring for interventional strategies, where users are partitioned, and model responses to input mutations are assessed, enabling the identification nuanced biases at runtime \cite{yangz2021biasrv,fengj2024designing}.

Regarding \emph{natural specifications}, \emph{latent projections} play a crucial role. By leveraging prior knowledge of the expected class distribution under specific conditions, performance can be estimated without direct access to ground-truth labels. This enables calibration of performance metrics and approximation of real-world model behavior \cite{godaup2023deployment,henzingerta2023monitoring,sibliniw2020master}. These techniques ensure that \emph{performance monitoring} remains effective even in scenarios without access to ground-truth labels.

\subsubsection{Model Validation}
Identifying issues in the real-world impact of the model is the primary goal of \textbf{model validation}. To accomplish this, \emph{technical observations}, particularly \emph{application assessments}, serve as downstream impact metrics that the ML model seeks to optimize \cite{bernardil2019150,binderf2022putting,csun2024ai,ebreck2017the,edsnascimento2019understanding,hanafimf2024machineassisted,onnesa2023bayesian,shergadwalamn2022a}. However, isolating the model’s influence on these assessments is nontrivial, and interventional monitoring, applied through A/B testing, addresses this by independently monitoring these assessments for both the primary and auxiliary models, disentangling the model’s actual contribution \cite{ebreck2017the,bernardil2019150}. For example, if a model is optimized for recommendation diversity under the hypothesis that it improves customer engagement, it may perform worse on other metrics like short-term click-through rate. A/B testing validates this by comparing the impact of predictions from the primary model (model A) to those from an auxiliary model (model B), which represent the hypothesis by explicitly optimizing for diversity. By measuring these effects, this approach identifies misaligned training requirements at runtime; offline quality metrics sometimes fail to reflect real-world impact.

User feedback captured within applications provides another crucial dimension for model validation. Through \emph{technical Signals}, particularly \emph{application comments}, studies aimed to identify failures related to invalid training assumptions by capturing textual feedback from end-users of downstream applications at runtime \cite{cabreraa2021discovering,jayalathh2022enhancing,jayalathh2023continual}. Natural language processing techniques are used to organize this feedback, such as clustering similar comments or analyzing sentiment, providing insights into model limitations and emergent failure patterns, enabling effective root-cause analysis \cite{jayalathh2023continual}.

Beyond metrics and feedback, establishing constraints ensures model behavior remains within acceptable boundaries. \emph{Technical specifications} are often applied for model validation in the form of \emph{inference guardrails} \cite{habdelkader2024mlonrails,kangdanielandguibasjohnandbail,myllyahol2022on,csun2024ai} or \emph{application guardrails} \cite{habdelkader2024mlonrails,sculley2015hidden}, triggering alerts when the model or downstream system deviates from predefined norms. A complementary interventional approach combines \emph{application assessments} with A/B testing to evaluate the downstream impact of the model against established standards \cite{csun2024ai}. Specifically, \emph{application guardrails} serve as rule-based baselines that encode normative expectations, providing reference points against which actual model behavior is compared.

\begin{walkthroughbox}{Contextualized detection and root-cause analysis}{red!15}

With the system maps already in place, the churn-risk model has run unchanged for several weeks. A fresh alert triggers the following investigation.

\begin{enumerate}
    \item \textbf{Alert:} Week-over-week retention offers ($a$) rise by 22\% \emph{(application guardrails)}.  
          Marketing can't explain the increase and notifies the ML team.

    \item \textbf{Model check:} Because churn labels arrive a month late, online performance is unknown. The staleness metric \emph{(inference diagnostics)} shows a successful retrain four days earlier, the model artefact that produced the churn-risk scores ($\hat{y}$) should be up-to-date -- though a faulty job is possible.  
          Meanwhile, the mean values in the user activity feature vector ($\tilde{X}$) has dropped, suggesting lower activity is driving the change.

    \item \textbf{External factors:}  
          The team rules out external explanations with three checks:  
          \textbf{(1)} The causal graph \emph{(reference diagram)} shows that user age ($x_1$) is upstream of activity ($x_2$), but the age distribution is unchanged; it does not explain the change in activity.
          \textbf{(2)} The regional holiday calendar \emph{(exogenous diagnostics)},  denoted as $u$, lists no public holidays during the period.
          \textbf{(3)} They skim through elicited potential failure modes \emph{(latent scenarios)} -- a competitor campaign ($z$) could explain a drop in activity in specific segments (e.g., young adults). But the change is consistent across all segments \emph{(reference slice)}; the shift appears global. External causes now seem unlikely.

    \item \textbf{Pipeline inspection:} Lineage \emph{(processing traces)} leads the user-activity feature back to the sessions table ($s_1$).  
          A Slack post forwarded by the bot reads: “SDK v5.3.0 now records \texttt{session\_duration\_ms}; legacy \texttt{session\_duration} kept” \emph{(processing notification)}.  
          The new SDK leaves the legacy field \texttt{NULL} on about 35\% of events. 
          Because the feature builder drops sessions with \texttt{NULL} duration.
        They run an offline check using a warehouse snapshot from before the SDK rollout, confirming that normal activity, churn scores, and offer counts are restored -- the SDK rollout caused a data quality issue.
\end{enumerate}

%  reusable “circle-with-label” macro
% \nodevar{<name>}{<x>}{<y>}{<colour>}{<centre symbol>}{<label>}
\newcommand{\nodevar}[6]{%
  \node[draw, circle, minimum size=15pt, inner sep=0pt,
        fill=#4!25,
        label=center:{\strut \normalsize #5},        % symbol in the middle
        label={[font=\normalsize]45:#6}]             % 45° label (left empty here)
        (#1) at (#2,#3) {};%
}

% ------------------------------------------------------------
%  distance “knobs”
\newcommand{\HorizontalDist}{2cm}
\newcommand{\VerticalDist}{0.5cm}
% ------------------------------------------------------------
\begin{center}
\begin{tikzpicture}[>=stealth,
                    x=\HorizontalDist, y=-\VerticalDist]   % row-0 on top
\usetikzlibrary{positioning,calc}

% ---------- NATURAL SYSTEM ----------
\nodevar{x1}{1}{0}{naturalColor}{$x_1$}{}
\nodevar{y}{2}{0}{naturalColor}{$y$}{}
\nodevar{z}{3}{0}{LatentColor}{$z$}{}

\nodevar{x2}{2}{2}{naturalColor}{$x_2$}{}
\nodevar{u}{3}{2}{externalColor}{$u$}{}

\draw[->] (x1) -- (y);
\draw[->] (x1) -- (x2);
\draw[->] (z)  -- (y);
\draw[->] (u)  -- (x2);
\draw[->] (y) -- (x2);

% ---------- TECHNICAL SYSTEM ----------
\nodevar{s1}{1}{2}{featureColor}{$s_1$}{}
\nodevar{s2}{1}{4}{featureColor}{$s_2$}{}
\nodevar{xh}{2}{4}{featureColor}{$\tilde{x}$}{}
\nodevar{yh}{3}{4}{inferenceColor}{$\hat{y}$}{}
\nodevar{a}{4}{4}{applicationColor}{$a$}{}

\draw[->] (s1) -- (xh);
\draw[->] (s2) -- (xh);
\draw[->] (xh) -- (yh);
\draw[->] (yh) -- (a);

% ---------- CROSS-LAYER LINK ----------
\draw[->, dashed] (x2) -- (xh);

% ---------- RED, OFFSET, SHORTENED ARROWS ----------
\begin{scope}[red, line width=0.9pt, ->,
              shorten <=10pt, shorten >=10pt,  % trims 5 pt off both ends
              draw=red,              % arrows & outlines only
              text=black, 
              font=\normalsize]

  % 1 — a → ŷ  (offset 4 pt ↑)
  \draw ($ (a)  + (0, 4pt) $) -- ($ (yh) + (0, 4pt) $)
        node[midway, above=1pt] {1};

  % 2 — ŷ → x̃  (offset 4 pt ↑)
  \draw ($ (yh) + (0, 4pt) $) -- ($ (xh) + (0, 4pt) $)
        node[midway, above=1pt] {2};

  % 3 — x̃ → s₁ (offset 4 pt ↑)
  \draw ($ (xh) + (0, 4pt) $) -- ($ (s1) + (0, 4pt) $)
        node[midway, above=1pt] {4};

  \draw ($ (y)  + (4pt,0) $) -- ($ (x2) + (4pt,0) $)
      node[midway,right=1pt]{};
      
    \draw ($ (x2)  + (4pt,0) $) -- ($ (xh) + (4pt,0) $)
          node[midway,right=1pt]{};

  % 5 — y → x₁  (offset 4 pt ↑)
  \draw ($ (z)  + (0, 4pt) $) -- ($ (y) + (0, 4pt) $)
        node[midway, above=1pt] {3.3};

        % 5 — y → x₁  (offset 4 pt ↑)
  \draw ($ (x1)  + (6pt, 0pt) $) -- ($ (x2) + (0, 4pt) $)
        node[midway, above=0pt] {3.1};

  % 6 — x₂ → x₁ (offset 4 pt ↑)
  \draw ($ (u) + (0, 4pt) $) -- ($ (x2) + (0, 4pt) $)
        node[midway, above=1pt] {3.2};

\end{scope}
\end{tikzpicture}
\end{center}

\begin{center}
\footnotesize\makebox[0.95\linewidth][c]{\textit{Simplified diagram with step numbers shown on red investigation arrows.}}
\end{center}

\end{walkthroughbox}

\subsection{Cross-domain Application of C-SAR Triplets}
\label{domain_cases}

This section draws from the primary studies in our review to illustrate the cross-domain application of the C-SAR framework. The following examples show how specific triplets are used for monitoring in \textit{finance}, \textit{healthcare}, and \textit{autonomous systems}, as motivated in the following for each of them.

\subsubsection{Finance}

Finance came up as a domain subject to various forms of data drift, driven by external conditions, evolving user behavior, and adversarial activity. Monitoring is further complicated by delayed or censored label feedback -- particularly in fraud prediction -- and by the need to evaluate models across sensitive subgroups.

To account for expected variation caused by shifting customer behavior, monitoring systems use \textbf{exogenous diagnostics} -- such as time of day and seasonality -- to distinguish natural fluctuations from true anomalies~\cite{zhoux2019a}. Variation also stems from heterogeneous user groups in the domain: \textbf{reference slices} based on attributes like account-family size help isolate high-value segments and reduce over-blocking~\cite{zhoux2019a}, while \textbf{exogenous identifiers} (e.g., gender, ethnicity) are used to expose subgroup disparities -- critical in regulated financial contexts~\cite{castelnovoa2021towards}.

To catch problematic changes in the data, domain knowledge is often encoded in two complementary ways. \textbf{Reference diagrams} capture expected relationships among entities -- such as invoice structures or administrative classifications -- and help flag semantic shifts that may violate model assumptions~\cite{sasthana2021ml}. \textbf{Reference assertions}, on the other hand, define domain-specific patterns in financial data (e.g., ``default risk increases with missed payments'') to identify implausible model outputs and support out-of-distribution (OOD) detection~\cite{dchen2022twostage}.

As reliable label feedback is often unavailable for fraud and default prediction tasks, monitoring can rely on auxiliary sources of supervision. \textbf{Inference assessments} use the outcomes of expert investigations -- such as risk officer reviews of suspected fraud cases -- to serve as proxy labels and provide supervisory signals on model predictions~\cite{vandervorstf2024claims, guann2022fila}. \textbf{Application assessments} extend monitoring to downstream outcomes -- such as the actual claim payouts based on claim classifications -- by analyzing whether changes in these outcomes can be attributed to the model’s behavior~\cite{paleyesandreiandlawrenceneilda}.

\subsubsection{Healthcare}

ML monitoring in healthcare must align with clinical workflows, institutional practices, and the need to ensure patient safety. Failures often arise from demographic shifts, inter-hospital variation, and feedback loops triggered when professionals act on model predictions -- factors that reduce the effectiveness of purely statistical indicators.

To support root-cause analysis and fairness diagnostics, \textbf{reference diagrams} are used to describe dependencies among features -- such as acquisition protocols and other clinical variables -- and to attribute changes in these dependencies to model degradation~\cite{zhangh2023why,schrouffj2022diagnosing}. Complementing this, \textbf{latent diagrams} help healthcare professionals trace such degradation back to upstream causes -- not just features, but specific latent factors like new treatments or emerging diseases~\cite{feng2022clinical}.

\textbf{Latent projections} are used to encode prior knowledge about likely performance shifts -- such as treatment changes or pandemic-induced population shifts -- to adjust monitoring assumptions and detect degradation, even when medical interventions censor part of the label space, limiting visibility into true model performance~\cite{feng2024monitoring}. Similarly, prior knowledge of the expected distribution of output classes after a distribution shift in medical imaging can support performance estimation even when labels are unavailable~\cite{godaup2023deployment}.

Accounting for variation across patient groups and clinical settings is essential in medical ML. Subgroup-level failures are surfaced using \textbf{reference slices} and \textbf{exogenous identifiers}~\cite{allenb2021evaluation,khoshravanazara2023the}, which enable stratified evaluations across known sources of variation -- such as patient demographics, hospital protocols. These identifiers are intentionally selected to support targeted oversight of clinically meaningful subpopulations.

Finally, monitoring can also include clinician behavior. Application assessments evaluate how model outputs influence clinical decision-making -- such as levels of trust in the model, responses to alerts, and choices of intervention -- to detect changes in clinician behavior and identify issues like off-label use~\cite{fengj2024designing}.

\subsubsection{Autonomous systems}

Autonomous systems often rely on perception pipelines built around computer vision models that process high-dimensional inputs such as camera and lidar data. These inputs are difficult to interpret directly, and ground-truth feedback is rarely available during deployment. As a result, monitoring focuses on detecting out-of-distribution (OOD) inputs -- data that fall outside the conditions seen during training. This is necessary in open-world settings, where unanticipated scenarios are common and may pose safety risks~\cite{klsm2019uncertainty, torfahh2022learning, langfordma2023modalas, bensalems2024continuous}.

A common strategy involves \textbf{exogenous diagnostics} -- such as weather, lighting, or time of day -- to identify the context in which an autonomous system operates~\cite{klsm2019uncertainty, torfahh2022learning, langfordma2023modalas, bensalems2024continuous}. Though not used directly for prediction, these variables help determine whether the system remains within its intended operational bounds. Complementary \textbf{application assessments} track behavioral signals -- such as speed or lane-following behavior -- to verify that actions align with pre-specified expectations of behavior~\cite{langfordma2023modalas, bensalems2024continuous}.

Prescriptive rules are further used to monitor the image model and downstream control system. \textbf{Inference guardrails} define expected patterns in model outputs -- for example, flagging temporal inconsistencies when tracked objects appear or disappear abruptly across frames~\cite{kangdanielandguibasjohnandbail}. At the system level, \textbf{application guardrails} specify constraints on final actions -- such as staying in lane or yielding to pedestrians -- informed by upstream predictions like object classification or tracking. These constraints are typically designed and tested in simulation to expose unsafe edge cases, then reused at runtime as monitors to enforce safe behavior during deployment~\cite{langfordma2023modalas, bensalems2024continuous}.

%% file: sections/figures/system_overview.tex
% Process dimensions
\newlength{\processwidth}
\newlength{\processheight}
\newlength{\subprocesswidth}
\newlength{\subprocessheight}
\newlength{\subprocessvspace}
\setlength{\processwidth}{0.49\textwidth}
\setlength{\processheight}{6.5cm}
\setlength{\subprocesswidth}{0.49\textwidth}
\setlength{\subprocessheight}{4.4cm}
\setlength{\subprocessvspace}{0.3cm}

% Colors
\definecolor{naturalColor}{RGB}{8, 230, 0}
\definecolor{LatentColor}{RGB}{69, 150, 66}
\definecolor{externalColor}{RGB}{164, 224, 34}
\definecolor{featureColor}{RGB}{71, 179, 255}
\definecolor{inferenceColor}{RGB}{56, 112, 232}
\definecolor{applicationColor}{RGB}{107, 47, 247}

% Counters for positioning
\newcounter{subprocesscount}
\newcounter{nodecount}

% TikZ styles
\tikzset{
    main/.style={
        draw=none,
        thick,
        rectangle,
        rounded corners=12pt,
        minimum width=\processwidth,
        minimum height=\processheight,
        inner sep=8pt,
        fill=#1!0,
        anchor=center
    },
    sub/.style={
        draw=none,
        rectangle,
        rounded corners=10pt,
        minimum width=\subprocesswidth,
        minimum height=\subprocessheight,
        inner sep=6pt,
        fill=#1!10
    },
    var/.style={
        circle,
        draw=black,
        minimum size=20pt,
        inner sep=2pt,
        fill=#1!40,
        font=\normalsize
    },
    func2/.style={
        rectangle,
        draw=red,
        line width=1pt,
        minimum width=20pt,
        minimum height=20pt,
        inner sep=2pt,
        fill=#1!40,
        font=\normalsize
    },
    func/.style={
        rectangle,
        draw=black,
        minimum width=20pt,
        minimum height=20pt,
        inner sep=2pt,
        fill=#1!40,
        font=\normalsize
    },
    conn/.style={
        ->,
        >=latex,
        thin,
        color=#1
    },
    conn/.default=black, % (Optional) default color if none is given
    dashedvarnode/.style={
    var=#1,     % If you pass something like dashedvarnode={blue}  
    dashed,
    fill=white
  },
  dashedconn/.style={
    ->,        % arrow tip (e.g., Stealth could be used instead)
    thick,     % or any line width
    dashed     % the key bit: make it dashed
  },
  staircasearrow/.style={
    ->,
    >=latex,
    thin,
    color=#1     % or any line width   
  }
}

\newcommand{\mainprocess}[4]{%
    \begin{pgfonlayer}{background}
    \node[main=#3] (#1) at (#4,0cm) {};
    % \node[font=\small, anchor=north] 
    %      at ($(#1.north)+(0.2,-0.2)$) {#2};
    \end{pgfonlayer}
    \setcounter{subprocesscount}{0}%
}

\newcommand{\subprocess}[5]{%
    \begin{pgfonlayer}{background}
    \stepcounter{subprocesscount}%
    \pgfmathsetmacro{\yshift}{(1.05*\subprocessheight)*\thesubprocesscount}%
    \node[sub=#3] (#1) at ($(#4.north)+(0,-\yshift+\subprocessheight-1.4cm)$) {};
    \ifnum\pdfstrcmp{#5}{right}=0
        \node[font=\normalsize\bfseries, text=#3, anchor=west] 
            at ($(#1.west)+(3.01cm,1.7cm)$) {#2};
    \else
        \node[font=\normalsize\bfseries, text=#3, anchor=west] 
            at ($(#1.west)+(0.34cm,1.7cm)$) {#2};
    \fi
    \end{pgfonlayer}
    \setcounter{nodecount}{0}%
}

% Rest of the commands remain the same
\newcommand{\varnode}[4][]{%
    \ifx\\#1\\%
        \node[var=#3] (#2) {#4};
    \else
        \node[var=#3, #1] (#2) {#4};
    \fi
}

\newcommand{\funcnodet}[4][]{%
    \ifx\\#1\\%
        \node[func2=#3] (#2) {#4};
    \else
        \node[func2=#3, #1] (#2) {#4};
    \fi
}

\newcommand{\dashedvarnode}[4][]{%
    % #1 = optional extra node options (e.g., at={...}, or text=..., etc.)
    % #2 = node name
    % #3 = var color/style argument
    % #4 = text/content of the node
    \ifx\\#1\\%
        % If no extra options, just use dashed + white fill
        \node[var=#3, dashed, fill=white] (#2) {#4};
    \else
        % If extra options are given, append them
        \node[var=#3, dashed, fill=white, #1] (#2) {#4};
    \fi
}

\newcommand{\funcnode}[4][]{%
    \ifx\\#1\\%
        \node[func=#3] (#2) {#4};
    \else
        \node[func=#3, #1] (#2) {#4};
    \fi
}

\newcommand{\connectnodes}[4][]{%
    % #1 = options (optional), #2 = from, #3 = to, #4 = color
    \ifx\\#1\\%
        \draw[conn=#4] (#2) -- (#3);
    \else
        \draw[conn=#4] (#2) to[#1] (#3);  % Changed from -- to to[]
    \fi
}

\newcommand{\makeloop}[3]{%
    \draw[conn=#2] ($(#1)+(#3:0.3)$) arc[start angle=#3, end angle=#3+300, radius=0.25];
}

\newcommand{\cornernodes}[5]{%
  % #1 = from node name
  % #2 = anchor side of "from" node (east, west, etc.)
  % #3 = to node name
  % #4 = anchor side of "to" node
  % #5 = color for the arrow
  %
  % We'll compute the midpoint of the x-coords so the vertical
  % segment is perfectly centered horizontally.
  \draw[->, thick, color=#5, dashed]  % or remove 'dashed' if desired
    let \p1 = (#1.#2),  % (x1, y1)
        \p2 = (#3.#4)   % (x2, y2)
    in
    (\p1) 
      -- (\x1 + 0.5*(\x2 - \x1), \y1)  % horizontal to midpoint (x-mid)
      -- (\x1 + 0.5*(\x2 - \x1), \y2)  % vertical to y2
      -- (\p2);                       % final horizontal into the second node
}

\newcommand{\connectdashednodes}[4][]{%
    % #1 = optional TikZ "to[...]" arguments (like out=45, in=135)
    % #2 = 'from' node
    % #3 = 'to' node
    % #4 = color
    \ifx\\#1\\%
        \draw[conn=#4, dashed] (#2) -- (#3);
    \else
        \draw[conn=#4, dashed] (#2) to[#1] (#3);
    \fi
}

\newcommand{\twoVarPill}[4][]{%
 \begin{pgfonlayer}{middle}
   \node[
     draw,
     thin,
     shape=rectangle,
     rounded corners=1.5em,
     inner sep=0.15cm,
     outer sep=0.0cm,
     #1,
     fit=(#3)(#4)
   ] (#2) {};
 \end{pgfonlayer}
}

\newcommand{\twoFuncRect}[4][]{%
 \begin{pgfonlayer}{middle}
   \node[
     draw,
     thin,
     shape=rectangle,
     inner sep=0.15cm,
     outer sep=0.15cm,
     #1,
     fit=(#3)(#4)
   ] (#2) {};
 \end{pgfonlayer}
}

% \staircasearrow{fromNode}{fromDir}{toNode}{toDir}{color}{elbow}{xShift}{yShift}
\newcommand{\staircasearrow}[9]{%
  % #1 = from node
  % #2 = from direction (east|west|north|south)
  % #3 = to node
  % #4 = to direction (east|west|north|south)
  % #5 = arrow color (used in conn)
  % #6 = elbow offset
  % #7 = x-shift for entire arrow
  % #8 = y-shift for entire arrow
  \draw[conn, rounded corners=5pt]
    %
    % 1) Start at (fromNode.fromDir), plus the (xShift, yShift)
    ($ (#1.#2) + (#7,#8) $)
    %
    % 2) Draw an elbow in the direction of #2 with length #6
    -- +(
      \ifnum\pdfstrcmp{#2}{east}=0   #6,0
      \else\ifnum\pdfstrcmp{#2}{west}=0  -#6,0
      \else\ifnum\pdfstrcmp{#2}{north}=0 0,#6
      \else 0,-#6
      \fi\fi\fi
    )
    %
    % 3) “Staircase” horizontally aligned with #1's direction,
    %    but at the level of (#3.#4), all shifted by (#7,#8)
    -- ($(#3.#4 -| #1.#2) + (#7,#8)$)
    %
    % 4) Finally go to (#3.#4) + the same shift
    -- ($ (#3.#4) + (#7,#8) $);
}

\newcommand{\processbarchart}[4]{%
    % #1 = subprocess node name (for positioning)
    % #2 = list of y-coordinates (categories)
    % #3 = list of x-coordinates (values)
    % #4 = alignment (l or r)
    \if r#4
        \node[anchor=west, text width=7cm] at ($(#1.west)+(6cm,-1.3cm)$) {
            \begin{axis}[
                width=2.9cm,
                height=3cm,
                xbar,
                xmin=0,
                bar width=0.15cm,
                nodes near coords,
                nodes near coords style={font=\scriptsize},
                tick label style={font=\scriptsize, align=left},
                ytick=data,
                symbolic y coords={#2},
                y dir=reverse,
                % Remove x ticks and labels
                xtick=\empty,
                % Remove right axis line
                axis y line*=left,
                % Remove x axis line
                axis x line=none,
                % Remove grid
                grid=none,
                % Add y axis padding
                ylabel style={align=center},
                y tick label style={align=left}
            ]
            \addplot[fill=white!30] coordinates {#3};
            \end{axis}
        };
    \else
        \node[anchor=east, text width=7cm] at ($(#1.east)+(2cm,-1.3cm)$) {
            \begin{axis}[
                width=2.9cm,
                height=3cm,
                xbar,
                xmin=0,
                bar width=0.15cm,
                nodes near coords,
                nodes near coords style={font=\scriptsize},
                tick label style={font=\scriptsize, align=right},
                ytick=data,
                symbolic y coords={#2},
                y dir=reverse,
                % Remove x ticks and labels
                xtick=\empty,
                % Remove right axis line
                axis y line*=left,
                % Remove x axis line
                axis x line=none,
                % Remove grid
                grid=none,
                % Add y axis padding
                ylabel style={align=center},
                y tick label style={align=right}
            ]
            \addplot[fill=white!30] coordinates {#3};
            \end{axis}
        };
    \fi
}

\newcommand{\bifuncnodes}[4]{%
    % #1 = left node name
    % #2 = right node name
    % #3 = function node name
    % #4 = color
    \def\arrowoffset{0.15cm}  % Vertical separation between arrows
    
    % Draw function node in center
     \node[minimum width=20pt, minimum height=20pt, inner sep=2pt] at ($(#1)!0.5!(#2)$) (#3) {};
    
    % Draw top arrows (all pointing right)
    \draw[conn=#4] ($(#1.east)+(0,\arrowoffset)$) -- ($(#3.west)+(0,\arrowoffset)$);
    \draw[conn=#4] ($(#3.east)+(0,\arrowoffset)$) -- ($(#2.west)+(0,\arrowoffset)$);
    
    % Draw bottom arrows (all pointing left)
    \draw[conn=#4] ($(#3.west)+(0,-\arrowoffset)$) -- ($(#1.east)+(0,-\arrowoffset)$);
    \draw[conn=#4] ($(#2.west)+(0,-\arrowoffset)$) -- ($(#3.east)+(0,-\arrowoffset)$);
}

\makeatother

% Define standard offsets as proportions of box width
\def\firstpos{0.52}    % 20% from edge
\def\midpos{0.82}      % Center
\def\lastpos{0.92}     % 80% from edge
\def\voffset{0.5cm}   % Vertical offset for stacked nodes
\def\leftpos{0.74}   
\def\rightpos{0.90}
\def\vdist{0.8}
\def\vdistt{1.1}

\def\tightleftpos{0.76}   
\def\tightrightpos{0.88}

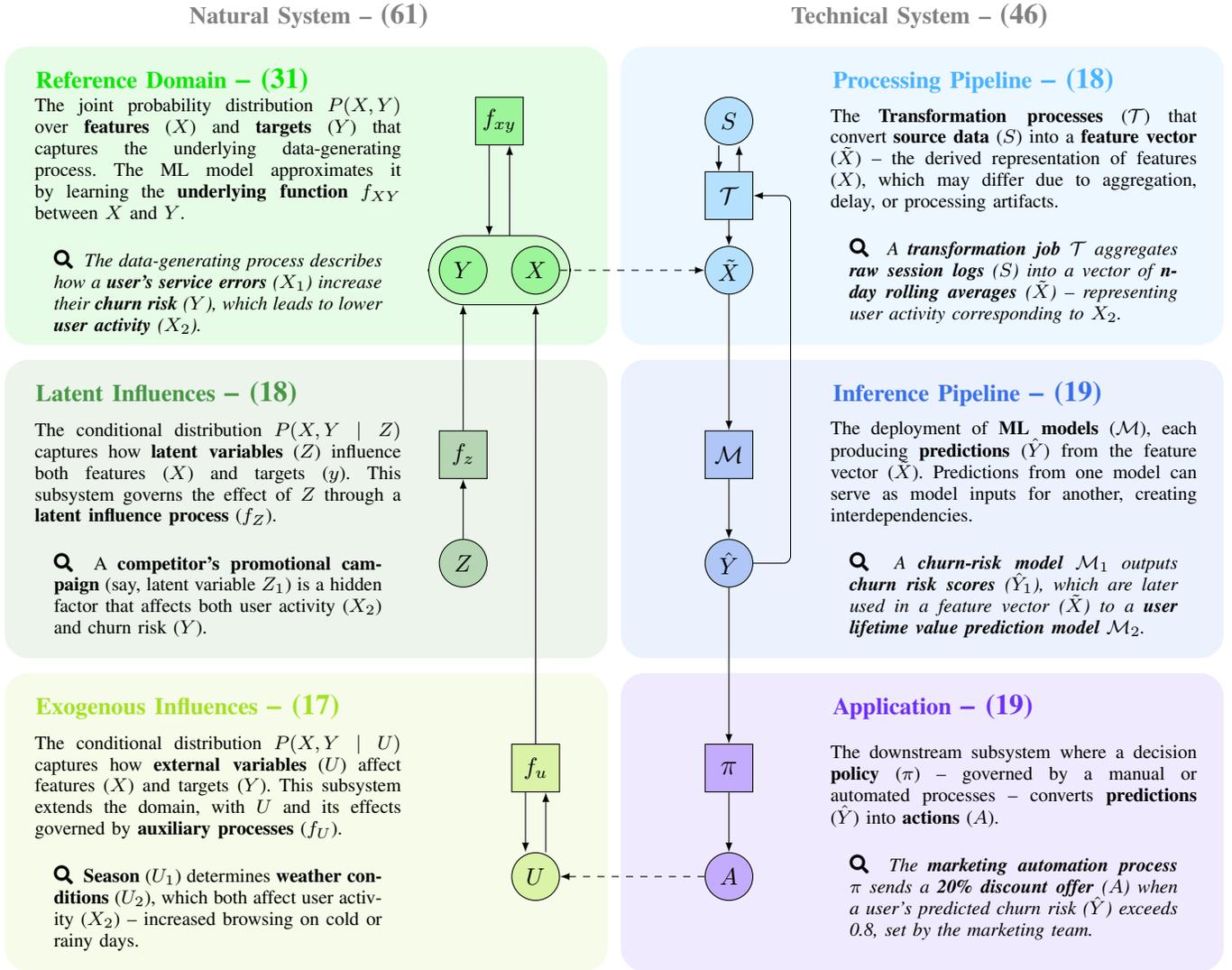
\begin{figure*}[b!]
    \centering

    % Define circle, square, and arrow with adjustments
    \newcommand{\VarSymbol}{\tikz[baseline=-0.6ex] \draw[fill=none, line width=0.5pt] (0,0) circle [radius=0.15cm];}
    \newcommand{\SquareSymbol}{\tikz[baseline=0.35ex] \draw[fill=none, line width=0.5pt] (0,0) rectangle (0.25,0.25);}
    
    % Adjust arrow length and alignment
    \newcommand{\FuncSymbol}{\tikz[baseline=-0.6ex] \draw[-latex, line width=0.8pt, black] (-0.1,0) -- (0.4,0);}

    % Adjust arrow length and alignment
    \newcommand{\FuncSymbolDashed}{\tikz[baseline=-0.6ex] \draw[-latex, line width=0.5pt, black, dashed] (-0.1,0) -- (0.4,0);}

    \vspace{0.2em}

    % Use these with relative calculations
    \begin{tikzpicture}[xshift=-0cm]
    \centering
        \pgfdeclarelayer{background}   % Bottom layer for process boxes
        \pgfdeclarelayer{middle}       % Middle layer for pills/rectangles
        \pgfdeclarelayer{main}         % Top layer for nodes and connections
        \pgfsetlayers{background,middle,main}

        \node[font=\bfseries, color=gray] at (-4.5cm,4.3cm) {Natural System -- \large(61)};
        \node[font=\bfseries, color=gray] at (4.5cm,4.3cm) {Technical System -- \large(46)};
    
        % Create main processes
        \mainprocess{natural_proc}{Natural Process}{gray}{-4.54cm}
        \mainprocess{system_proc}{System Process}{gray}{4.54cm}
        
         % Create subprocesses in Natural Process (left labels)
        \subprocess{primary_proc}{Reference Domain {\large %$P(X,Y)$
        -- (31)
        }}{naturalColor}{natural_proc}{left}
        \subprocess{hidden_inf}{Latent Influences {\large %$P(X,Y \mid L)$
        -- (18)
        }}{LatentColor}{natural_proc}{left}
        \subprocess{aux_inf}{Exogenous Influences {\large %$P(X, Y \mid E, A)$
        -- (17)
        }}{externalColor}{natural_proc}{left}
        \setcounter{subprocesscount}{0}
        % Create subprocesses in System Process (right labels)
        \subprocess{pre_proc}{Processing Pipeline  {\large %$\mathcal{T}(X,Y,S \mid S, \hat{Y}_a)$
        -- (18)
        }}{featureColor}{system_proc}{right}
        \subprocess{inference}{Inference Pipeline {\large %$\mathcal{M}(\hat{Y} \mid X)$
        -- (19)
        }} {inferenceColor}{system_proc}{right}
        \subprocess{application}{Application {\large 
        %$\pi(A \mid \hat{Y},E)$
        -- (19)
        }}{applicationColor}{system_proc}{right}

        % Add nodes to Hidden Influences - using relative positioning
        \path let \p1 = ($(hidden_inf.east)-(hidden_inf.west)$) in \pgfextra{
            \varnode[at={($(hidden_inf.west)+(\tightleftpos*\x1,-\vdist)$)}]{h}{LatentColor}{$Z$}
            \funcnode[at={($(hidden_inf.west)+(\tightleftpos*\x1,\vdist)$)}]{h_func}{LatentColor}{$f_z$}
        };
        
        % Add nodes to Primary Process
        \path let \p1 = ($(primary_proc.east)-(primary_proc.west)$) in \pgfextra{
            \varnode[at={($(primary_proc.west)+(\tightleftpos*\x1,-\vdistt)$)}]{y}{naturalColor}{$Y$}
            \varnode[at={($(primary_proc.west)+(\tightrightpos*\x1,-\vdistt)$)}]{x}{naturalColor}{$X$}
            \funcnode[at={($(primary_proc.west)+(\midpos*\x1,\vdistt)$)}]{yx_func}{naturalColor}{ $f_{xy}$}
        };
        
        % Add nodes to Auxiliary Influences
        \path let \p1 = ($(aux_inf.east)-(aux_inf.west)$) in \pgfextra{
            \varnode[at={($(aux_inf.west)+(\tightrightpos*\x1,-\vdist)$)}]{e}{externalColor}{$U$}
            \funcnode[at={($(aux_inf.west)+(\tightrightpos*\x1,\vdist)$)}]{e_func}{externalColor}{$f_u$}
        };
        
        % Add nodes to Pre-Processing (mirror from right)
        \path let \p1 = ($(pre_proc.east)-(pre_proc.west)$) in \pgfextra{
            \varnode[at={($(pre_proc.east)+(-\midpos*\x1,\vdistt)$)}]{d}{featureColor}{$S$}
            \funcnode[at={($(pre_proc.east)+(-\midpos*\x1,0)$)}]{p_func}{featureColor}{$\mathcal{T}$}
            \varnode[at={($(pre_proc.east)+(-\midpos*\x1,-\vdistt)$)}]{fs}{featureColor}{$\tilde{X}$}
        };
        
        % Add nodes to Inference (mirror from right)
        \path let \p1 = ($(inference.east)-(inference.west)$) in \pgfextra{
            \varnode[at={($(inference.east)+(-\midpos*\x1,-\vdist)$)}]{yp}{inferenceColor}{$\hat{Y}$}
            \funcnode[at={($(inference.east)+(-\midpos*\x1,\vdist)$)}]{m_func}{inferenceColor}{$\mathcal{M}$}
        };
        
        % Add nodes to Application (mirror from right)
        \path let \p1 = ($(application.east)-(application.west)$) in \pgfextra{
            \varnode[at={($(application.east)+(-\midpos*\x1,-\vdist)$)}]{a}{applicationColor}{$A$}
            \funcnode[at={($(application.east)+(-\midpos*\x1,\vdist)$)}]{a_func}{applicationColor}{\large$\pi$}
        };
        
        \twoVarPill[fill=naturalColor!20]{pill1}{y}{x}

        \draw[conn] ([shift={(0.15,0.0)}]p_func.north) -- ([shift={(0.15,0)}]d.south);
        \draw[conn] ([shift={(-0.15,0.0)}]d.south) -- ([shift={(-0.15,0)}]p_func.north);
        
        \draw[conn] ([shift={(0.15,0.0)}]pill1.north) -- ([shift={(0.15,0)}]yx_func.south);
        \draw[conn] ([shift={(-0.15,0.0)}]yx_func.south) -- ([shift={(-0.15,0)}]pill1.north);
    
        \draw[conn] (h.north) -- (h_func.south);
        
        \draw[conn] ([shift={(0.15,0.0)}]e.north) -- ([shift={(0.15,0)}]e_func.south);
        \draw[conn] ([shift={(-0.15,0.0)}]e_func.south) -- ([shift={(-0.15,0)}]e.north);
        
        \draw[conn] (h_func.north) -- ([shift={(-0.535,0)}]pill1.south);
        \draw[conn] (e_func.north) -- ([shift={(0.535,0)}]pill1.south);

        % \draw[conn, rounded corners=6pt] ([xshift=-0.0cm]x.south) -- +(0,-0.35) -- ++(2.96,-0.35) -- ([yshift=-0.1cm]pill3.north);

        \draw[conn, rounded corners=6pt] ([xshift=0cm]p_func.south)  -- +(0,-0.44) -- ([xshift=0cm]fs.north);
    
        \draw[conn] ([xshift=-0.0cm]m_func.south) -- ([xshift=-0.0cm]yp.north);

        \draw[conn] (fs.south) -- (m_func.north);
        \draw[conn] (yp.south) -- (a_func.north);
    
        \draw[conn, rounded corners=3pt] 
            ([xshift=0cm]yp.east)              % Start at the east of yp
            -- ++(0.55,0)                      % Step 1: go right
            -- ++(0,5.42)                       % Step 2: go up
            -- ([xshift=0cm, yshift=0.0cm]p_func.east); % Step 3: go left to fs.east with a slight offset

        \draw[conn] (a_func.south) -- (a.north);
    
        % \node[font=\footnotesize, gray] 
        %   at ($(a.west)!0.5!(e.east)+(0,0.25)$) {Intervention};
        
        % \node[font=\footnotesize, gray] 
        %   at ($(x.east)!0.5!(fs.west)+(0,0.25)$) {Encoding};

        \draw[conn,dashed] (a.west) -- (e.east);
        \draw[conn,dashed] (x.east) -- (fs.west);
        
         % Modeled System - P(X,Y)
        \node[text width=5.4cm, align=left, color=black, font=\footnotesize, align=justify] 
            at ($(primary_proc.west)+(3.15cm,-0.3cm)$) 
            {
            The joint probability distribution $P(X,Y)$ over \textbf{features} ($X$) and \textbf{targets} ($Y$) that captures the underlying data-generating process. The ML model approximates it by learning the \textbf{underlying function} $f_{XY}$ between $X$ and $Y$.
            \\
           \begingroup
            \renewenvironment{quote}
              {\list{}{\leftmargin=1em \rightmargin=1em}\item\relax}
              {\endlist}
            \begin{quote}

            \textit{\textbf{\faSearch\ } The data-generating process describes how a \textbf{user's service errors} ($X_1$) increase their \textbf{churn risk} ($Y$), which leads to lower \textbf{user activity} ($X_2$).}
            \end{quote}
            
            \endgroup
            };
        
        % Hidden Influences – H(X,Y)
        \node[text width=5.4cm, align=left, color=black, font=\footnotesize, align=justify] 
            at ($(hidden_inf.west)+(3.15cm,-0.3cm)$) 
            {
            The conditional distribution $P(X, Y \mid Z)$ captures how \textbf{latent variables} ($Z$) influence both features ($X$) and targets ($y$). This subsystem governs the effect of $Z$ through a \textbf{latent influence process} ($f_Z$).
            \\
            \begingroup
              \renewenvironment{quote}
                {\list{}{\leftmargin=1em \rightmargin=1em}\item\relax}
                {\endlist}
              \begin{quote}
              \textbf{\faSearch\ } A \textbf{competitor’s promotional campaign} (say, latent variable $Z_1$) is a hidden factor that affects both user activity ($X_2$) and churn risk ($Y$).
              \end{quote}
            \endgroup
            };
        
        % Auxiliary Influences – E(X,Y)
        \node[text width=5.4cm, align=left, color=black, font=\footnotesize, align=justify] 
            at ($(aux_inf.west)+(3.15cm,-0.3cm)$) 
            {
            The conditional distribution $P(X, Y \mid U)$ captures how \textbf{external variables} ($U$) affect features ($X$) and targets ($Y$). This subsystem extends the domain, with $U$ and its effects governed by \textbf{auxiliary processes} ($f_U$).
            \\
            \begingroup
              \renewenvironment{quote}
                {\list{}{\leftmargin=1em \rightmargin=1em}\item\relax}
                {\endlist}
              \begin{quote}
              \textbf{\faSearch\ } \textbf{Season} ($U_1$) determines \textbf{weather conditions} ($U_2$), which both affect user activity ($X_2$) -- increased browsing on cold or rainy days.
              \end{quote}
            \endgroup
            };
        
        % Data Pipeline – T(X̃)
        \node[text width=5.4cm, align=left, color=black, font=\footnotesize, align=justify] 
            at ($(pre_proc.west)+(5.8cm,-0.3cm)$) 
            {
            The \textbf{Transformation processes} ($\mathcal{T}$) that convert \textbf{source data} ($S$) into a \textbf{feature vector} ($\tilde{X}$) -- the derived representation of features ($X$), which may differ due to aggregation, delay, or processing artifacts.
            \\
            \begingroup
              \renewenvironment{quote}
                {\list{}{\leftmargin=1em \rightmargin=1em}\item\relax}
                {\endlist}
              \begin{quote}
              \textit{\textbf{\faSearch\ } A \textbf{transformation job} $\mathcal{T}$ aggregates \textbf{raw session logs} ($S$) into a vector of \textbf{n-day rolling averages} ($\tilde{X}$) -- representing user activity corresponding to $X_2$.}
              \end{quote}
            \endgroup
            };
        
        % Inference Pipeline – M(Ŷ, Ŷₐ)
        \node[text width=5.4cm, align=left, color=black, font=\footnotesize, align=justify] 
            at ($(inference.west)+(5.8cm,-0.3cm)$) 
            {
            The deployment of \textbf{ML models} ($\mathcal{M}$), each producing \textbf{predictions} ($\hat{Y}$) from the feature vector ($\tilde{X}$). Predictions from one model can serve as model inputs for another, creating interdependencies.
            \\
            \begingroup
              \renewenvironment{quote}
                {\list{}{\leftmargin=1em \rightmargin=1em}\item\relax}
                {\endlist}
              \begin{quote}
              \textit{\textbf{\faSearch\ } A \textbf{churn-risk model} $\mathcal{M}_1$ outputs \textbf{churn risk scores} ($\hat{Y}_1$), which are later used in a feature vector ($\tilde{X}$) to a \textbf{user lifetime value prediction model} $\mathcal{M}_2$.}
              \end{quote}
            \endgroup
            };
        
        % Application System – A(Ŷ, E)
        \node[text width=5.4cm, align=left, color=black, font=\footnotesize, align=justify] 
            at ($(application.west)+(5.8cm,-0.3cm)$) 
            {
            The downstream subsystem where a decision \textbf{policy} ($\pi$) -- governed by a manual or automated processes -- converts \textbf{predictions} ($\hat{Y}$) into \textbf{actions} ($A$).
            \\
            \begingroup
              \renewenvironment{quote}
                {\list{}{\leftmargin=1em \rightmargin=1em}\item\relax}
                {\endlist}
              \begin{quote}
              \textit{\textbf{\faSearch\ } The \textbf{marketing automation process} $\pi$ sends a \textbf{20\% discount offer} ($A$) when a user's predicted churn risk ($\hat{Y}$) exceeds 0.8, set by the marketing team.}
              \end{quote}
            \endgroup
            };

        \end{tikzpicture}
        % \caption{System diagram with aligned nodes showing environment and system processes.}

\caption{Overview of the subject \textit{systems} -- \emph{natural} (1) and \emph{technical} (2) -- considered relevant to ML monitoring, with (sub)systems annotated by their unique occurrences in the primary studies. Systems are composed of data elements ( \protect\VarSymbol \ ), processes ( \protect\SquareSymbol \ ), and solid connectors ( \protect\FuncSymbol ). These connectors are defined within the process nodes' code specifications; they can be interpreted as either data flows between processes (annotated by data elements) or functional links between data elements (defined by processes). Dashed connectors ( \protect\FuncSymbolDashed ) indicate conceptual mappings: \textit{intervention} -- actions $A$ becoming external variables $U$ at t+1; \textit{encoding} -- the engineered vector $\tilde{X}$ is a processed view of the natural feature $X$.}

\label{fig:system_diagram}
\end{figure*}

%% file: sections/figures/context_info_taxonomy.tex
\begin{table*}[!b]
\centering
\setlength{\tabcolsep}{6pt}
\renewcommand{\arraystretch}{1.4} % Increased for better spacing with row colors
\normalsize
\begin{adjustbox}{max width=\textwidth}

% We now have 5 columns: Aspect, Definition, Example, #, References
\begin{tabular}{
  p{0.15\textwidth} % aspect
  p{0.33\textwidth} % definition
  p{0.33\textwidth} % example
  c                 % #
  >{\footnotesize}p{0.24\textwidth} % references
}
\toprule
\textbf{Aspect} & \textbf{Definition} & \textbf{\faSearch\ Example} & \textbf{\#} & \textbf{References} \\
\midrule

%----------------------------------------------------------------------------------
% STATE (top level, colored row)
%----------------------------------------------------------------------------------
\rowcolor{featureColor!20} % Apply color to the entire row
\LevelOne{State}
  & A system’s observed situation at runtime.
  & --
  & 39
  & -- \\
\cmidrule{1-5}

% Conditions (second level)
\LevelTwo{Conditions}
  & External or internal circumstances that influence the behavior of the system.
  & The status of a data transformation job computing user activity features.
  & 26
  & \cite{allenb2021evaluation,baquon2022concept,bensalems2024continuous,%
    cobbo2022contextaware,ebreck2017the,fengj2024designing,foidlh2019riskbased,%
    gomesjb2010calds,heynhm2023automotive,jbgomes2014mining,kirchheimk2023towards,%
    kkirchheim2024outofdistributio,klsm2019uncertainty,langfordma2023modalas,%
    muirurid2022practices,nguyenmt2024novel,onnesa2023bayesian,%
    paleyesandreiandlawrenceneilda,sculley2015hidden,swami2020data,%
    torfahh2022learning,tzoppi2021detect,wangs2021a,xuz2023alertiger,%
    xxu2022dependency,zhoux2019a} \\
\cmidrule{1-5}

% Behaviour / Evaluation (second level)
\LevelTwo{Evaluations}
  & Observed outcomes of runtime system behavior.
  & The impact of retention offers, informed by churn predictions, on targeted users.
  & 17
  & \cite{bensalems2024continuous,bernardil2019150,binderf2022putting,%
    cabreraa2021discovering,csun2024ai,ebreck2017the,edsnascimento2019understanding,%
    ginartaa2022mldemon,guann2022fila,hanafimf2024machineassisted,%
    jayalathh2022enhancing,jayalathh2023continual,langfordma2023modalas,%
    paleyesandreiandlawrenceneilda,shergadwalamn2022a,vandervorstf2024claims,%
    vishwakarmah2024taming} \\
\cmidrule{1-5}

%----------------------------------------------------------------------------------
% STRUCTURE (top level, colored row)
%----------------------------------------------------------------------------------
\rowcolor{inferenceColor!20} % Apply color to the entire row
\LevelOne{Structure}
  & Segmentations and interdependencies.
  & --
  & 34
  & -- \\
\cmidrule{1-5}

% Entities (2nd level)
\LevelTwo{Entities}
  & Specific data-generating segments tied to specific actors or objects.
  & A segment of young adult users with distinct churn behavior.
  & 19
  & \cite{azonoozi2016contrack,castelnovoa2021towards,drevesm2020from,%
    fedelea2024the,fengj2024designing,ghosha2022fair,henzingerta2023monitoring,%
    henzingerthomasandkarimimahyar,khoshravanazara2023the,mammanh2024biastrap,%
    manerikerp2023online,rc2020overton,sackerman2021machine,schelter2018automating,%
    schrouffj2022diagnosing,swami2020data,vasudevans2020lift,yangz2021biasrv,%
    zhoux2019a} \\
\cmidrule{1-5}

% Relations (2nd level)
\LevelTwo{Relations}
  & Structured associations between elements within the system.
  & A hypothesized causal link between the quality of service and churn.
  & 17
  & \cite{baquon2022concept,bontempellia2022humanintheloop,%
    borchanih2015modeling,budhathokik2021why,dreyfuspa2022databased,ebreck2017the,%
    feng2022clinical,fengj2024designing,heynhm2023automotive,%
    kkirchheim2024outofdistributio,namakimh2020vamsa,paleyesandreiandlawrenceneilda,%
    sasthana2021ml,schrouffj2022diagnosing,shankars2022towards,xxu2022dependency,%
    zhangh2023why} \\
\cmidrule{1-5}

%----------------------------------------------------------------------------------
% PROPERTIES (top level, colored row)
%----------------------------------------------------------------------------------
\rowcolor{applicationColor!20} % Apply color to the entire row
\LevelOne{Properties}
  & Anticipated data characteristics.
  & --
  & 45
  & -- \\
\cmidrule{1-5}
\cmidrule{1-5}

% Nominal (2nd level)
\LevelTwo{Nominal}
  & Assumed or stable data characteristics under normal conditions.
  & A user's weekly activity cannot exceed the total number of hours in a week (168).
  & 13
  & \cite{bachingerf2024data,cavenesse2020tensorflow,chenl2022estimating,%
    dchen2022twostage,ehrlingerl2019a,kirchheimk2023towards,lelwakatare2021on,%
    liu2020towards,myllyahol2022on,pichlerg2024on,schelter2018automating,%
    shankars2024we,swami2020data} \\
\cmidrule{1-5}

% Event Properties (2nd level)
\LevelTwo{Event}
  & Expected characteristics of a predictable event, describing anticipated changes in the data.
  & An upcoming product launch expected to shift user attention and affect churn patterns.
  & 21
  & \cite{apaul2024mlops,chenm2021mandoline,dreyfuspa2022databased,%
    feng2022clinical,feng2024monitoring,galewis2022augur,godaup2023deployment,%
    haidert2021domain,henzingerta2023monitoring,leestj2024expert,%
    masegosaar2020analyzing,pichlerg2024on,pwelinder2013a,schelters2020learning,%
    schelters2021jenga,sibliniw2020master,sobolewskip2017scr,%
    xuanjunyuandlujieandzhangguang,xuz2023alertiger,%
    zhoux2019a,borchanih2015modeling} \\
\cmidrule{1-5}

% Normative Properties (2nd level)
\LevelTwo{Normative}
  & Prescribed limits or compliance on system conditions and behavior.
  & Retention offers should not be sent to more than 10\% of our user base.
  & 13
  & \cite{barkerm2023feedbacklogs,henrikssonj2023outofdistributi,%
    klsm2019uncertainty,amarco2023out,bensalems2024continuous,csun2024ai,%
    habdelkader2024mlonrails,kangdanielandguibasjohnandbail,langfordma2023modalas,%
    myllyahol2022on,onnesa2022monitoring,onnesa2023bayesian,sculley2015hidden} \\
\bottomrule
\end{tabular}
\end{adjustbox}

\vspace{0.6em}
\caption{Taxonomy of system aspects, with definitions, examples, unique study count (\#), and references.}
\label{fig:context_taxonomy_updated}
\end{table*}

%% file: sections/figures/representation_taxonomy.tex
\begin{table*}[!b]
\centering
\setlength{\tabcolsep}{6pt}
\renewcommand{\arraystretch}{1.4}
\normalsize
\begin{adjustbox}{max width=\textwidth}
\begin{tabular}{
  p{0.14\textwidth} % Representation
  p{0.32\textwidth} % Definition
  p{0.31\textwidth} % Definition
  c                 % #
  >{\footnotesize}p{0.23\textwidth} % References
}
\toprule
\textbf{Representation} & \textbf{Definition} & \textbf{\faSearch\ Example} & \textbf{\#} & \textbf{References} \\
\midrule

%----------------------------------------------------------------------------------
% FORMAL REPRESENTATION (top level)
%----------------------------------------------------------------------------------
\rowcolor{featureColor!20}
\LevelOne{Formal}
  & Encoding using syntax and semantics.
  & -
  & 78
  & - \\
\cmidrule{1-5}

% Numerical Representation (second level)
\LevelTwo{Numerical}
  & Quantitative measurements representing values.
  & Last week's retention offer acceptance rate was 0.28.
  & 40
  & \cite{zhoux2019a,klsm2019uncertainty,castelnovoa2021towards,binderf2022putting,torfahh2022learning,khoshravanazara2023the,kirchheimk2023towards,bensalems2024continuous,langfordma2023modalas,vandervorstf2024claims,allenb2021evaluation,tzoppi2021detect,wangs2021a,jbgomes2014mining,kkirchheim2024outofdistributio,gomesjb2010calds,vasudevans2020lift,yangz2021biasrv,xxu2022dependency,guann2022fila,paleyesandreiandlawrenceneilda,manerikerp2023online,onnesa2023bayesian,ginartaa2022mldemon,cobbo2022contextaware,vishwakarmah2024taming,fengj2024designing,schrouffj2022diagnosing,ebreck2017the,baquon2022concept,edsnascimento2019understanding,muirurid2022practices,csun2024ai,bernardil2019150,xuz2023alertiger,hanafimf2024machineassisted,nguyenmt2024novel,shergadwalamn2022a,sculley2015hidden,foidlh2019riskbased} \\
\cmidrule{1-5}

% Probabilistic Representation (second level)
\LevelTwo{Probabilistic}
  & Statistical representations of a data distribution using parameters.
  & Session duration with a mean ($\mu$) of 300s and standard deviation ($\sigma$) of 50s.
  & 15
  & \cite{sibliniw2020master,henzingerta2023monitoring,borchanih2015modeling,sobolewskip2017scr,schelters2020learning,galewis2022augur,masegosaar2020analyzing,schelters2021jenga,pichlerg2024on,godaup2023deployment,amarco2023out,pwelinder2013a,leestj2024expert,onnesa2023bayesian,feng2024monitoring} \\
\cmidrule{1-5}

% Logical Representation (second level)
\LevelTwo{Logical}
  & Formal statements, expressions and rules.
  & IF user country equals 'US' THEN user state IS NOT NULL.
  & 38
  & \cite{zhoux2019a,klsm2019uncertainty,ehrlingerl2019a,henrikssonj2023outofdistributi,kirchheimk2023towards,bensalems2024continuous,langfordma2023modalas,myllyahol2022on,sackerman2021machine,azonoozi2016contrack,lelwakatare2021on,csun2024ai,cavenesse2020tensorflow,drevesm2020from,xuanjunyuandlujieandzhangguang,ghosha2022fair,xuz2023alertiger,henzingerthomasandkarimimahyar,kangdanielandguibasjohnandbail,shankars2024we,habdelkader2024mlonrails,chenm2021mandoline,rc2020overton,fengj2024designing,swami2020data,schelter2018automating,liu2020towards,sculley2015hidden,dchen2022twostage,bachingerf2024data,onnesa2022monitoring,chenl2022estimating,fedelea2024the, henzingerta2023monitoring, mammanh2024biastrap,heynhm2023automotive,shankars2022towards,namakimh2020vamsa} \\
\cmidrule{1-5}

%----------------------------------------------------------------------------------
% INFORMAL REPRESENTATION (top level)
%----------------------------------------------------------------------------------
\rowcolor{applicationColor!20}
\LevelOne{Informal}
  & Formats that lack a strict syntax.
  & -
  & 26
  & - \\
\cmidrule{1-5}

% Graphical Representation (second level)
\LevelTwo{Graphical}
  & Visual and graph-based representations.
  & A dependency graph linking data elements: users, sessions $\rightarrow$ user activity.
  & 11
  & \cite{borchanih2015modeling,dreyfuspa2022databased,paleyesandreiandlawrenceneilda,fengj2024designing,zhangh2023why,schrouffj2022diagnosing,budhathokik2021why,feng2022clinical,bontempellia2022humanintheloop,sasthana2021ml,kkirchheim2024outofdistributio} \\
\cmidrule{1-5}

% Semi-Structured Representation (second level)
\LevelTwo{Semi-Structured}
  & Partially organized representations of context information.
  & Textual note on an upcoming campaign expected to increase traffic.
  & 17
  & \cite{dreyfuspa2022databased,ebreck2017the,heynhm2023automotive,apaul2024mlops,cabreraa2021discovering,jayalathh2022enhancing,xuz2023alertiger,barkerm2023feedbacklogs,leestj2024expert,jayalathh2023continual,haidert2021domain,liu2020towards,feng2022clinical,foidlh2019riskbased,xxu2022dependency,swami2020data,baquon2022concept} \\

\bottomrule
\end{tabular}
\end{adjustbox}

\vspace{0.6em}
\caption{Taxonomy of contextual representations, with definitions, examples, unique study count (\#), and references.}
\label{fig:representation_taxonomy}
\end{table*}

%% file: sections/figures/3d_matrix.tex
% === New parameters for pie sizes and outline thickness ===
\definecolor{naturalColor}{RGB}{8,230,0}
\definecolor{LatentColor}{RGB}{69,150,66}
\definecolor{externalColor}{RGB}{164,224,34}
\definecolor{featureColor}{RGB}{71,179,255}
\definecolor{inferenceColor}{RGB}{56,112,232}
\definecolor{applicationColor}{RGB}{107,47,247}

\begin{figure*}[!b]
\newcommand{\PieMinSize}{0.25cm}         % Minimum pie (and circle) size
\newcommand{\PieMaxSize}{1cm}            % Maximum pie (and circle) size
\newcommand{\PieOutlineThickness}{0.3mm} % Outline thickness for all pies and circles

% Global scale for pies/circles (to control their size)
\newcommand{\pieScale}{1.5} % Adjust this value to scale pies and circles

   % Define our colors
   \definecolor{naturalColor}{RGB}{8, 230, 0}
   \definecolor{LatentColor}{RGB}{69, 150, 66}
   \definecolor{externalColor}{RGB}{164, 224, 34}
   \definecolor{featureColor}{RGB}{71, 179, 255}
   \definecolor{inferenceColor}{RGB}{56, 112, 232}
   \definecolor{applicationColor}{RGB}{107, 47, 247}

    \centering

    \hbox{
    \begin{minipage}{0.26\textwidth}  % Reduced from 0.45
        \begin{tikzpicture}[scale=1,
               xscale=0.44,
               yscale=1.35,
            % Base style with two parameters: height and shift
            level1/.style 2 args={
                draw=none,
                minimum width=0.15\textwidth, 
                minimum height=2.25cm,
                fill=none, 
                font=\normalsize
            },
            level2/.style 2 args={
                draw=none,
                minimum width=0.42\textwidth, 
                minimum height=#1,
                fill=none, 
                font=\normalsize,
                path picture={
                    \draw[line width=0.8mm] (path picture bounding box.north east) ++(0,#2) -- ++(0,-#1);
                    \draw[thin] (path picture bounding box.east) -- ++(0cm,0);
                }
            },
            level3/.style 2 args={
                draw=none,
                minimum width=0.55\textwidth, 
                minimum height=#1,
                fill=none,
                align=center,
                font=\normalsize,
                path picture={
                    \draw[white] (path picture bounding box.north east) ++(0,0) -- ++(0,0);
                }
            },
           level4/.style n args={2}{
                draw=none,
                minimum width=0.3\textwidth, 
                minimum height=1cm,
                align=center,
                font=\footnotesize
            },
           level5/.style n args={3}{
                draw=none,
                minimum width=0.4\textwidth, 
                minimum height=1cm,
                rounded corners=2pt,
                align=center,
                fill=none,
                font=\normalsize
            }
        ]
        
        % \draw[white, very thick]{(-5.5cm,6.5cm) -- (5.4cm,6.5cm)};
        
            \node[level1={2.65cm}{-0cm}, align=center, font=\large] (artifacts) at (-2cm,5.6cm) {%
                System%
            };

            \node[level1={2.65cm}{-0cm}, align=right] (artifacts) at (3.5cm,5.6cm) {%
                \textit{Section \ref{contextual_system}}%
            };      
            \node[level1={2.65cm}{-0cm}, font=\LARGE, align=right] (artifacts) at (-4.7cm,5.6cm) {%
                (\bfseries{S})%
            };     
            
            \draw[-, very thick]{(-5.5cm,5.3cm) -- (5.4cm,5.3cm)};
            % \draw[dashed]{(6.3cm,4.4cm) -- (6.3cm,6.1cm)};

            \draw (-5.5cm,-0.45cm) -- (-5.5cm,5.3cm);

            \draw[dashed, gray] (6.15cm,4cm) -- (6.15cm,5.8cm);
            
            \draw[-, thin]{(-5.5cm,2.65cm) -- (-4.5cm,2.65cm)};
            \draw[-, thin]{(-5.5cm,-0.45cm) -- (-4.5cm,-0.45cm)};
        
           % Define vertical and horizontal positions
        \def\topY{3.6cm}
        \def\secondY{2.65cm}
        \def\thirdY{1.7cm}
        \def\fourthY{0.45cm}
        \def\fifthY{-0.45cm}
        \def\bottomY{-1.35cm}
        
        % Level 2 parameters and nodes
        \def\leftX{-2.65cm}
        \def\levelTwoStyle{3.61cm}{-02.28cm}
        
        \node[level2={2.6cm}{0cm}, align=center] (runtime) at (-2.65cm,2.65cm) {Natural \\ System};
        \node[level2={2.58cm}{-0.15cm}, align=center] (behavior) at (-2.65cm,-0.4cm) {Technical \\ System};
        
        % Level 5 parameters and color boxes
        \def\colorBoxX{3.8cm}
        \def\levelFiveStyle{0.4cm}{0cm}
        \node[level5={\levelFiveStyle}{naturalColor!30}] at (\colorBoxX,\topY) {};
        \node[level5={\levelFiveStyle}{LatentColor!30}] at (\colorBoxX,\secondY) {};
        \node[level5={\levelFiveStyle}{externalColor!30}] at (\colorBoxX,\thirdY) {};
        \node[level5={\levelFiveStyle}{featureColor!30}] at (\colorBoxX,\fourthY) {};
        \node[level5={\levelFiveStyle}{inferenceColor!30}] at (\colorBoxX,\fifthY) {};
        \node[level5={\levelFiveStyle}{applicationColor!30}] at (\colorBoxX,\bottomY) {};
        
        % Level 3 parameters and nodes
        \def\middleX{3.1cm}
        \def\levelThreeStyle{1cm}{0cm}
        \node[level3={\levelThreeStyle}] (measurements) at (\middleX,\topY) {Reference \\ Domain};
        \node[level3={\levelThreeStyle}] (feedback) at (\middleX,\secondY) {Latent \\ Influences};
        \node[level3={\levelThreeStyle}] (partitions) at (\middleX,\thirdY) {Exogenous \\ Domain};
        \node[level3={\levelThreeStyle}] (schematics) at (\middleX,\fourthY) {Processing \\ Pipeline};
        \node[level3={\levelThreeStyle}] (models) at (\middleX,\fifthY) {Inference \\ Pipeline};
        \node[level3={\levelThreeStyle}] (playbook) at (\middleX,\bottomY) {Application};
        
        % Level 4 parameters and nodes
        \def\rightX{3.7cm}
        \def\levelFourStyle{0.45cm}{0cm}
        \node[level4={\levelFourStyle}, text=gray] (measurementsc) at (\rightX,\topY) {};
        \node[level4={\levelFourStyle}, text=gray] (feedbackc) at (\rightX,\secondY) {};
        \node[level4={\levelFourStyle}, text=gray] (partitionsc) at (\rightX,\thirdY) {};
        \node[level4={\levelFourStyle}, text=gray] (schematicsc) at (\rightX,\fourthY) {};
        \node[level4={\levelFourStyle}, text=gray] (modelsc) at (\rightX,\fifthY) {};
        \node[level4={\levelFourStyle}, text=gray] (playbookc) at (\rightX,\bottomY) {};
        
        % Line parameters and connecting lines
        \def\lineStartX{-0.45cm}
        \def\lineEndX{0.55cm}
        \draw (\lineStartX,\topY) -- (\lineEndX,\topY);
        \draw (\lineStartX,\secondY) -- (\lineEndX,\secondY);
        \draw (\lineStartX,\thirdY) -- (\lineEndX,\thirdY);
        \draw (\lineStartX,\fourthY) -- (\lineEndX,\fourthY);
        \draw (\lineStartX,\fifthY) -- (\lineEndX,\fifthY);
        \draw (\lineStartX,\bottomY) -- (\lineEndX,\bottomY);
        
        % Gray line parameters and lines
        \def\grayLineExtension{27.6}
        \draw[color=gray]{(measurementsc.east) -- ++(\grayLineExtension,0)};
        \draw[color=gray]{(feedbackc.east) -- ++(\grayLineExtension,0)};
        \draw[color=gray]{(partitionsc.east) -- ++(\grayLineExtension,0)};
        \draw[color=gray]{(schematicsc.east) -- ++(\grayLineExtension,0)};
        \draw[color=gray]{(modelsc.east) -- ++(\grayLineExtension,0)};
        \draw[color=gray]{(playbookc.east) -- ++(\grayLineExtension,0)};

        \end{tikzpicture}
    \end{minipage}%  % Add space between
    }
    
% --- Right (matrix) diagram ---
\begin{minipage}[t]{0.38\textwidth}
    % Define grid scaling parameters (to control overall grid dimensions)
    \newcommand{\gridXscale}{1.8}
    \newcommand{\gridYscale}{1.33}
    % A helper macro that creates an inner tikzpicture (with no extra scaling)
    \newcommand{\FixedTikZ}[2][]{%
       \begin{tikzpicture}[#1, baseline]
       #2
       \end{tikzpicture}%
    }
    \vspace{-10.25cm}
    \begin{tikzpicture}[scale=1, xscale=\gridXscale, yscale=\gridYscale]
    % Offsets for the grid (these coordinates are scaled)
    \pgfmathsetmacro{\yOffset}{2cm}
    \pgfmathsetmacro{\yOffsettwo}{2cm}

    % Define sizes for each value using our new parameters (for potential use)
    \pgfmathsetmacro{\minSize}{\PieMinSize}
    \pgfmathsetmacro{\maxSize}{\PieMaxSize}
    \pgfmathsetmacro{\stepSize}{(\maxSize-\minSize)/13}
    
    % (Pre-calculations for numbers 1--14 omitted for brevity)
    
    % (Re)define some colors for this minipage:
   \definecolor{featureColor}{RGB}{8, 230, 0}
   \definecolor{inferenceColor}{RGB}{164, 224, 34}
   \definecolor{naturalColor}{RGB}{71, 179, 255}
   \definecolor{LatentColor}{RGB}{56, 112, 232}
   \definecolor{externalColor}{RGB}{107, 47, 247}
    
    \begin{scope}[
        every node/.style={
            anchor=center,
            font=\normalsize
        }
    ]
    
    % Column headers
    \node[anchor=center] at (1.4,7.5+\yOffset) {Condition};
    \node[anchor=center] at (2.35,7.5+\yOffset) {Evaluation};
    \node[anchor=center] at (3.45,7.5+\yOffset) {Entities};
    \node[anchor=center] at (4.4,7.5+\yOffset) {Relations};
    \node[anchor=center] at (5.5,7.5+\yOffset) {Nominal};
    \node[anchor=center] at (6.45,7.5+\yOffset) {Event};
    \node[anchor=center] at (7.4,7.5+\yOffset) {Normative};

    \draw[thick] (1,7.75+\yOffset) -- (2.75,7.75+\yOffset);
    \draw[thick] (3.05,7.75+\yOffset) -- (4.8,7.75+\yOffset);
    \draw[thick] (5.1,7.75+\yOffset) -- (7.8,7.75+\yOffset);

    \node[anchor=center] at (1.875,8+\yOffset) {State};
    \node[anchor=center] at (3.95,8+\yOffset) {Structure};
    \node[anchor=center] at (6.45,8+\yOffset) {Properties};

    \draw[very thick] (1,8.25+\yOffset) -- (7.8,8.25+\yOffset);
    
    \node[anchor=center, font=\LARGE, align=center] at (1.4,8.55+\yOffset) {(\textbf{A})};
    \node[anchor=center, font=\normalsize, align=center] at (7.2,8.55+\yOffset) {\textit{Section \ref{contextual_aspects}}};
    \node[anchor=center, font=\normalsize, align=center, font=\large] at (2.1,8.55+\yOffset) {Aspect};

    \newcommand{\startY}{1.6}  % Starting y coordinate
    \newcommand{\segLen}{0.71}  % Length of regular segments
    \newcommand{\segLenBig}{1.11}  % Length of the group-separating segments
    \newcommand{\withinGroupDist}{0.2}  % Distance between all segments
    
    % Draw vertical segments for each column
    \foreach \x in {1.4, 2.35, 3.45, 4.4, 5.5, 6.45, 7.4} {
        \draw[line width=0.2pt, gray] (\x,{\startY+\yOffset}) -- (\x,{\startY+\segLen+\yOffset});
        \draw[line width=0.2pt, gray] (\x,{\startY+\segLen+\withinGroupDist+\yOffset}) -- (\x,{\startY+2*\segLen+\withinGroupDist+\yOffset});
        \draw[line width=0.2pt, gray] (\x,{\startY+2*\segLen+2*\withinGroupDist+\yOffset}) -- (\x,{\startY+2*\segLen+\segLenBig+2*\withinGroupDist+\yOffset});
        \draw[line width=0.2pt, gray] (\x,{\startY+2*\segLen+\segLenBig+3*\withinGroupDist+\yOffset}) -- (\x,{\startY+3*\segLen+\segLenBig+3*\withinGroupDist+\yOffset});
        \draw[line width=0.2pt, gray] (\x,{\startY+3*\segLen+\segLenBig+4*\withinGroupDist+\yOffset}) -- (\x,{\startY+4*\segLen+\segLenBig+4*\withinGroupDist+\yOffset+0.1});
        \draw[line width=0.2pt, gray] (\x,{\startY+4*\segLen+\segLenBig+5*\withinGroupDist+\yOffset+0.05}) -- (\x,{7.3+\yOffset});
    }

    % --- Set default outline thickness for pies globally ---
    \pgfkeys{/pgfpie/every number/.append style={anchor=center,align=center,font=\scriptsize}}
    \pgfkeys{/pgfpie/every slice/.append style={draw=black,thin}}

    % Now each grid cell is drawn in a fixed (unscaled) inner picture using \FixedTikZ.
    % The radii (or pie radii) are computed as (base radius * \pieScale).
\begin{scope}[shift={(3.45,6.5+\yOffsettwo)}]
  \node[inner sep=0, anchor=center] {%
    \FixedTikZ{%
      \pgfmathsetmacro{\r}{\pieScale * 0.36}%
      \path[use as bounding box] (-\r cm,-\r cm) rectangle (\r cm,\r cm);
      \draw[draw=black,line width=\PieOutlineThickness,fill=externalColor!30] (0,0) circle (\r cm);
      \node[inner sep=0pt] at (0,0) {13};
    }%
  };
\end{scope}
\begin{scope}[shift={(4.4,6.5+\yOffsettwo)}]
  \node[inner sep=0, anchor=center] {%
    \FixedTikZ{%
      \pgfmathsetmacro{\r}{\pieScale * 0.26}%
      \path[use as bounding box] (-\r cm,-\r cm) rectangle (\r cm,\r cm);
      \draw[draw=black,line width=\PieOutlineThickness,fill=featureColor!30] (0,0) circle (\r cm);
      \node[inner sep=0pt] at (0,0) {7};
    }%
  };
\end{scope}
\begin{scope}[shift={(5.5,6.5+\yOffsettwo)}]
  \node[inner sep=0, anchor=center] {%
    \FixedTikZ{%
      \pgfmathsetmacro{\r}{\pieScale * 0.35}%
      \path[use as bounding box] (-\r cm,-\r cm) rectangle (\r cm,\r cm);
      \pie[
        radius=\r cm,
        sum=auto,
        hide number,
        line width=\PieOutlineThickness,
        color={LatentColor!30, externalColor!30, inferenceColor!30}
      ]{ 1/, 11/, 1/ }
      \pgfmathsetmacro{\xA}{\pieScale * 0.14}%
      \pgfmathsetmacro{\yA}{\pieScale * 0.03}%
      \pgfmathsetmacro{\xB}{-\pieScale * 0.14}%
      \pgfmathsetmacro{\yC}{\pieScale * 0.03}%
      \node[inner sep=0pt] at (\xA cm, \yA cm) {};
      \node[inner sep=0pt] at (\xB cm, 0cm) {11};
      \node[inner sep=0pt] at (\xA cm, -\yC cm) {};
    }%
  };
\end{scope}
\begin{scope}[shift={(7.4,6.5+\yOffsettwo)}]
  \node[inner sep=0, anchor=center] {%
    \FixedTikZ{%
      \pgfmathsetmacro{\r}{\pieScale * 0.14}%
      \path[use as bounding box] (-\r cm,-\r cm) rectangle (\r cm,\r cm);
      \pie[
        radius=\r cm,
        sum=auto,
        hide number,
        line width=\PieOutlineThickness,
        color={externalColor!30, inferenceColor!30}
      ]{ 1/, 1/ }
      \pgfmathsetmacro{\yD}{\pieScale * 0.06}%
      \node[inner sep=0pt] at (0, \yD/2 cm) {};
      \node[inner sep=0pt] at (0, -\yD/2 cm) {};
    }%
  };
\end{scope}
\begin{scope}[shift={(4.4,5.6+\yOffsettwo)}]
  \node[inner sep=0, anchor=center] {%
    \FixedTikZ{%
      \pgfmathsetmacro{\r}{\pieScale * 0.17}%
      \path[use as bounding box] (-\r cm,-\r cm) rectangle (\r cm,\r cm);
      \draw[draw=black,line width=\PieOutlineThickness,fill=featureColor!30] (0,0) circle (\r cm);
      \node[inner sep=0pt] at (0,0) {3};
    }%
  };
\end{scope}
\begin{scope}[shift={(6.45,5.6+\yOffsettwo)}]
  \node[inner sep=0, anchor=center] {%
    \FixedTikZ{%
      \pgfmathsetmacro{\r}{\pieScale * 0.45}%
      \path[use as bounding box] (-\r cm,-\r cm) rectangle (\r cm,\r cm);
      \pie[
        radius=\r cm,
        sum=auto,
        hide number,
        line width=\PieOutlineThickness,
        color={LatentColor!30, externalColor!30, inferenceColor!30}
      ]{ 11/, 3/, 6/ }
      \pgfmathsetmacro{\xE}{\pieScale * 0.04}%
      \pgfmathsetmacro{\yE}{\pieScale * 0.18}%
      \pgfmathsetmacro{\xF}{-\pieScale * 0.11}%
      \pgfmathsetmacro{\xG}{\pieScale * 0.11}%
      \node[inner sep=0pt] at (\xE cm, \yE cm) {11};
      \node[inner sep=0pt] at (-0.3 cm, -0.35 cm) {3};
      \node[inner sep=0pt] at (\xG cm, -\pieScale*0.14 cm) {6};
    }%
  };
\end{scope}
\begin{scope}[shift={(1.4,4.6+\yOffsettwo)}]
  \node[inner sep=0, anchor=center] {%
    \FixedTikZ{%
      \pgfmathsetmacro{\r}{\pieScale * 0.31}%
      \path[use as bounding box] (-\r cm,-\r cm) rectangle (\r cm,\r cm);
      \draw[draw=black,line width=\PieOutlineThickness,fill=naturalColor!30] (0,0) circle (\r cm);
      \node[inner sep=0pt] at (0,0) {10};
    }%
  };
\end{scope}
\begin{scope}[shift={(3.45,4.6+\yOffsettwo)}]
  \node[inner sep=0, anchor=center] {%
    \FixedTikZ{%
      \pgfmathsetmacro{\r}{\pieScale * 0.24}%
      \path[use as bounding box] (-\r cm,-\r cm) rectangle (\r cm,\r cm);
      \draw[draw=black,line width=\PieOutlineThickness,fill=naturalColor!30] (0,0) circle (\r cm);
      \node[inner sep=0pt] at (0,0) {6};
    }%
  };
\end{scope}
\begin{scope}[shift={(4.4,4.6+\yOffsettwo)}]
  \node[inner sep=0, anchor=center] {%
    \FixedTikZ{%
      \pgfmathsetmacro{\r}{\pieScale * 0.14}%
      \path[use as bounding box] (-\r cm,-\r cm) rectangle (\r cm,\r cm);
      \draw[draw=black,line width=\PieOutlineThickness,fill=featureColor!30] (0,0) circle (\r cm);
      \node[inner sep=0pt] at (0,0) {2};
    }%
  };
\end{scope}
\begin{scope}[shift={(5.5,4.6+\yOffsettwo)}]
  \node[inner sep=0, anchor=center] {%
    \FixedTikZ{%
      \pgfmathsetmacro{\r}{\pieScale * 0.10}%
      \path[use as bounding box] (-\r cm,-\r cm) rectangle (\r cm,\r cm);
      \draw[draw=black,line width=\PieOutlineThickness,fill=externalColor!30] (0,0) circle (\r cm);
      \node[inner sep=0pt] at (0,0) {};
    }%
  };
\end{scope}
\begin{scope}[shift={(6.45,4.6+\yOffsettwo)}]
  \node[inner sep=0, anchor=center] {%
    \FixedTikZ{%
      \pgfmathsetmacro{\r}{\pieScale * 0.10}%
      \path[use as bounding box] (-\r cm,-\r cm) rectangle (\r cm,\r cm);
      \draw[draw=black,line width=\PieOutlineThickness,fill=externalColor!30] (0,0) circle (\r cm);
      \node[inner sep=0pt] at (0,0) {};
    }%
  };
\end{scope}
\begin{scope}[shift={(7.4,4.6+\yOffsettwo)}]
  \node[inner sep=0, anchor=center] {%
    \FixedTikZ{%
      \pgfmathsetmacro{\r}{\pieScale * 0.10}%
      \path[use as bounding box] (-\r cm,-\r cm) rectangle (\r cm,\r cm);
      \draw[draw=black,line width=\PieOutlineThickness,fill=externalColor!30] (0,0) circle (\r cm);
      \node[inner sep=0pt] at (0,0) {};
    }%
  };
\end{scope}
\begin{scope}[shift={(1.4,3.3+\yOffsettwo)}]
  \node[inner sep=0, anchor=center] {%
    \FixedTikZ{%
      \pgfmathsetmacro{\r}{\pieScale * 0.29}%
      \path[use as bounding box] (-\r cm,-\r cm) rectangle (\r cm,\r cm);
      \pie[
        radius=\r cm,
        sum=auto,
        hide number,
        line width=\PieOutlineThickness,
        color={naturalColor!30, inferenceColor!30}
      ]{ 5/, 4/ }
      \pgfmathsetmacro{\xH}{\pieScale * 0.02}%
      \pgfmathsetmacro{\yH}{\pieScale * 0.12}%
      \node[inner sep=0pt] at (-\xH cm, \yH cm) {5};
      \node[inner sep=0pt] at (\xH cm, -\yH cm) {4};
    }%
  };
\end{scope}
\begin{scope}[shift={(4.4,3.3+\yOffsettwo)}]
  \node[inner sep=0, anchor=center] {%
    \FixedTikZ{%
      \pgfmathsetmacro{\r}{\pieScale * 0.22}%
      \path[use as bounding box] (-\r cm,-\r cm) rectangle (\r cm,\r cm);
      \pie[
        radius=\r cm,
        sum=auto,
        hide number,
        line width=\PieOutlineThickness,
        % Add a third color for the new segment
        color={featureColor!30, inferenceColor!30, applicationColor!30}
      ]
      % Add the value for the third segment here
      { 1/, 1/, 3/ } % Example: added a segment with value 5
      \pgfmathsetmacro{\xI}{\pieScale * 0.07}%
      \pgfmathsetmacro{\yI}{\pieScale * 0.05}%
      \node[inner sep=0pt] at (\xI cm, \yI cm) {};
      \node[inner sep=0pt] at (-\xI cm, -\yI cm - 0.05cm) {3};
    }%
  };
\end{scope}
\begin{scope}[shift={(5.5,3.3+\yOffsettwo)}]
  \node[inner sep=0, anchor=center] {%
    \FixedTikZ{%
      \pgfmathsetmacro{\r}{\pieScale * 0.22}%
      \path[use as bounding box] (-\r cm,-\r cm) rectangle (\r cm,\r cm);
      \draw[draw=black,line width=\PieOutlineThickness,fill=applicationColor!30] (0,0) circle (\r cm);
      \node[inner sep=0pt] at (0,0) {6};
    }%
  };
\end{scope}
\begin{scope}[shift={(6.45,3.3+\yOffsettwo)}]
  \node[inner sep=0, anchor=center] {%
    \FixedTikZ{%
      \pgfmathsetmacro{\r}{\pieScale * 0.17}%
      \path[use as bounding box] (-\r cm,-\r cm) rectangle (\r cm,\r cm);
      \pie[
        radius=\r cm,
        sum=auto,
        hide number,
        line width=\PieOutlineThickness,
        color={LatentColor!30, inferenceColor!30}
      ]{ 2/, 1/ }
      \pgfmathsetmacro{\xI}{\pieScale * 0.03}%
      \pgfmathsetmacro{\yI}{\pieScale * 0.06}%
      \node[inner sep=0pt] at (-\xI cm, \yI cm) {2};
      \node[inner sep=0pt] at (\xI cm, -\yI cm) {};
    }%
  };
\end{scope}
\begin{scope}[shift={(1.4,2.4+\yOffsettwo)}]
  \node[inner sep=0, anchor=center] {%
    \FixedTikZ{%
      \pgfmathsetmacro{\r}{\pieScale * 0.28}%
      \path[use as bounding box] (-\r cm,-\r cm) rectangle (\r cm,\r cm);
      \draw[draw=black,line width=\PieOutlineThickness,fill=naturalColor!30] (0,0) circle (\r cm);
      \node[inner sep=0pt] at (0,0) {8};
    }%
  };
\end{scope}
\begin{scope}[shift={(2.35,2.4+\yOffsettwo)}]
  \node[inner sep=0, anchor=center] {%
    \FixedTikZ{%
      \pgfmathsetmacro{\r}{\pieScale * 0.22}%
      \path[use as bounding box] (-\r cm,-\r cm) rectangle (\r cm,\r cm);
      \draw[draw=black,line width=\PieOutlineThickness,fill=naturalColor!30] (0,0) circle (\r cm);
      \node[inner sep=0pt] at (0,0) {5};
    }%
  };
\end{scope}
\begin{scope}[shift={(4.4,2.4+\yOffsettwo)}]
  \node[inner sep=0, anchor=center] {%
    \FixedTikZ{%
      \pgfmathsetmacro{\r}{\pieScale * 0.20}%
      \path[use as bounding box] (-\r cm,-\r cm) rectangle (\r cm,\r cm);
      \pie[
        radius=\r cm,
        sum=auto,
        hide number,
        line width=\PieOutlineThickness,
        % 1. Add a third color for the new segment
        color={featureColor!30, applicationColor!30, inferenceColor!30}
      ]
      % 2. Add the value for the new segment to the list
      { 1/, 1/, 2/ } % Example: added a segment with value 2
      \pgfmathsetmacro{\xJ}{\pieScale * 0.06}%
      \pgfmathsetmacro{\yJ}{\pieScale * 0.06}%
      \node[inner sep=0pt] at (\xJ cm, \yJ cm) {};
      \node[inner sep=0pt] at (-\xJ cm + 0.05cm, -\yJ cm - 0.05cm) {2};
    }%
  };
\end{scope}
\begin{scope}[shift={(7.4,2.4+\yOffsettwo)}]
  \node[inner sep=0, anchor=center] {%
    \FixedTikZ{%
      \pgfmathsetmacro{\r}{\pieScale * 0.22}%
      \path[use as bounding box] (-\r cm,-\r cm) rectangle (\r cm,\r cm);
      \pie[
        radius=\r cm,
        sum=auto,
        hide number,
        line width=\PieOutlineThickness,
        color={LatentColor!30, externalColor!30}
      ]{ 1/, 4/ }
      \pgfmathsetmacro{\xK}{\pieScale * 0.07}%
      \pgfmathsetmacro{\yK}{\pieScale * 0.05}%
      \node[inner sep=0pt] at (\xK cm, \yK cm) {};
      \node[inner sep=0pt] at (-\xK cm, -\yK cm) {4};
    }%
  };
\end{scope}
\begin{scope}[shift={(1.4,1.5+\yOffsettwo)}]
  \node[inner sep=0, anchor=center] {%
    \FixedTikZ{%
      \pgfmathsetmacro{\r}{\pieScale * 0.17}%
      \path[use as bounding box] (-\r cm,-\r cm) rectangle (\r cm,\r cm);
      \draw[draw=black,line width=\PieOutlineThickness,fill=naturalColor!30] (0,0) circle (\r cm);
      \node[inner sep=0pt] at (0,0) {3};
    }%
  };
\end{scope}
\begin{scope}[shift={(2.35,1.5+\yOffsettwo)}]
  \node[inner sep=0, anchor=center] {%
    \FixedTikZ{%
      \pgfmathsetmacro{\r}{\pieScale * 0.35}%
      \path[use as bounding box] (-\r cm,-\r cm) rectangle (\r cm,\r cm);
      \pie[
        radius=\r cm,
        sum=auto,
        hide number,
        line width=\PieOutlineThickness,
        color={naturalColor!30, inferenceColor!30}
      ]{ 10/, 3/ }
      \pgfmathsetmacro{\xL}{\pieScale * 0.11}%
      \pgfmathsetmacro{\yL}{\pieScale * 0.09}%
      \node[inner sep=0pt] at (-\xL cm, \yL cm) {10};
      \node[inner sep=0pt] at (0.2 cm, -0.2 cm) {3};
    }%
  };
\end{scope}
\begin{scope}[shift={(4.4,1.5+\yOffsettwo)}]
  \node[inner sep=0, anchor=center] {%
    \FixedTikZ{%
      \pgfmathsetmacro{\r}{\pieScale * 0.14}%
      \path[use as bounding box] (-\r cm,-\r cm) rectangle (\r cm,\r cm);
      \draw[draw=black,line width=\PieOutlineThickness,fill=featureColor!30] (0,0) circle (\r cm);
      \node[inner sep=0pt] at (0,0) {2};
    }%
  };
\end{scope}
\begin{scope}[shift={(7.4,1.5+\yOffsettwo)}]
  \node[inner sep=0, anchor=center] {%
    \FixedTikZ{%
      \pgfmathsetmacro{\r}{\pieScale * 0.24}%
      \path[use as bounding box] (-\r cm,-\r cm) rectangle (\r cm,\r cm);
      \pie[
        radius=\r cm,
        sum=auto,
        hide number,
        line width=\PieOutlineThickness,
        color={LatentColor!30, externalColor!30}
      ]{ 1/, 5/ }
      \pgfmathsetmacro{\xM}{\pieScale * 0.08}%
      \pgfmathsetmacro{\yM}{\pieScale * 0.05}%
      \node[inner sep=0pt] at (\xM cm, \yM cm) {};
      \node[inner sep=0pt] at (-\xM cm, -\yM cm) {5};
    }%
  };
\end{scope}

    \end{scope}
            
    \end{tikzpicture}
\end{minipage}

\vspace{1em} % space between main content and legend

% --- Legend layout ---
\newlength{\legendWidth}
\setlength{\legendWidth}{\linewidth} % entire available width

\newlength{\gapLength}
\setlength{\gapLength}{1cm} % fixed gap between the two group lines

% Formal gets 3/5 and Informal 2/5 of the width (after subtracting the gap)
\newlength{\formalLength}
\setlength{\formalLength}{\dimexpr (\legendWidth - \gapLength)*3/5\relax}
\newlength{\informalLength}
\setlength{\informalLength}{\dimexpr (\legendWidth - \gapLength)*2/5\relax}

\vspace{-0.3cm}

\begin{tikzpicture}[x=1cm,y=1cm]
   \definecolor{featureColor}{RGB}{8, 230, 0}
   \definecolor{inferenceColor}{RGB}{164, 224, 34}
   \definecolor{naturalColor}{RGB}{71, 179, 255}
   \definecolor{LatentColor}{RGB}{56, 112, 232}
   \definecolor{externalColor}{RGB}{107, 47, 247}

    \normalsize
    
  % Draw the outer horizontal line
  \draw[thick] (0,0) -- (\legendWidth,0);
  \node[anchor=south west] at (0,0.05cm) {\LARGE (R) \large Representation};
  \node[anchor=center, font=\normalsize, align=center] at (17cm,0.38cm) {\textit{Section \ref{contextual_representations}}};

  % Draw two group lines at y = -1
  \draw[thick] (0,-0.7) -- (\formalLength,-0.7);
  \draw[thick] (\formalLength+\gapLength,-0.7) -- (\legendWidth,-0.7);
  
  % Group labels
  \node[anchor=south,font=\normalsize] at ($ (0,-1)!0.5!(\formalLength,-1)+(0,0.4) $) {Formal};
  \node[anchor=south,font=\normalsize] at ($ (\formalLength+\gapLength,-1)!0.5!(\legendWidth,-1)+(0,0.4) $) {Informal};

  % Parameters for color swatches
  \def\boxWidth{3cm}
  \def\boxHeight{0.5cm}
  \def\boxYOffset{0cm}

  % Formal group: Numerical, Probabilistic, Logical
  \coordinate (formalCenter1) at ($ (0,-1)!0.16667!(\formalLength,-1)$);
  \draw[fill=naturalColor!30, draw=black, rounded corners=2pt] 
       ($(formalCenter1)+(-0.5*\boxWidth, -\boxHeight-\boxYOffset)$)
       rectangle 
       ($(formalCenter1)+(0.5*\boxWidth, -\boxYOffset)$);
  \node[anchor=north,font=\normalsize] at 
       ($(formalCenter1)+(0,0)$) {Numerical};

  \coordinate (formalCenter2) at ($ (0,-1)!0.5!(\formalLength,-1)$);
  \draw[fill=LatentColor!30, draw=black, rounded corners=2pt] 
       ($(formalCenter2)+(-0.5*\boxWidth, -\boxHeight-\boxYOffset)$)
       rectangle 
       ($(formalCenter2)+(0.5*\boxWidth, -\boxYOffset)$);
  \node[anchor=north,font=\normalsize] at 
       ($(formalCenter2)+(0,0)$) {Probabilistic};

  \coordinate (formalCenter3) at ($ (0,-1)!0.83333!(\formalLength,-1)$);
  \draw[fill=externalColor!30, draw=black, rounded corners=2pt] 
       ($(formalCenter3)+(-0.5*\boxWidth, -\boxHeight-\boxYOffset)$)
       rectangle 
       ($(formalCenter3)+(0.5*\boxWidth, -\boxYOffset)$);
  \node[anchor=north,font=\normalsize] at 
       ($(formalCenter3)+(0,0)$) {Logical};

  % Informal group: Graphical, Sem-structured
  \coordinate (informalCenter1) at ($ (\formalLength+\gapLength,-1)!0.25!(\legendWidth,-1)$);
  \draw[fill=featureColor!30, draw=black, rounded corners=2pt] 
       ($(informalCenter1)+(-0.5*\boxWidth, -\boxHeight-\boxYOffset)$)
       rectangle 
       ($(informalCenter1)+(0.5*\boxWidth, -\boxYOffset)$);
  \node[anchor=north,font=\normalsize] at 
       ($(informalCenter1)+(0,0)$) {Graphical};

  \coordinate (informalCenter2) at ($ (\formalLength+\gapLength,-1)!0.75!(\legendWidth,-1)$);
  \draw[fill=inferenceColor!30, draw=black, rounded corners=2pt] 
       ($(informalCenter2)+(-0.5*\boxWidth, -\boxHeight-\boxYOffset)$)
       rectangle 
       ($(informalCenter2)+(0.5*\boxWidth, -\boxYOffset)$);
  \node[anchor=north,font=\normalsize] at 
       ($(informalCenter2)+(0,0)$) {Sem-structured};

\end{tikzpicture}

    \caption{Cross-dimensional distribution between \textit{system}, \textit{aspect}, and \textit{representation} dimensions. Each cell represents a combination of \textit{system} and \textit{aspect}, with pie segments corresponding to different \textit{representation} types. Circle sizes indicate occurrence frequency; cells without a number represent a single occurrence.}
    
   \label{fig:system-properties-dist}
\end{figure*}
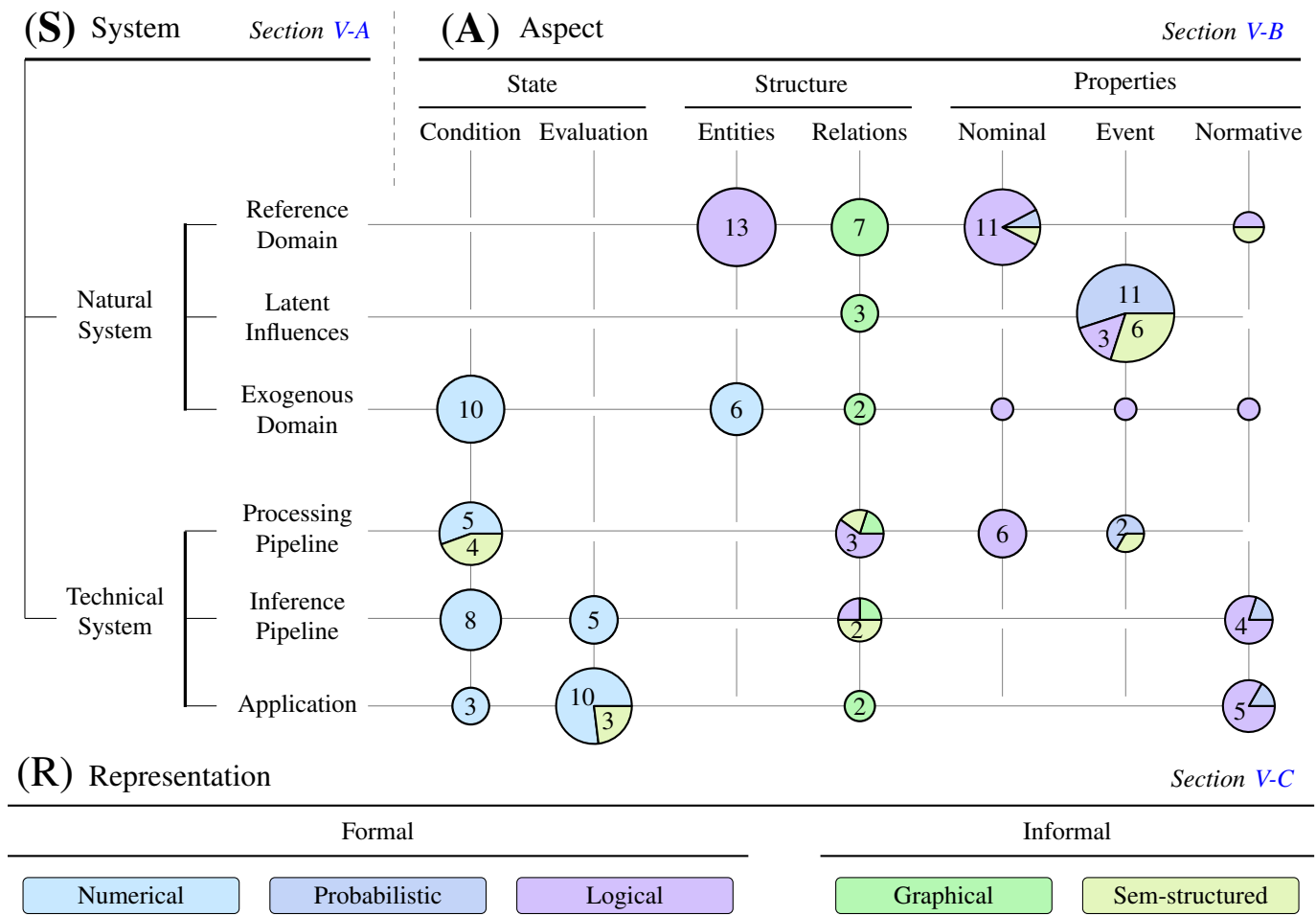

%% file: sections/figures/3d_table.tex
\begin{table*}[!b]
\small
\centering
\renewcommand{\arraystretch}{1.2}
\begin{tabular}{|p{3.8cm}|p{3cm} l l | c p{4.6cm}|}
\toprule[1.5pt]
\textbf{Pattern Name} 
 & \textbf{System} 
 & \textbf{Aspect} 
 & \textbf{Representation} 
 & \textbf{\#} 
 & \textbf{Support} \\
\midrule

% ==================== Domain Map label in single cell =====================
\rowcolor{naturalColor!10}
\multicolumn{6}{|p{17.3cm}|}{\textbf{\textit{Natural System Map}} -- \S\ref{domain_map}} \\
\midrule

% -------------------- Domain Partitions ----------------------
\cellcolor{naturalColor!10}\textbf{\textit{Natural Subgroups}}
  & Natural System* % <-- Added asterisk
  & Entities 
  & Formal* % <-- Added asterisk
  & 19 
  &  \\
 
\quad \cellcolor{naturalColor!10}\textit{Reference Slices}
  & Reference Domain
  & Entities 
  & Logical 
  & 13 
  & {\scriptsize \cite{azonoozi2016contrack,drevesm2020from,fengj2024designing,ghosha2022fair,henzingerthomasandkarimimahyar,rc2020overton,sackerman2021machine,schelter2018automating,swami2020data,zhoux2019a,fedelea2024the,henzingerta2023monitoring,mammanh2024biastrap}} \\

\quad \cellcolor{naturalColor!10}\textit{Exogenous Identifiers}
  & Exogenous Domain
  & Entities 
  & Numerical 
  & 6 
  & {\scriptsize \cite{castelnovoa2021towards,khoshravanazara2023the,manerikerp2023online,schrouffj2022diagnosing,vasudevans2020lift,yangz2021biasrv}} \\
\midrule

% -------------------- Domain Diagrams ------------------------
\cellcolor{naturalColor!10}\textbf{\textit{Natural Diagrams}}
  & Natural System* % <-- Added asterisk
  & Relations 
  & Graphical 
  & 10 
  &  \\

\quad \cellcolor{naturalColor!10}\textit{Reference Diagrams}
  & Reference Domain
  & Relations 
  & Graphical 
  & 7 
  & {\scriptsize \cite{bontempellia2022humanintheloop,budhathokik2021why,dreyfuspa2022databased,fengj2024designing,sasthana2021ml,schrouffj2022diagnosing,zhangh2023why}} \\

\quad \cellcolor{naturalColor!10}\textit{Latent Diagrams}
  & Latent Influences
  & Relations 
  & Graphical  
  & 3 
  & {\scriptsize \cite{borchanih2015modeling,dreyfuspa2022databased,feng2022clinical}} \\

% \quad \cellcolor{naturalColor!10}\textit{Exogenous Diagrams}
%   & Exogenous Domain
%   & Relations 
%   & Graphical  
%   & 2 
%   & {\scriptsize \cite{fengj2024designing,kkirchheim2024outofdistributio}} \\
\midrule

% -------------------- Domain Specifications -------------------
\cellcolor{naturalColor!10}\textbf{\textit{Natural Specifications}}
  & Natural System* % <-- Added asterisk
  & Properties* % <-- Added asterisk
  & Formal* % <-- Added asterisk
  & 25
  &  \\

\quad \cellcolor{naturalColor!10}\textit{Reference Assertions}
  & Reference Domain
  & Nominal 
  & Logical 
  & 11 
  & {\scriptsize \cite{bachingerf2024data,cavenesse2020tensorflow,chenl2022estimating,dchen2022twostage,ehrlingerl2019a,lelwakatare2021on,liu2020towards,myllyahol2022on,schelter2018automating,shankars2024we,swami2020data}} \\

\quad \cellcolor{naturalColor!10}\textit{Latent Projections}
  & Latent Influences
  & Event 
  & Probabilistic 
  & 11 
  & {\scriptsize \cite{borchanih2015modeling,feng2024monitoring,galewis2022augur,godaup2023deployment,henzingerta2023monitoring,leestj2024expert,masegosaar2020analyzing,pichlerg2024on,pwelinder2013a,sibliniw2020master,sobolewskip2017scr}} \\ 

\quad \cellcolor{naturalColor!10}\textit{Latent Expectations}
  & Latent Influences
  & Event 
  & Logical 
  & 3 
  & {\scriptsize \cite{chenm2021mandoline,xuanjunyuandlujieandzhangguang,xuz2023alertiger}} \\
\midrule

\cellcolor{naturalColor!10}\textbf{\textit{Latent Scenarios}}
  & Latent Influences
  & Event 
  & Semi-structured 
  & 6 
  & {\scriptsize \cite{apaul2024mlops,dreyfuspa2022databased,feng2022clinical,haidert2021domain,leestj2024expert,xuz2023alertiger}} \\
\midrule

% ------------------- Exogenous Measurements ------------------
\cellcolor{naturalColor!10}\textit{\textbf{Exogenous Diagnostics}}
  & Exogenous Domain
  & Condition 
  & Numerical 
  & 10 
  & {\scriptsize \cite{bensalems2024continuous,cobbo2022contextaware,gomesjb2010calds,jbgomes2014mining,kirchheimk2023towards,kkirchheim2024outofdistributio,klsm2019uncertainty,langfordma2023modalas,torfahh2022learning,zhoux2019a}} \\
\midrule

% ================== Pipeline Map label in single cell =====================
\rowcolor{inferenceColor!10}
\multicolumn{6}{|p{17.3cm}|}{\textbf{\textit{Technical System Map}} -- \S\ref{pipeline_map}} \\
\midrule

% ------------------- Pipeline Observations --------------------
\cellcolor{inferenceColor!10}\textbf{\textit{Technical Observations}}
  & Technical System* % <-- Added asterisk
  & State* % <-- Added asterisk
  & Numerical
  & 25
  &  \\
\quad \cellcolor{inferenceColor!10}\textit{Processing Diagnostics}
  & Processing Pipeline
  & Condition 
  & Numerical
  & 5
  & {\scriptsize \cite{ebreck2017the,muirurid2022practices,nguyenmt2024novel,paleyesandreiandlawrenceneilda,sculley2015hidden}} \\
\quad \cellcolor{inferenceColor!10}\textit{Inference Diagnostics}
  & Inference Pipeline
  & Condition
  & Numerical
  & 8
  & {\scriptsize \cite{baquon2022concept,ebreck2017the,paleyesandreiandlawrenceneilda,sculley2015hidden,tzoppi2021detect,wangs2021a,xuz2023alertiger,xxu2022dependency}} \\
\quad \cellcolor{inferenceColor!10}\textit{Inference Assessments}
  & Inference Pipeline
  & Evaluation
  & Numerical
  & 5
  & {\scriptsize \cite{binderf2022putting,ginartaa2022mldemon,guann2022fila,vandervorstf2024claims,vishwakarmah2024taming}} \\
\quad \cellcolor{inferenceColor!10}\textit{Application Diagnostics}
  & Application
  & Condition
  & Numerical
  & 3
  & {\scriptsize \cite{allenb2021evaluation,fengj2024designing,onnesa2023bayesian}} \\
\quad \cellcolor{inferenceColor!10}\textit{Application Assessments}
  & Application
  & Evaluation
  & Numerical
  & 10
  & {\scriptsize \cite{bensalems2024continuous,bernardil2019150,binderf2022putting,csun2024ai,ebreck2017the,edsnascimento2019understanding,hanafimf2024machineassisted,langfordma2023modalas,paleyesandreiandlawrenceneilda,shergadwalamn2022a}} \\
\midrule

% -------------------- Pipeline Signals ------------------------
\cellcolor{inferenceColor!10}\textbf{\textit{Technical Signals}}
  & Technical System* % <-- Added asterisk
  & State* % <-- Added asterisk
  & Informal* % <-- Added asterisk
  & 7
  &  \\

\quad \cellcolor{inferenceColor!10}\textit{Processing Notifications}
  & Processing Pipeline
  & Condition
  & Semi-structured
  & 4
  & {\scriptsize \cite{ebreck2017the,foidlh2019riskbased,heynhm2023automotive,swami2020data}} \\
\quad \cellcolor{inferenceColor!10}\textit{Application Comments}
  & Application
  & Evaluation
  & Semi-structured
  & 3
  & {\scriptsize \cite{cabreraa2021discovering,jayalathh2022enhancing,jayalathh2023continual}} \\
\midrule

% -------------------- Pipeline Descriptions -----------------------
\cellcolor{inferenceColor!10}\textbf{\textit{Processing Traces}}
  & Processing Pipeline
  & Relations
  & Logical
  & 3
  & {\scriptsize \cite{heynhm2023automotive,namakimh2020vamsa,shankars2022towards} } \\

% \quad \cellcolor{inferenceColor!10}\textit{Application Diagrams}
%   & Application
%   & Relations
%   & Graphical
%   & 2
%   & {\scriptsize \cite{fengj2024designing,paleyesandreiandlawrenceneilda}} \\
\midrule

% ----------------- Pipeline Specifications --------------------
\cellcolor{inferenceColor!10}\textbf{\textit{Technical Specifications}}
  & Technical System* % <-- Added asterisk
  & Properties* % <-- Added asterisk
  & Formal* % <-- Added asterisk
  & 15
  &  \\

% \quad \cellcolor{inferenceColor!10}\textit{Processing Projections}
%   & Processing Pipeline
%   & Event
%   & Probabilistic
%   & 2
%   & {\scriptsize \cite{schelters2020learning,schelters2021jenga}} \\

\quad \cellcolor{inferenceColor!10}\textit{Processing Assertions}
  & Processing Pipeline
  & Nominal
  & Logical
  & 6
  & {\scriptsize \cite{ehrlingerl2019a, myllyahol2022on, lelwakatare2021on, cavenesse2020tensorflow,schelter2018automating, swami2020data}} \\

\quad \cellcolor{inferenceColor!10}\textit{Inference Guardrails}
  & Inference Pipeline
  & Normative
  & Logical
  & 4
  & {\scriptsize \cite{csun2024ai,habdelkader2024mlonrails,kangdanielandguibasjohnandbail,myllyahol2022on}} \\

\quad \cellcolor{inferenceColor!10}\textit{Application Guardrails}
  & Application
  & Normative
  & Logical
  & 5
  & {\scriptsize \cite{bensalems2024continuous,habdelkader2024mlonrails,langfordma2023modalas,onnesa2022monitoring,sculley2015hidden}} \\

\bottomrule[1.5pt]
\end{tabular}

\vspace{0.5em}
{\footnotesize * Denotes a higher-level category as defined in the respective \textit{System-Aspect-Representation} taxonomies in prior sections.}

\caption{Catalog of Contextual \textit{System} (\S\ref{contextual_system}), \textit{Aspect} (\S\ref{contextual_aspects}), and \textit{Representation} (\S\ref{contextual_representations}) (C-SAR) triplet patterns. Each pattern is uniquely named and coded. The \# column indicates the number of appearances across primary studies (only patterns with three or more occurrences are included), and \textit{Support} lists the references to the primary studies implementing the pattern.}
\label{tabpatterns_reorg}
\end{table*}

%% file: sections/figures/activity_mappings.tex
% Define custom colors
\definecolor{featureColor}{RGB}{71,179,255}
\definecolor{inferenceColor}{RGB}{56,112,232}
\definecolor{applicationColor}{RGB}{107,47,247}
\definecolor{naturalColor}{RGB}{8,230,0}
\definecolor{externalColor}{RGB}{164,224,34}

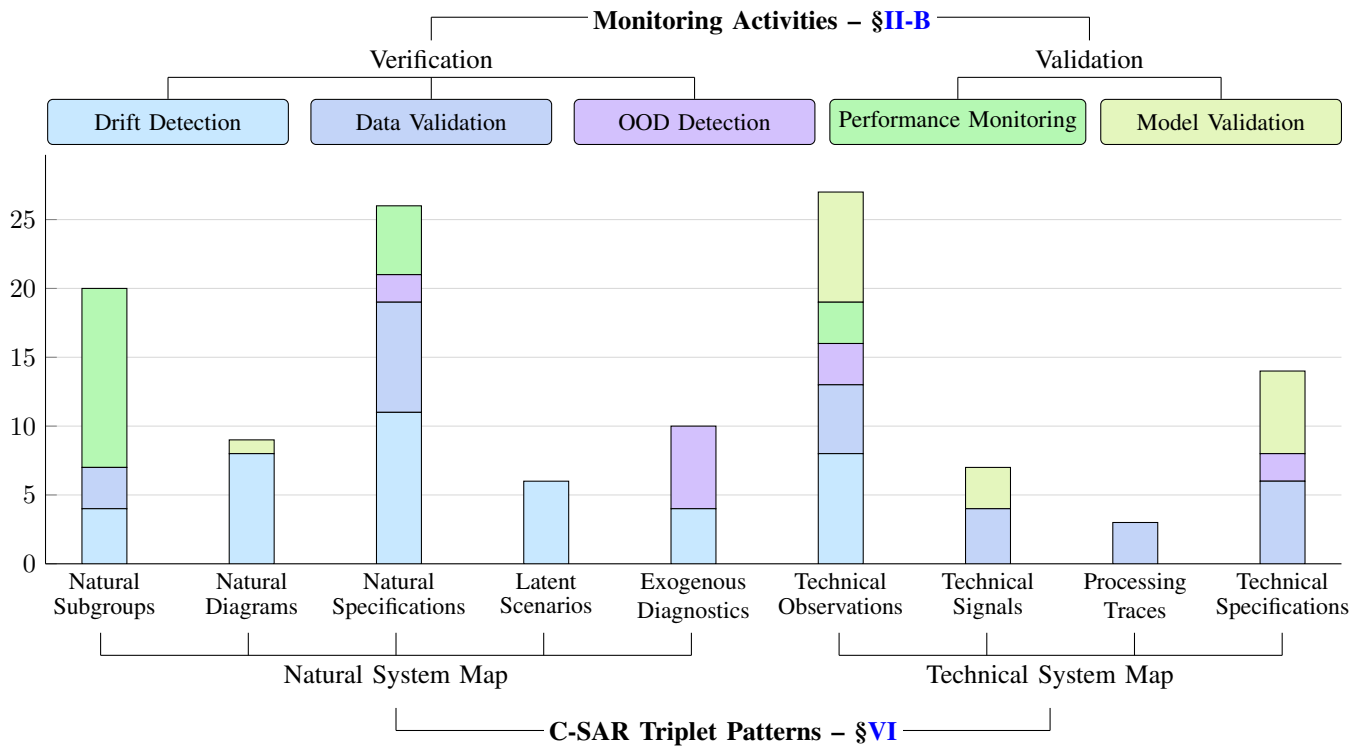
\begin{figure*}[!b]
\centering
\begin{minipage}{\textwidth}
\begin{tikzpicture}
\begin{axis}[
    width=0.95\textwidth,
    height=0.6\textwidth,
    yscale=0.5,
    scale only axis,
    ybar stacked,
    bar width=17pt,
    symbolic x coords={\shortstack{Natural\\Subgroups},\shortstack{Natural\\Diagrams},\shortstack{Natural\\Specifications},\shortstack{Latent\\Scenarios},\shortstack{Exogenous\\Diagnostics},\shortstack{Technical\\Observations},\shortstack{Technical\\Signals},\shortstack{Processing\\Traces},\shortstack{Technical\\Specifications}},
    xtick=data,
    xlabel=,
    ylabel=,
    enlarge x limits=0.05,
    x tick label style={font=\small},
    ymin=0,
    axis x line*=bottom,
    axis y line*=left,
    grid=major,
    grid style={gray!30},
    ymajorgrids=true,
    xmajorgrids=false,
]

\addplot+[draw=black, fill=featureColor!30, forget plot] coordinates {(\shortstack{Natural\\Subgroups},4) (\shortstack{Natural\\Diagrams},8) (\shortstack{Natural\\Specifications},11) (\shortstack{Latent\\Scenarios},6) (\shortstack{Exogenous\\Diagnostics},4) (\shortstack{Technical\\Observations},8) (\shortstack{Technical\\Signals},0) (\shortstack{Processing\\Traces},0) (\shortstack{Technical\\Specifications},0)};
\addplot+[draw=black, fill=inferenceColor!30, forget plot] coordinates {(\shortstack{Natural\\Subgroups},3) (\shortstack{Natural\\Diagrams},0) (\shortstack{Natural\\Specifications},8) (\shortstack{Latent\\Scenarios},0) (\shortstack{Exogenous\\Diagnostics},0) (\shortstack{Technical\\Observations},5) (\shortstack{Technical\\Signals},4) (\shortstack{Processing\\Traces},3) (\shortstack{Technical\\Specifications},6)};
\addplot+[draw=black, fill=applicationColor!30, forget plot] coordinates {(\shortstack{Natural\\Subgroups},0) (\shortstack{Natural\\Diagrams},0) (\shortstack{Natural\\Specifications},2) (\shortstack{Latent\\Scenarios},0) (\shortstack{Exogenous\\Diagnostics},6) (\shortstack{Technical\\Observations},3) (\shortstack{Technical\\Signals},0) (\shortstack{Processing\\Traces},0) (\shortstack{Technical\\Specifications},2)};
\addplot+[draw=black, fill=naturalColor!30, forget plot] coordinates {(\shortstack{Natural\\Subgroups},13) (\shortstack{Natural\\Diagrams},0) (\shortstack{Natural\\Specifications},5) (\shortstack{Latent\\Scenarios},0) (\shortstack{Exogenous\\Diagnostics},0) (\shortstack{Technical\\Observations},3) (\shortstack{Technical\\Signals},0) (\shortstack{Processing\\Traces},0) (\shortstack{Technical\\Specifications},0)};
\addplot+[draw=black, fill=externalColor!30, forget plot] coordinates {(\shortstack{Natural\\Subgroups},0) (\shortstack{Natural\\Diagrams},1) (\shortstack{Natural\\Specifications},0) (\shortstack{Latent\\Scenarios},0) (\shortstack{Exogenous\\Diagnostics},0) (\shortstack{Technical\\Observations},8) (\shortstack{Technical\\Signals},3) (\shortstack{Processing\\Traces},0) (\shortstack{Technical\\Specifications},6)};
\end{axis}

% Custom legend with rounded corners
\hspace{0.2cm}
\node[above] at (current bounding box.north) {
    \begin{tikzpicture}[every node/.style={font=\small}]

    % Calculate total width (0.95\textwidth to match the axis width)
    \def\totalwidth{17.5cm}  % This should match your 0.95\textwidth
    \def\spacing{0.3cm}     % Space between boxes
    
    % Calculate individual box widths (total width minus 4*spacing, divided by 5)
    \def\boxwidth{3.2cm}    % (\totalwidth - 4*\spacing) / 5
    % Position boxes with even spacing using the new color order and rounded corners
    \node[draw=black, rounded corners=2pt, fill=featureColor!30, minimum width=\boxwidth, minimum height=0.6cm] (drift) 
        at (0,0) {Drift Detection};
    \node[draw=black, rounded corners=2pt, fill=inferenceColor!30, minimum width=\boxwidth, minimum height=0.6cm] (validation)
        at (\boxwidth + \spacing,0) {Data Validation};
    \node[draw=black, rounded corners=2pt, fill=applicationColor!30, minimum width=\boxwidth, minimum height=0.6cm] (ood)
        at (2*\boxwidth + 2*\spacing,0) {OOD Detection};
    \node[draw=black, rounded corners=2pt, fill=naturalColor!30, minimum width=\boxwidth, minimum height=0.6cm] (performance)
        at (3*\boxwidth + 3*\spacing,0) {Performance Monitoring};
    \node[draw=black, rounded corners=2pt, fill=externalColor!30, minimum width=\boxwidth, minimum height=0.6cm] (model)
        at (4*\boxwidth + 4*\spacing,0) {Model Validation};

    % Draw vertical lines from the center of each box
    \draw[black] (drift.north) -- (drift.north |- {0,0.9});
    \draw[black] (validation.north) -- (validation.north |- {0,0.9});
    \draw[black] (ood.north) -- (ood.north |- {0,0.9});
    \draw[black] (performance.north) -- (performance.north |- {0,0.9});
    \draw[black] (model.north) -- (model.north |- {0,0.9});

    \draw[black] (0,0.9) -- (7,0.9);
    
    \draw[black] (10.5,0.9) -- (14,0.9);

    \draw[black] (3.5,1.7) -- (12.25,1.7);
    
    \node[above, fill=white, draw=none, inner sep=2pt] at (7.9,1.4) {\normalsize\textbf{Monitoring Activities -- \S\ref{sec:monitoring_activities}}};

    \node[above] (var) at (3.5,0.9) {\normalsize Verification};
    \node[above] (val) at (12.25,0.9) {\normalsize Validation};

     \draw[black] (var.north) -- (var.north |- {0,1.7});
    \draw[black] (val.north) -- (val.north |- {0,1.7});

    \end{tikzpicture}
};
\end{tikzpicture}
\end{minipage}

\begin{minipage}{\textwidth}
\centering
\hspace{0.1cm} 
\begin{tikzpicture}[xscale=1.87]
% Draw 9 vertical lines spaced evenly
\foreach \x in {0,1,2,3,4,5,6,7,8} {
    \draw[black] (\x*1.05cm + 1.8cm, 0) -- (\x*1.05cm + 1.8cm, -0.3cm);
}
% First line connecting first 5 vertical lines
\draw[black] (1.8cm, -0.3cm) -- (6.0cm, -0.3cm);
% Second line connecting last 4 vertical lines
\draw[black] (7.05cm, -0.3cm) -- (10.2cm, -0.3cm);
% Add text below the lines
\node[below] at (3.9cm, -0.3cm) {Natural System Map};
\node[below] at (8.55cm, -0.3cm) {Technical System Map};
% Add second level vertical lines (0.3cm long)
\draw[black] (3.9cm, -1.0cm) -- (3.9cm, -1.3cm);
\draw[black] (8.55cm, -1.0cm) -- (8.55cm, -1.3cm);
% Add horizontal line connecting second level verticals
\draw[black] (3.9cm, -1.3cm) -- (8.55cm, -1.3cm);
% Add text below the second horizontal line
\node[below, fill=white, draw=none, inner sep=2pt] at (6.225cm, -1.1cm) {\textbf{C-SAR Triplet Patterns -- \S\ref{sec:analysis}}};

\end{tikzpicture}
\end{minipage}
\caption{Distribution of C-SAR triplet patterns across monitoring activities. The x-axis depicts the patterns, and the stacked bars indicate how often each appears in different monitoring activities, with each color corresponding to a specific activity.}
\label{fig:bar-chart}
\end{figure*}

%% file: sections/discussion.tex
\section{Discussion} \label{discussion}

In this section, we revisit our key contributions and examine their broader implications for future research on context-aware ML monitoring. In particular, we consider how the C-SAR framework can serve as both an analytical tool and a conceptual model for structuring and integrating contextual information in ML monitoring. We reflect on our study’s contributions, validity, limitations, and implications for tool support. We then conclude with a discussion of future directions, including the need for empirical evaluations.

\subsection{Reflection}
\label{sec:key_contributions}

This study presented three core contributions, each addressing one of our research questions (RQ) stated in Section \ref{research_questions}.

\textbf{We developed a transparent and interpretable systematic review approach to examine literature across heterogenous research communities.} 
We developed a systematic review method that maps terminology across domains \ref{sec:search_string_and_terms}, and uses NLP-based semantic filtering with cluster-specific thresholds for study selection\ref{sec:study_select}. While large language models in systematic reviews have recently gained attention \cite{huaaccelerating}, our method offers a practical approach through its two-step process and deliberate threshold calibration.

\textbf{We introduced the C-SAR framework that systematically describes and organizes contextual information for ML monitoring along three complementary dimensions.} 
As detailed in Section~\ref{sec:theory}, these dimensions are: (1) \emph{system} -- identifying where the contextual information originates, whether in the natural data-generating environment or the technical implementation; (2) \emph{aspect} -- specifying what type of information is captured, from runtime states to structural relationships to prescriptive properties; and (3) \emph{representation} -- describing how this information is encoded, ranging from formal mathematical constructs to semi-structured formats.

\textbf{We identified recurring C-SAR triplet patterns and analyzed how they support the various activities in ML monitoring workflows}.
We analyzed many-to-many relationships between the dimensions of the C-SAR framework, resulting in recurring triplet patterns that capture how context is structured in ML monitoring (Section~\ref{sec:analysis}). These patterns, in turn, map onto a range of monitoring activities -- such as drift detection, data validation, and model evaluation -- showing how different configurations of contextual information support different monitoring goals. This mapping provides a unifying structure for understanding context-aware ML monitoring approaches.

These three contributions summarize the main outcomes of our study. Below, we briefly reflect on the process that led to the formulation of the framework.

The C-SAR framework emerged through iterative synthesis during our review, grounded in the recognition that the ML monitoring literature is highly fragmented. Across initial pilot studies, we observed that different communities emphasized isolated system elements -- some focusing on input features, others on data processing or downstream effects -- often shaped by domain and subsystem-specific concerns. While all studies incorporated some form of contextual information, there was no shared structure to explain how these related. We realized early on that a descriptive model would need to capture this variance, but also to connect it, in order to support a holistic understanding of context information in ML monitoring.

Our thinking was informed by prior work in context modeling within software engineering, particularly the distinction between internal system components and external environmental factors \cite{bedjeti2017modeling,dey2001conceptual,cabrera2017ontology}. Initial attempts to capture contextual elements within a single taxonomy proved insufficient, as differences in what information was being described (e.g., states, properties, relations) were entangled with where in the system it occurred, obscuring meaningful variation. Separating these concerns led us to introduce the \textit{system} and \textit{aspect} dimensions, allowing contextual elements to be described independently of their location.

Later in the synthesis process, we noticed that even when the same type of information described the same part of a system, it could be expressed in fundamentally different ways -- such as a formal metric, an engineering heuristic, or a textual note. This observation motivated the addition of the \textit{representation} dimension, which added a practicable layer to the framework that captured the operationalization of context information in monitoring workflows.

As we refined the framework, we drew inspiration from systems thinking, which centers around interdependence, behavior, and holistic modeling \cite{cabrera2023systems}. However, systems thinking offered limited operational guidance for distinguishing types and forms of information. C-SAR inherits this systems orientation, but adds the granularity and descriptive structure needed for practical application in ML monitoring workflows.

\subsection{Implications for Tooling Support}

The C-SAR framework provides a conceptual model that can guide the design of next-generation MLOps tooling. By analyzing current tools through the C-SAR lens, we can identify opportunities for immediate integration as well as limitations that suggest a need for a new architectural approach.

\subsubsection{Integration with modern MLOps platforms}
Modern MLOps platforms offer a starting point for implementing C-SAR patterns, as their core function is to structure and manage the exact workflows defined by the \textbf{processing pipeline} and \textbf{inference pipeline} subsystems of C-SAR. For example, TensorFlow Extended (TFX) \cite{tfx} natively represents these pipelines as a Directed Acyclic Graph (DAG) of components that produce versioned artifacts. In contrast, MLflow \cite{mlflow}, which focuses on tracking experiment artifacts and metrics, is often paired with an external orchestrator like Apache Airflow \cite{airflow} to define and execute such pipeline graphs. Within these systems, contextual information is treated as metadata attached to specific artifacts or execution runs.

Several C-SAR patterns can be mapped onto their existing metadata models. \textbf{Technical observations} such as parameters or component versions can be logged as metadata, directly corresponding to \textbf{processing} and \textbf{inference diagnostics}, while model quality metrics can be mapped to \textbf{inference assessments}. \textbf{Natural} and \textbf{technical specifications} can be integrated into the workflow. TFX schemas, for example, can encode both technical constraints (\textbf{processing assertions}) and semantic constraints (\textbf{reference assertions}). While MLflow lacks native schema enforcement, validation rules can be logged as artifacts and applied via custom wrappers. These platforms can automate monitoring by comparing runtime data against this stored context, thereby operationalizing a subset of C-SAR patterns.

\subsubsection{Systems as first-class entities in tooling design}
A core limitation of current tooling is that contextual information is treated as ephemeral metadata scoped to a specific execution or artifact. The C-SAR framework suggests that system elements -- from the \textbf{natural system} to downstream \textbf{applications} -- should be treated as persistent, first-class entities that exist independently of any single model or pipeline run.

While TFX schemas are expressive, they remain tightly coupled to datasets, and the platform lacks native mechanisms for natural-system relationships, like causality, or for the inclusion of external an latent variables. Custom tags in MLflow are free-form -- no enforced semantics or schema -- so the resulting metadata isn’t machine-actionable for automation at scale. Similarly, metadata platforms like Amundsen \cite{amundsen} and DataHub \cite{datahub} reflect the growing adoption of graph-based metadata tracking to support observability of upstream dependencies. However, their primary focus remains on data management rather than holistic ML system tracking.

\textcolor{blue}{C-SAR suggests redesigning tooling around holistic graph-based metadata models of the ML system and its environment.} In this model: Nodes would represent persistent system elements (e.g., source data, features, models, downstream policies), while attributes on these nodes would explicitly encode C-SAR triplets, capturing their contextual properties (e.g., nominal constraints, structural relations).

This shift from a run-centric to a system-centric architecture would create a persistent, meta-level model of the system, which is populated and updated by the outputs of individual runs. This enables automation driven by holistic contextual relationships. For example, monitors could be configured to automatically inherit \textbf{reference assertions} for new models using the same features, adapt drift thresholds based on version changes in the processing pipeline, or apply constraints only under specific \textbf{exogenous diagnostics}. Such an architecture could form a more foundational basis for systematic, reusable, and interpretable ML monitoring.

\subsection{Scaling Context Modelling in Practice}
While our study focused on identifying and structuring contextual information, we did not evaluate where context modeling is most beneficial or how its use should scale. However, in large organizations, scaling context modeling can be expected to present practical challenges. Contextual knowledge is often distributed across roles -- data engineers handle ETL processes, ML engineers manage model pipelines, domain experts understand underlying processes, and application teams interpret outcomes. Coordinating this knowledge across teams and systems may be difficult, especially as organizational complexity increases.

Efforts to operationalize context at scale are likely to depend on shared practices and infrastructure -- such as metadata management, elicitation tooling, and consolidation of system maps that span team boundaries. While these strategies lie beyond the scope of this review, we see them as important enablers of scalable context-aware monitoring. As operational concerns become more complex, context modeling efforts may need to grow accordingly, but this potential benefit should be weighed against the costs of effort. In some settings, simpler heuristic-based approaches may remain sufficient.

\subsection{Risks and Limitations for Adoption}

The C-SAR framework is descriptive and synthesized from contextual information reported in the literature, not necessarily what constitutes a prescriptive engineering standard. It should not be interpreted as a prescriptive guide. The effectiveness of the identified patterns has not been empirically evaluated or compared, which is a necessary step for future work. In Section \ref{eval_csar}, we outline a research agenda to provide the empirical evidence required before C-SAR can be recommended as best practice.

One limitation concerns the treatment of time. While the \textit{aspect} taxonomy implicitly includes temporal properties of events—such as whether a change is gradual or abrupt, and how long it persists -- most studies focused on static descriptions of the event itself, without explicitly modeling its temporal dynamics. Monitoring, however, is inherently a time-series task. We do not view this as a shortcoming of C-SAR’s current, categorization-focused scope, but we note that temporal handling is not yet made explicit -- something one would expect in a more prescriptive methodology. For C-SAR to move beyond description, we expect it will need a supplemental temporal vocabulary and convention.

A second limitation concerns the boundary C-SAR draws between the \textit{natural system} -- the real-world processes that generate data -- and the \textit{technical system} that hosts the model. In cyber-physical domains, this boundary can become counter-intuitive. Imagine a robot whose drive wheel breaks and slips, because the failure alters the robot’s motion -- the process that generates its sensor inputs -- C-SAR classifies the event as a \textit{natural system} change, even though the wheel is part of the robot’s hardware. This reflects that C-SAR classifies context from the monitored model’s perspective: wheel slip arises in the vehicle’s physical hardware and is therefore part of the external (natural) environment, not a fault in the internal (technical) system -- the model's serving infrastructure or the downstream ML-enabled navigation application. In multi-agent systems, similar ambiguity can arise. C-SAR does not prescribe whether it should be instantiated at the level of a single agent or across multiple agents. From one agent’s perspective, another agent’s behavior appears as part of the environment and is treated as natural context -- even if both agents are part of a shared architecture. These edge cases do not violate the framework but suggest that additional guidance may be needed for some domains.

\section{Threats to Validity}
\label{sec:validity_reflection}

In line with recent calls to consider validity explicitly  as part of the research design \cite{lago2024threats}, we addressed threats directly within our study protocol. As such, threats related to the review process itself, namely to \textbf{study selection validity} and \textbf{data validity} \cite{ampatzoglou2019identifying}, were primarily mitigated within the search, selection, and synthesis protocol as described in Section~\ref{methodology}.

This section reflects on the remaining overarching threats to \textbf{research validity} \cite{ampatzoglou2019identifying}, which concerns the overall research design and the generalizability of our findings.

\subsection{Soundness of Synthesis}
A threat to \textit{internal validity} is the soundness of our synthesis process -- the confidence that our C-SAR framework is a credible conclusion derived from the primary studies. The primary concern is researcher bias, where subjective interpretation during qualitative coding could lead to an arbitrary framework. We mitigated this through: (1) a protocol that included a systematic open coding process, with categories refined by multiple authors through multiple iterations to ensure mutual exclusivity; and (2) ensuring the traceability of our findings by grounding every element of the C-SAR framework and its patterns in specific examples from the primary studies. Despite these measures, qualitative synthesis is inherently subjective, and this threat cannot be entirely eliminated.

\subsection{Generalizability of Findings}
External validity concerns whether the C-SAR framework and its patterns are applicable beyond the primary studies analyzed. To mitigate this threat, we designed a high-coverage search across multiple databases (Scopus, IEEEXplore, ACM) and research communities (e.g., data mining, software engineering, and ML), ensuring our sample was not confined to a single research silo. The robustness of this strategy is confirmed by our findings: the patterns identified recur across different monitoring activities, research communities, and diverse application domains such as finance, healthcare, and autonomous driving (see Section \ref{csar_activities} and \ref{domain_cases}). This cross-cutting evidence strengthens the framework's claim to generalizability across the broader field of ML monitoring as studied in the academic literature.

\subsection{Limitations}
Some limitations to our study remain, which include the following:

\textbf{Tooling and reproducibility.} A limitation of this study is the use of a proprietary model for semantic filtering. At the time of selection, it was the most capable option available, offering strong performance and long input support. We compared it to the strongest open-source alternative then available, \textit{BGE-M3}, and found a meaningful gap in performance that supported our choice. As open-source models continue to improve, we recommend that future work explicitly compare open and closed alternatives, as the trade-offs may shift significantly over time. Benchmarks such as \textit{MTEB} \cite{mteb} provide valuable guidance for such decisions.

\textbf{Traceability of exploratory search.}
While our initial exploratory search process to collect relevant ML monitoring terminology was iterative and guided by saturation, we did not maintain a structured or version-controlled record of how the search terms evolved. The refinements emerged through backward snowballing and thematic consolidation, but only the final set of terms and their rationale are reported (Fig. \ref{fig:search_string_and_terms}). This limits traceability, and we encourage future reviews to document and archive such term evolution.

\textbf{Exclusion of grey literature}
Our focus on peer-reviewed publications excludes grey literature such as practitioner reports, technical blogs, industry white papers, and experience reports from practical settings. Consequently, while our framework systematically maps the state-of-the-art in academic research, it may not fully represent the current state of practice documented in these sources.

\section{Future Work}
\label{sec:implications_future}

The C-SAR framework provides a conceptual model for understanding how contextual information is structured and utilized in ML monitoring. As a literature-derived conceptual model, its practical value must first be validated before further applications can be developed. This section outlines a research agenda, starting with the required empirical evaluation and then discusses promising future directions.

\subsection{Evaluation of the C-SAR framework} \label{eval_csar}
The C-SAR framework is a conceptual model designed to structure the use of contextual information in ML monitoring. To transition it into a practical engineering tool, we propose a three-phase validation agenda that will first validate it as a descriptive model (Phases 1-2) and then prove its value as an applicable tool for systematic monitoring practice (Phase 3).

\textbf{Phase 1: conceptual grounding and validation.} The initial phase will validate the framework's conceptual integrity as a descriptive model through structured interviews and focus groups. Practitioners will map real-world monitoring scenarios and the contextual information they relied upon, allowing us to assess the framework's \textit{completeness}, \textit{usability}, and \textit{granularity}. This phase yields a refined and practitioner-grounded version of the C-SAR framework, confirming its concepts accurately reflect the realities of ML monitoring.

\textbf{Phase 2: system mapping and guideline development.} Building on the refined framework from Phase 1, this step uses retrospective case studies for holistic validation and to elicit practice-informed guidelines. By instantiating holistic C-SAR system maps of real-world cases across diverse domains and analyzing historical monitoring scenarios, we will map successful practitioner workflows to specific C-SAR patterns. This process yields twofold results: (1) validated, real-world system maps that confirm the framework's descriptive power; (2) an initial catalog of \textit{prescriptive guidelines} for applying C-SAR to common monitoring challenges.

\textbf{Phase 3: application and framework utility testing.} The final phase assesses the framework's prescriptive value by testing whether C-SAR enables practitioners to more systematically structure, identify, and configure a monitoring system. In a prospective action study, a team will be guided by the guidelines from Phase 2 to systematically elicit their C-SAR system maps -- collaborating with domain experts for the \textit{natural system map} and analyzing technical artifacts for the \textit{technical system map}. The framework's practical utility will then be validated by its impact on monitoring outcomes, such as alert quality and diagnostic speed. This will provide the empirical validation of the framework's practical utility, confirming C-SAR's role as an actionable engineering tool.

\subsection{Future Research Directions}
In addition to empirical evaluation, we identify several future research directions to extend the C-SAR framework and support its practical adoption.

\textbf{Tooling for context-aware monitoring.} The structured nature of C-SAR triplets provides a foundation for reusable and goal-driven tooling. A Domain-Specific Language (DSL) could formalize the specification of contextual elements — such as system components, feature slices, expected behaviors, and alerting conditions -- into a machine-readable format. Combined with a Model-Driven Engineering (MDE) approach, this would support transparent, declarative monitoring configurations that can be compiled into executable monitors aligned with system goals.

A causal monitoring engine is a natural candidate to interpret such specifications. Structural relationships define causal pathways between technical components and natural variables; logical assertions constrain relationships to ensure identifiability; and probabilistic specifications define priors over parameters or outcomes. Together, these contribute to a causal model that unifies the technical data flow with the structure of the natural system. We consider C-SAR a strong foundation for developing such a unified methodology.

\textbf{Benchmarking and testbed development.} To evaluate context-aware monitoring more effectively, benchmarks should include structured metadata -- such as distributional properties and causal relationships -- to make underlying assumptions explicit. Beyond static datasets, holistic testbeds should simulate full system dynamics: a natural data generator (e.g., a Bayesian network modeling latent and external factors \cite{poropudas2011simulation}) connected to a technical pipeline for processing, inference, and application. Model outputs should feed back into the generator, reflecting how deployed systems shape future inputs. Such setups enable controlled, reproducible experiments that assess how monitoring strategies respond to context-rich conditions.

\textbf{Cataloging monitoring challenges and failure modes.} Mapping real-world failure modes to recurring C-SAR triplets can inform targeted monitoring strategies. For example, delayed label feedback may be addressed using auxiliary supervision via \textit{inference assessments}; ambiguous drift signals can be clarified through \textit{reference diagrams} and \textit{assertions}; and fairness concerns require \textit{exogenous identifiers} to support subgroup analysis. A structured catalog of such mappings would aid risk-informed monitor design and highlight gaps in current monitoring practices.

\textbf{Structured elicitation of C-SAR triplets.} Since contextual knowledge is often fragmented, structured methods are needed to elicit and maintain C-SAR-based system maps. Different triplet components align with different organizational roles: domain experts hold natural-system knowledge, data platform teams manage preprocessing logic, and MLOps tools (e.g., MLflow) trace inference pipelines. Future work could explore elicitation templates, collaborative workflows, and integration mechanisms to support context modeling across teams.

\textbf{Context-informed testing and evaluation.} Significant overlaps exist between monitoring, testing, and evaluation practices. C-SAR could inform more systematic test generation by embedding contextual assumptions — such as causal diagrams or processing logic — into test scenarios. For example, “what-if” tests \cite{biswas2021fair} can be derived from these specifications to assess model robustness. Reference assertions and failure patterns could also be reused in offline testing to simulate plausible failure conditions observed at runtime.

%% file: sections/conclusion.tex
\section{Conclusion} \label{conclusion}

The C-SAR framework offers a structured way to think about context in ML monitoring -- not just as metadata around features and predictions, but as a reflection of the systems models operate within. Our review shows that context can be understood as the relevant systems surrounding a model, and the information we have about their elements. By organizing this information, C-SAR supports more systematic monitoring practices that go beyond intuition and tribal knowledge. While we see clear potential for practical benefits -- like reducing alert fatigue, supporting root-cause analysis, and improving knowledge sharing -- these need validation in real-world settings. Our next steps focus on validating C-SAR in practice, towards developing more reusable, interpretable monitoring tools.